\documentclass[12pt]{article}%
\usepackage{amsmath,amssymb,amsthm,amsfonts}
\usepackage{wasysym}
\usepackage{graphicx}
\usepackage{xcolor}
\usepackage{stackengine}

\newcounter{subfigure}
\usepackage[colorlinks]{hyperref}
\usepackage{multirow}

\usepackage{geometry}
\usepackage{appendix}

\theoremstyle{definition}
\newtheorem{definition}{Definition}

\newcommand{\ea}{\textit{et al. }}

\newcommand{\rd}{r_\text{death}}
\newcommand{\pd}{P_\text{dead}}

\renewcommand{\epsilon}{\varepsilon}
\newcommand{\OD}{\text{OD}}

\begin{document}

\title{Modeling of drug diffusion in a solid tumor leading to tumor cell death}

\author{Aminur Rahman\thanks{Corresponding Author, \url{amin.rahman@ttu.edu}}
\thanks{Department of Mathematics and Statistics, Texas Tech University},
Souparno Ghosh\footnotemark[2], Ranadip Pal\thanks{Department of Electrical and Computer Engineering,
Texas Tech University}}

\date{}
\maketitle

\begin{abstract}
It has been shown recently that changing the fluidic properties of a drug can improve its efficacy
in ablating solid tumors.  We develop a modeling framework for tumor ablation, and present the simplest
possible model for drug diffusion in a spherical tumor with leaky boundaries and assuming cell death
eventually leads to ablation of that cell effectively making the two quantities numerically equivalent.  The
death of a cell after a given exposure time depends on both the concentration of
the drug and the amount of oxygen available to the cell.  Higher oxygen availability leads to cell
death at lower drug concentrations.  It can be assumed that a minimum concentration is
required for a cell to die, effectively connecting diffusion with efficacy.  The concentration threshold
decreases as exposure time increases, which allows us to compute dose-response curves.  Furthermore,
these curves can be plotted at much
finer time intervals compared to that of experiments, which is used to produce a
dose-threshold-response surface giving an observer a complete picture of the drug's efficacy for an individual. 
In addition, since
the diffusion, leak coefficients, and the availability of oxygen is different for different individuals and tumors,
we produce artificial replication data through bootstrapping to simulate error.
While the usual data-driven model with Sigmoidal curves use 12 free parameters, our
mechanistic model only has two free parameters, allowing it to be open to scrutiny rather than
forcing agreement with data.  Even so, the simplest model in our framework, derived here, shows
close agreement with the bootstrapped curves, and reproduces well established relations, such
as Haber's rule.

\end{abstract}



\section{Introduction}
\label{Sec: Intro}

Historically, cancer has been treated using either generic-global drugs or by cutting away the infected
cells via surgery.  While success rates have increased, these types of treatments tend to have unwanted
side-effects and are often quite expensive.  This necessitates a new paradigm for cancer treatment.
In recent years there has been a shift towards developing individualized and targeted drug therapy
\cite{Schork2015, Ciardiello2016}.  However, there are still many technical and financial hurdles \cite{Hayes2013}
to overcome before truly personalized medicine can be implemented in real-life situations.  In an
effort to alleviate these hurdles, data analysis techniques with statistical models have been employed \cite{HRGP2015}.
With enough accurate data these models are able to make remarkable predictions, however one cannot expect
the patient specific data to always be available and accurate.  Furthermore, the mechanisms driving
the phenomena are not captured in these statistical models. A reliable mechanistic modeling framework
may alleviate the financial hurdles by simulating various treatment options, based on the physical rather than statistical
properties of a patient's tumor, before administering drugs.  While data analysis has yielded promising
results and should still be employed whenever possible, mechanistic models serve as another
weapon in the fight against cancer.

In the seminal work of Sugiura \ea \cite{SOOH1983} it was shown that ethanol can successfully
deteriorate malignant tissue.  In fact, Ryu \ea \cite{Ryu1997} showed a comparable survival rate
to surgery in a statistical survey of hepatocellular carcinoma patients treated with ethanol
injections.  As with any type of treatment, safety supersedes efficacy.  An early work on the safety
of injecting ethanol into parathyroid tumors was conducted by Solbiati \ea \cite{Solbiati1985}.
For the past decade there have been numerous studies on the safety and efficacy of ethanol
ablation for different types of tumors \cite{Jurgensen2006, Artifon2007, Sorajja2008,
KLXXXLYHL09, DeWittMohamadnejad2011}.
Some major drawbacks of the technique include the need for multiphase treatments, large
amounts of fluid, and the rapid escape of ethanol in non-capsulated tumors \cite{Morhard2017}.
Traditionally these drawbacks were greater for larger tumors, but more recently it has been
shown by Kuang \ea \cite{KLXXXLYHL09} that a single-session high-dose ethanol injection
can ablate hepatocellular carcinoma tumors of diameters up to $5$ cm.

Recently, Morhard \ea \cite{Morhard2017} developed a new method employing ethyl cellulose
for which single-phase small-volume treatments suffice to trigger ablation.  Their goal is to maximize
distribution of the ethanol throughout the tumor while minimizing the necessary solution volume.
In essence, by changing the fluidic properties of the drug, they are able to increase efficacy and
decrease toxicity -- the goal of cancer treatments.  While this type of optimization of fluidic
properties has only been used with ethanol injections, one may see the potential of using it for
other drugs as well.  After all, there is a wide variety of drug injection therapies \cite{CSRYL13, BMS14, SCLAMGL18}.

Although it may seem obvious that drug transport occurs as a diffusive process, it is still
worth noting that there is an abundance of drug diffusion models.  Recent review articles have
discussed diffusion-based models of drug distribution from the blood stream into solid tumors
\cite{WaiteRoth12, KGR2013}.  In \cite{KGR2013}, Kim \ea advocate integrating mechanistic
models with clinical data to improve understanding and prediction.  In addition, quite sophisticated transport
models have been developed for penetration into a solid tumor \cite{SoltaniChen2011, SoltaniChen2012, SSRSBBM15}.
However, it is evident that drug distribution from an internal injection has not been investigated
mathematically.  Further, it is often the case that mathematical models describe either drug
transport or cell death, but seldom both.  In this work we endeavor to build a new modeling framework for injected drug
distribution in solid tumors and its effect on tumor cell death, which we hope will facilitate
individualized treatment.  This is done by producing the simplest possible model in this framework,
and laying bare the simplifying assumptions in order to open it to scrutiny and improvement.

The remainder of the paper is organized as follows: we start by discussing the modeling procedure,
the focus of the article, in Sec. \ref{Sec: Diffusion} and \ref{Sec: Cell Death}.  Section \ref{Sec: Diffusion}
contains the diffusion model for drug distribution, and Sec. \ref{Sec: Cell Death} relates the concentration
profile from diffusion to cell death.  While the analytical solutions to the diffusion equation would
generally be a linear combination of Bessel functions, since our initial condition is not a Bessel function
a numerical code would be required to match the coefficients of the Bessel functions.  Instead we
use finite difference methods in Sec. \ref{Sec: Numerics} to numerically solve the equation from
the start, and using this solution for the concentration profile the cell death is computed.
In Sec. \ref{Sec: Comparison} we compare our dose-response curves with data points from experiments
different from that of Morhard \ea \cite{Morhard2017} since they did not include dose-response
curves in their article.  Furthermore, using a different experiment shows the robustness of this
type of modeling.  Finally, we conclude with a discussion about applying our modeling framework to individualized
treatment in Sec. \ref{Sec: Conclusion}.

\section{Radially symmetric concentration diffusion model}\label{Sec: Diffusion}

In this section we derive the simplest possible model for drug diffusion in a tumor with leaky boundaries where
injection occurs at a much faster timescale than diffusion.  First, let us assume a spherical tumor
with constant density.  The drug concentration is given as the ratio of the volume of the drug
to that of the tumor.  Since this is a diffusive process the initial condition is taken as a compact
Gaussian.  Then the leak at the boundary is modeled as ``Newton's law of cooling''\cite{NewtonCooling}.  Finally, we
nondimensionalize the equation and conditions to further simplify the model.

\begin{figure}[htbp]
\centering
\includegraphics[width = 0.9\textwidth]{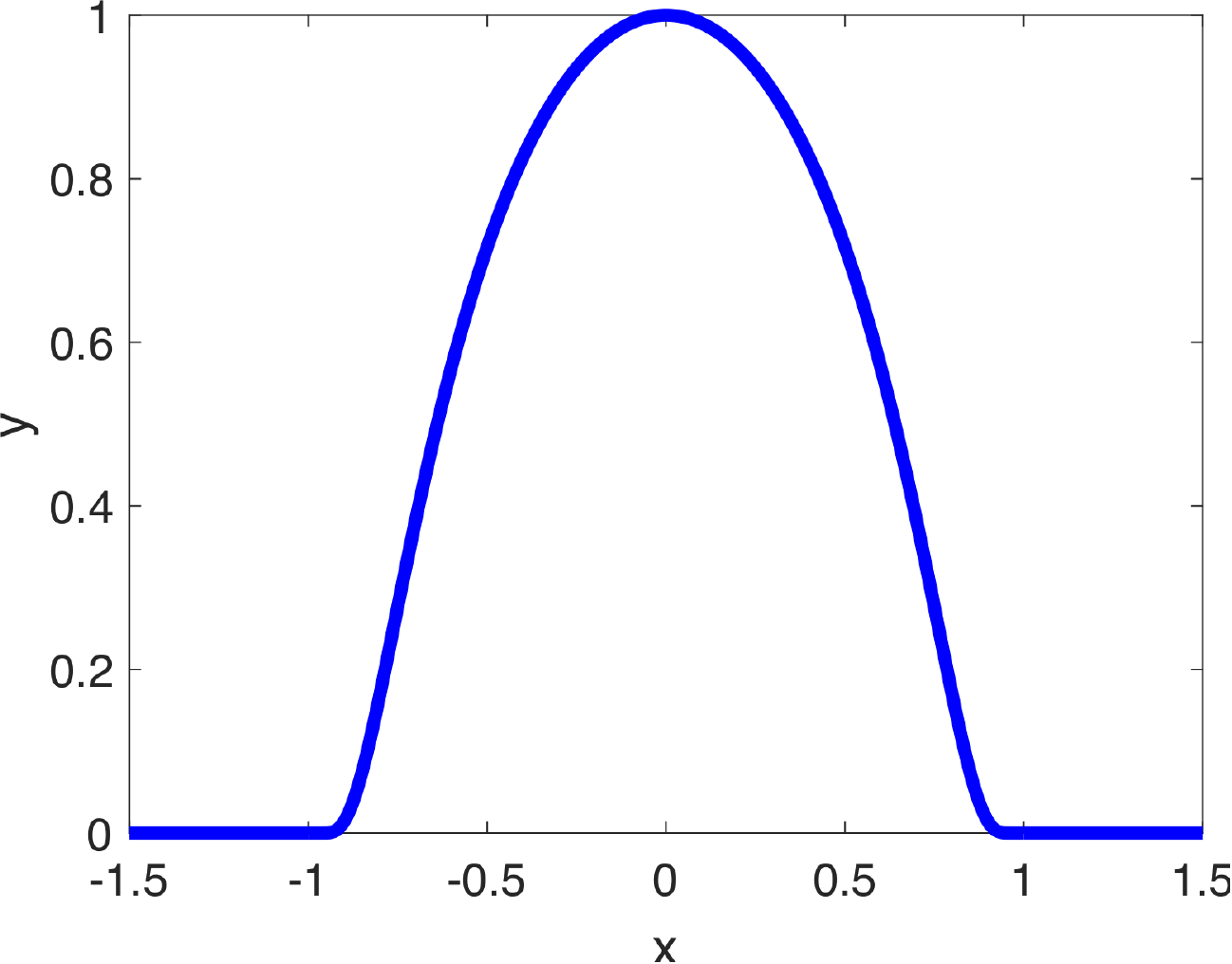}
\caption{A generic bump function with normalized height.}
\label{Fig: Bump}
\end{figure}

\begin{figure}[htbp]
\centering
\includegraphics[width = 0.9\textwidth]{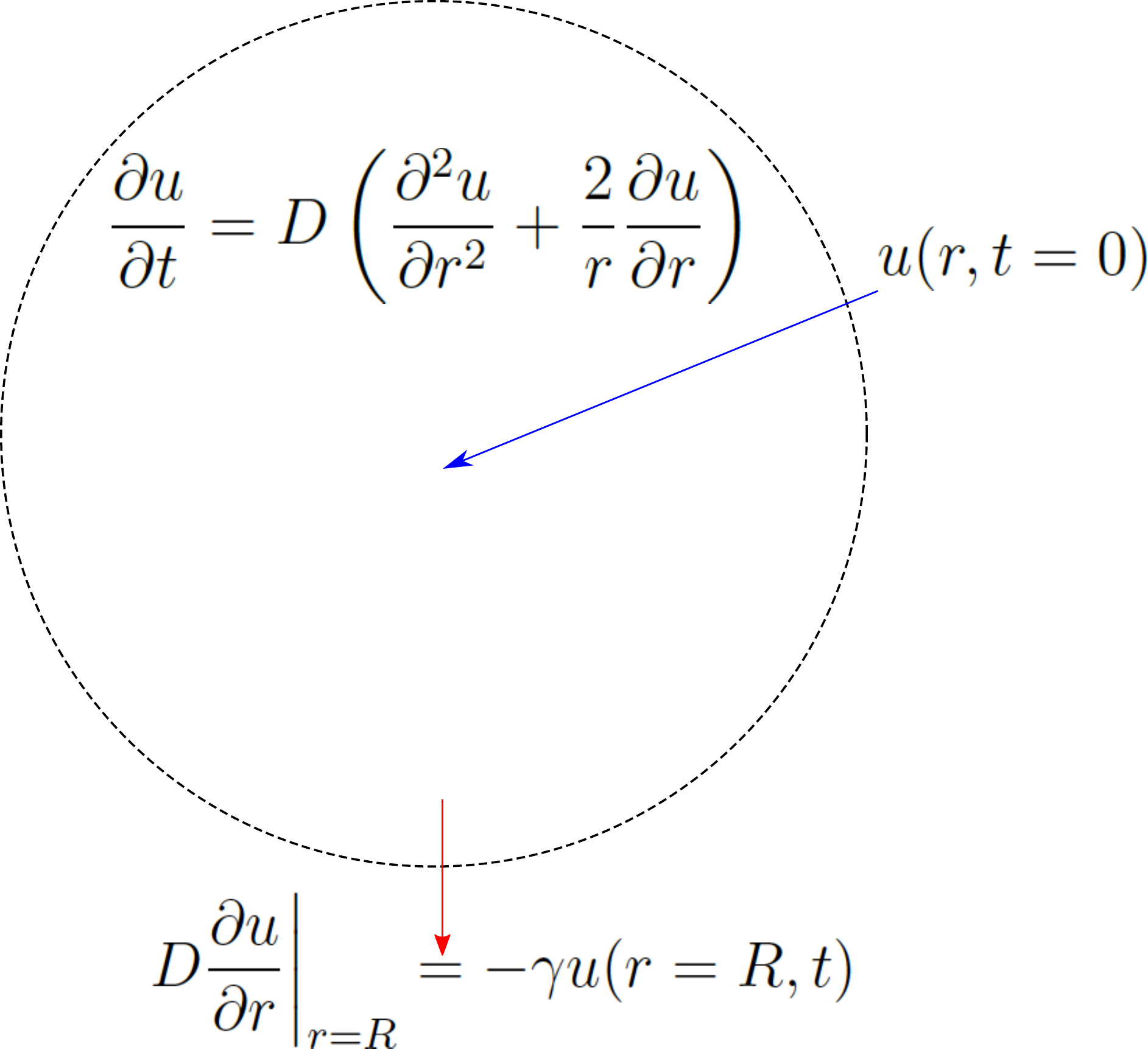}
\caption{Diagram of the diffusion model where the blue arrow represents injection into the tumor and
the red arrow represents leakage out of the tumor.  The tumor is illustrated as a projection of the
sphere into 2-dimensions.}\label{Fig: Model}
\end{figure}

Consider the radially symmetric diffusion equation in spherical coordinates with constant diffusivity
\begin{equation}
\frac{\partial u}{\partial t} = D\left(\frac{\partial^2 u}{\partial r^2} + \frac{2}{r}\frac{\partial u}{\partial r}\right)
\end{equation}
where $u$ is the concentration of the drug and $D$ is the constant diffusivity.
Since we are injecting a volume into
the center of the tumor (blue arrow in Fig. \ref{Fig: Model}) 
and the diffusive timescale is fast, it is reasonable to assume an initial compact
Gaussian distribution (i.e., a \emph{bump function}) where the tail just reaches the boundary
\begin{equation}
u(r,t=0)=
\frac{V_0}{V_b}\begin{cases}
\exp\left(1 - \frac{R^2}{R^2-r^{2}}\right) & \text{for  $r<R$},\\
0 & \text{for  $r\geq R$};
\end{cases}
\label{Eq: simple bump}
\end{equation}
where $R$ is the tumor radius, $V_0$ is the injected volume, and $V_b$ is the volume of the
bump function in spherical coordinates used for the purpose of normalization; that is,
\begin{align*}
V_0 &= \int_0^\pi\int_0^{2\pi}\int_0^R u(r,t=0)dr d\theta d\phi, \text{  when}\\
V_b &= \int_0^\pi\int_0^{2\pi}\int_0^R \exp\left(1 - \frac{R^2}{R^2-r^{2}}\right)dr d\theta d\phi
\end{align*}
The generic bump function takes the 1-dimensional shape shown in Fig. \ref{Fig: Bump}, and by
radial symmetry this can be extended to 3-dimensions.
At the boundary we know the drug leaks due to the porosity of the tumor (red arrow in
Fig. \ref{Fig: Model}).  However, diffusion in the tumor is different from diffusion outside
of the tumor.  In fact, the tumor is far denser than healthy cells, and therefore will hold
onto the drug and let it accumulate as it leaks at a much slower rate.  We assume the drugs start leaking
at the boundary of the tumor according to ``Newton's law of cooling''\cite{NewtonCooling},
\begin{equation}
D\frac{\partial u}{\partial r}\bigg|_{r = R} = -\gamma u(r=R,t),
\end{equation}
where $\gamma$ is the \emph{leak coefficient}.

\subsection{Nondimensionalization}

In order to simplify the model and reduce the number of parameters into only the essential
relations we nondimensionalize the problem.  The concentration $u$ (ratio of drug volume
to tumor cell volume) is by definition nondimensional.  Therefore, we need only do a change
of variables on $t$ and $r$.  Let $\hat{r} = r/R$ and $\hat{t} = t/T$, which gives us
the derivatives
\begin{equation*}
\frac{\partial}{\partial t} = \frac{1}{T}\frac{\partial}{\partial \hat{t}};\qquad
\frac{\partial}{\partial r} = \frac{1}{R}\frac{\partial}{\partial \hat{r}}
\Rightarrow \frac{\partial^2}{\partial r^2} = \frac{1}{R^2}\frac{\partial^2}{\partial \hat{r}^2}.
\end{equation*}
Then our PDE becomes
\begin{equation*}
\frac{\partial u}{\partial t} = \frac{DT}{R^2}
\left(\frac{\partial^2 u}{\partial r^2} + \frac{2}{r}\frac{\partial u}{\partial r}\right),
\end{equation*}
and our initial condition becomes
\begin{align*}
u(\hat{r},t=0) &= \frac{\hat{V_0}}{\hat{V_b}}\begin{cases}
\exp\left(1 - \frac{R^2}{R^2-\hat{r}^{2}R^2}\right) & \text{for  $\hat{r}<R/R$},\\
0 & \text{for  $\hat{r}\geq R/R$};
\end{cases}\\
&= \frac{\hat{V_0}}{\hat{V_b}}\begin{cases}
\exp\left(1 - \frac{1}{1-\hat{r}^{2}}\right) & \text{for  $\hat{r}<1$},\\
0 & \text{for  $\hat{r}\geq 1$};
\end{cases}
\end{align*}
where $\hat{V_0}$ is the injected nondimensional volume and $\hat{V_b}$ is the volume of the nondimensional bump
function.  Furthermore, the boundary condition becomes
\begin{equation*}
\frac{D}{R}\frac{\partial u}{\partial r}\bigg|_{\hat{r}=1} = -\gamma u(\hat{r} = 1, t) 
\Rightarrow \frac{\partial u}{\partial r}\bigg|_{\hat{r}=1} = -\frac{R\gamma}{D}u(\hat{r} = 1, t)
= -\epsilon u(\hat{r} = 1, t)
\end{equation*}
where $\epsilon = R\gamma/D$ is the nondimensional leak coefficient.
Removing the hats gives us the full nondimensional problem
\begin{subequations}
\begin{align}
\frac{\partial u}{\partial t} = \frac{\partial^2 u}{\partial r^2} + \frac{2}{r}\frac{\partial u}{\partial r};\qquad
\frac{\partial u}{\partial r}\bigg|_{r=1} = -\epsilon u(r=1,t);\\
u(r,0) = \frac{V_0}{V_b}\begin{cases}
\exp\left(1- \frac{1}{1 - r^2}\right) & \text{for  $r < 1$},\\
0 & \text{for  $r \geq 1$};
\end{cases}\label{Eq: IC}
\end{align}
\label{Eq: Model}
\end{subequations}

\section{Relating concentration to cell death}\label{Sec: Cell Death}

Once we have a model for drug distribution, we may use it to analyze cell death.  In biological
experiments this is done through \emph{dose-response} curves (amount of drug administered
vs percent cell death at a specific time after administration) \cite{Hill1910, Altshuler1981}.  The dose-response
is recorded by measuring the effect of a drug at various doses.  A mathematical model allows us to
do the same, but for far more finely spaced doses and exposure times, thereby allowing an observer to find an optimal
treatment strategy for a specific individual.

While varying oxygen levels are a concern \cite{Pappas2016, GAP16, KHP14, IRP13}, as a first
approximation and ``proof of concept'' model, we assume that the oxygen concentration throughout
the tumor is constant for a given time.  Therefore, the same amount of drug will kill a tumor cell no matter where
in the tumor it is.  We define a drug concentration threshold that is required to kill a cell for a given
time, which is dependent on both the oxygen concentration and dose strength.

\begin{definition}
The minimum concentration $u_T$ required to kill one cell after $T$ hours is said to be the
\emph{concentration threshold} of the tumor at time $T$.
\end{definition}

\begin{figure}[htbp]
\includegraphics[width = 0.9\textwidth]{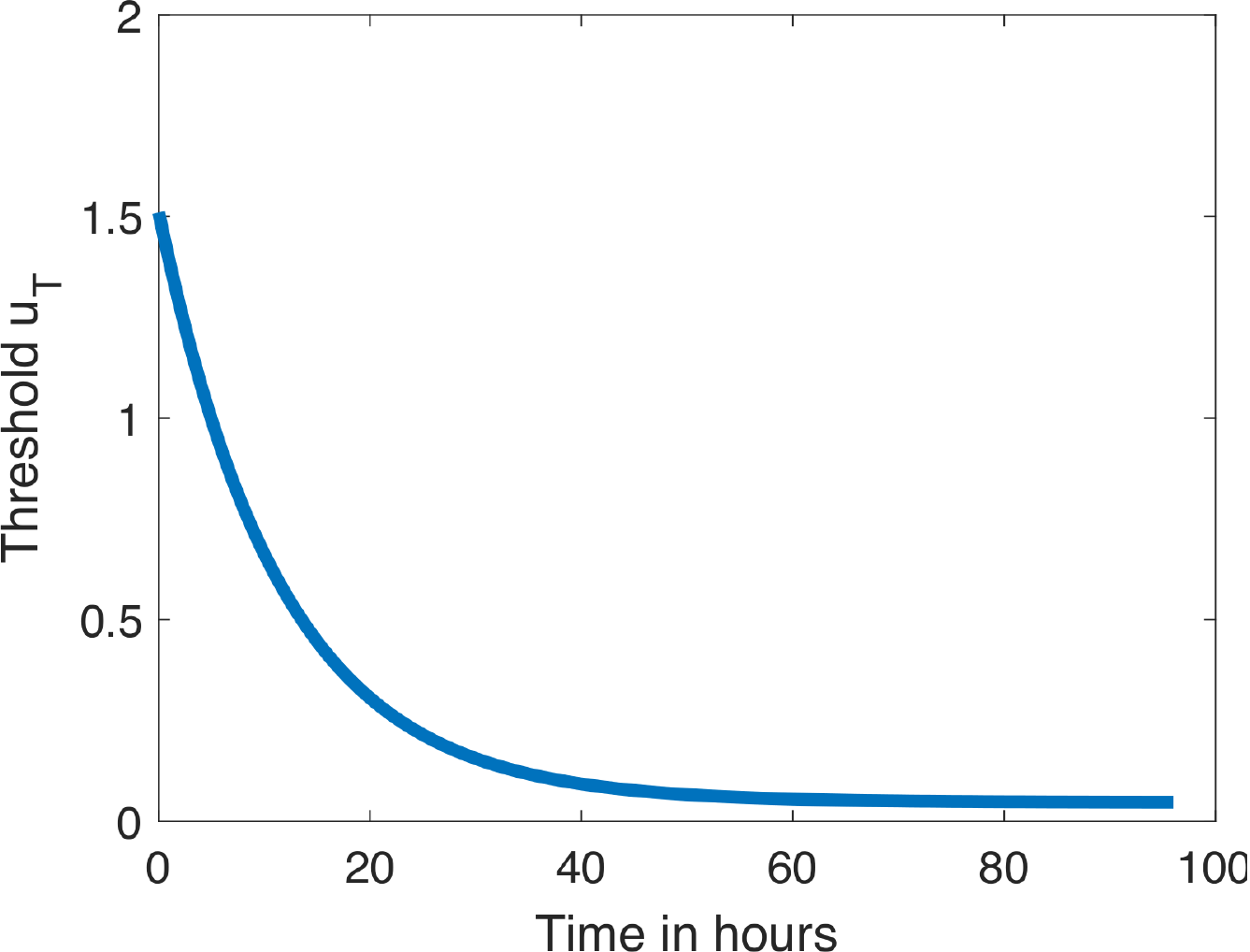}
\caption{Relation between time and concentration threshold from \eqref{Eq: time-threshold}.}
\label{Fig: time-threshold}
\end{figure}

Now we must relate the threshold $u_T$ to time explicitly in order to compare against experimental
data.  Since log time is often used for time-response curves and time-dose-response surfaces, we
assume the threshold to be related to time in a negative exponential manner, $u_T = a - b\exp(-cT)$.
Employing the three most used time points in oncology, $T = 24, 48, 72$, we solve for the constants
at a fixed initial concentration to get
\begin{equation}
u_T = a - be^{-cT};\qquad
a = \frac{u_{24}u_{72} - u_{48}^2}{u_{24} + u_{72} - 2u_{48}};\qquad
b = \frac{(a - u_{24})^2}{a - u_{48}}; \qquad
c = \frac{1}{24}\ln\left(\frac{a - u_{24}}{a - u_{48}}\right).
\label{Eq: time-threshold}
\end{equation}
The plot of \eqref{Eq: time-threshold} is shown in Fig. \ref{Fig: time-threshold}.

This allows us to produce dose-response curves at those three threshold values; i.e. curves that
go through those data points.  Then the other data points are plotted at appropriate concentrations.

\section{Numerical simulation of single-time dose-response curve}\label{Sec: Numerics}

In order to simulate the dose-response curves one must first solve \eqref{Eq: Model}.  Since our problem
is radially symmetric, we may solve it on the spacial domain $r \in [0,1]$ by adding an additional
``boundary'' condition at the origin, namely $u_r(0,t) = 0$.  Although a Bessel function solution can be
found, matching it with the initial conditions in the model would not yield analytical solutions, and it would be
necessary to solve the integrals from the inner products numerically.  Therefore, we implement finite difference
schemes to solve \eqref{Eq: Model} instead of solving for the constants in the Bessel function solution.

We compare the solutions from three stencils: Forwards in Time Centered in Space (FTCS),
Backwards in Time Centered in Space (BTCS), and Crank-Nicholson.  For diffusion problems, Crank-Nicholson
is often used, however diffusion is only the base of our model, and we often need to run diffusion $10^5$
to $10^8$ times for one plot, which renders Crank-Nicholson inefficient.  This application requires a fast, accurate,
and stable solver.  FTCS is certainly fast, but it is at the mercy of the CFL condition \cite{CFL1928}; i.e.,
we require $\Delta t/(\Delta x)^2 \leq 1/2$.  For the videos illustrating the scheme we do not need
high accuracy, but we do need a fast scheme for high spatial resolution, so FTCS is employed.  To produce
the dose-response curves in Sec. \ref{Sec: Comparison} we employ BTCS since it is not much slower than
FTCS and almost as accurate as Crank-Nicholson.  This way we get high temporal resolution, which has the
added benefit of providing accurate dose-response curves, without sacrificing stability.

\begin{figure}[htbp]
\includegraphics[width = 0.9\textwidth]{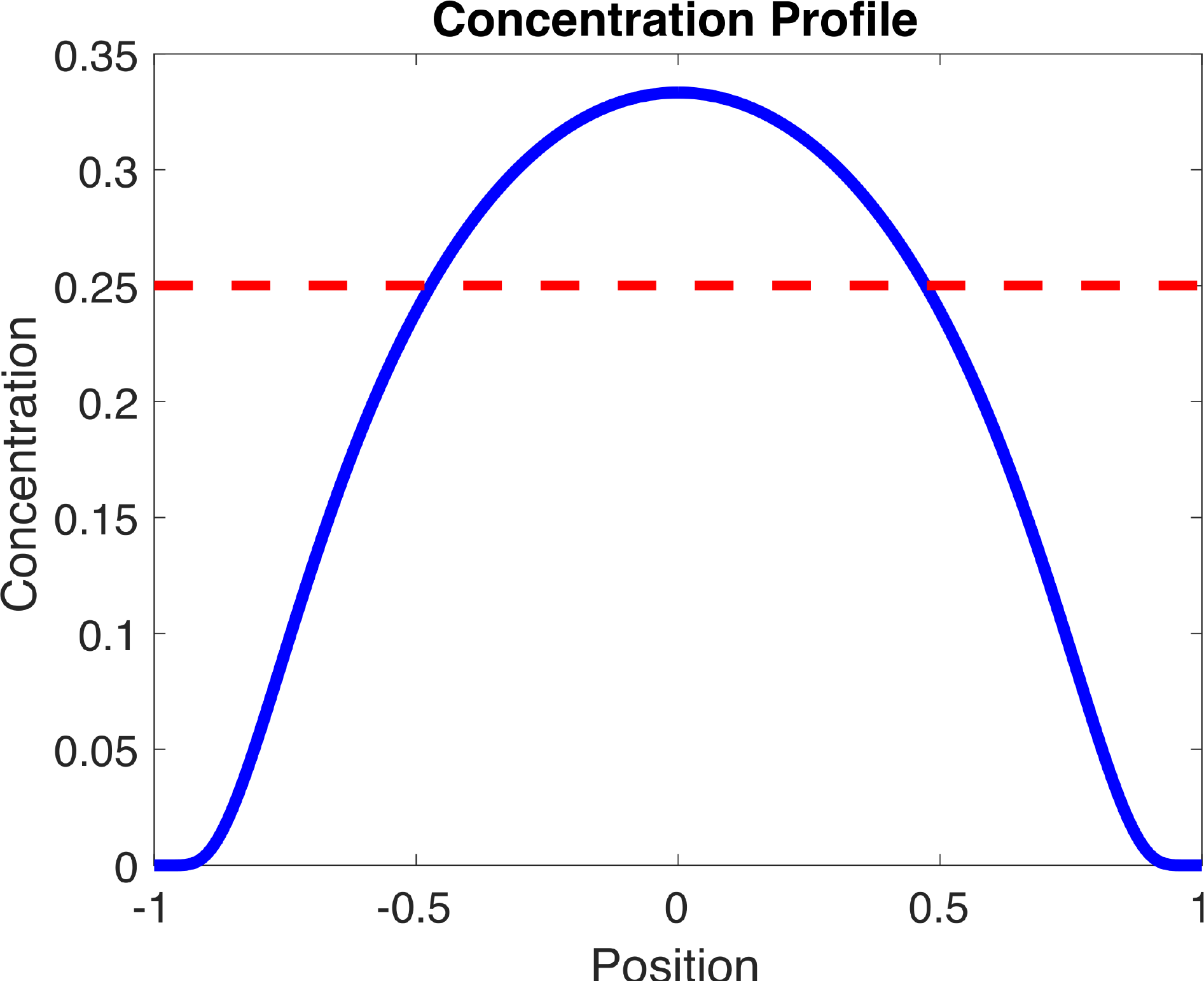}
\caption{Hypothetical concentration profile used to demonstrate percent tumor death calculation.
The blue curve represents the concentration profile of the drug, and the red dashed line represents
the concentration threshold required to kill a cell.  The radius $\rd$ is taken at the intersection of the
two curves and is used to calculate the percent volume and hence percent cell death.}
\label{Fig: PercentKill}
\end{figure}

Once a solution for $u$ from the finite difference schemes is obtained, the dose-response curves
can be plotted.  As outlined in Sec. \ref{Sec: Cell Death}, in biological experiments the dose-response
is recorded by measuring the effect of a drug at various doses.  In our model we implement a threshold
$u_T$ defined as the drug concentration required to kill a single cell.  When the concentration $u$ is
above this threshold on some interval $r \in [0,\rd]$, we may calculate the volume of tumor death,
which we assume eventually ablates.
Dividing this volume by the total nondimensional volume of $4\pi/3$ outputs the fraction of
dead tumor cells, namely
\begin{equation}
\text{Fraction of Dead Cells:  }  \pd := \frac{\text{Volume Dead}}{\text{Total Volume}}
= \frac{4\pi\rd^3/3}{4\pi(r=1)^3/3} = \rd^3.
\end{equation}
This is demonstrated in Fig. \ref{Fig: PercentKill}.  Where the concentration profile (blue curve)
intersects the threshold $u_T$ is the radius of cells killed after time $T$, which we denoted
as $\rd$.

\begin{figure}[htbp]
\centering
\includegraphics[width = 0.9\textwidth]{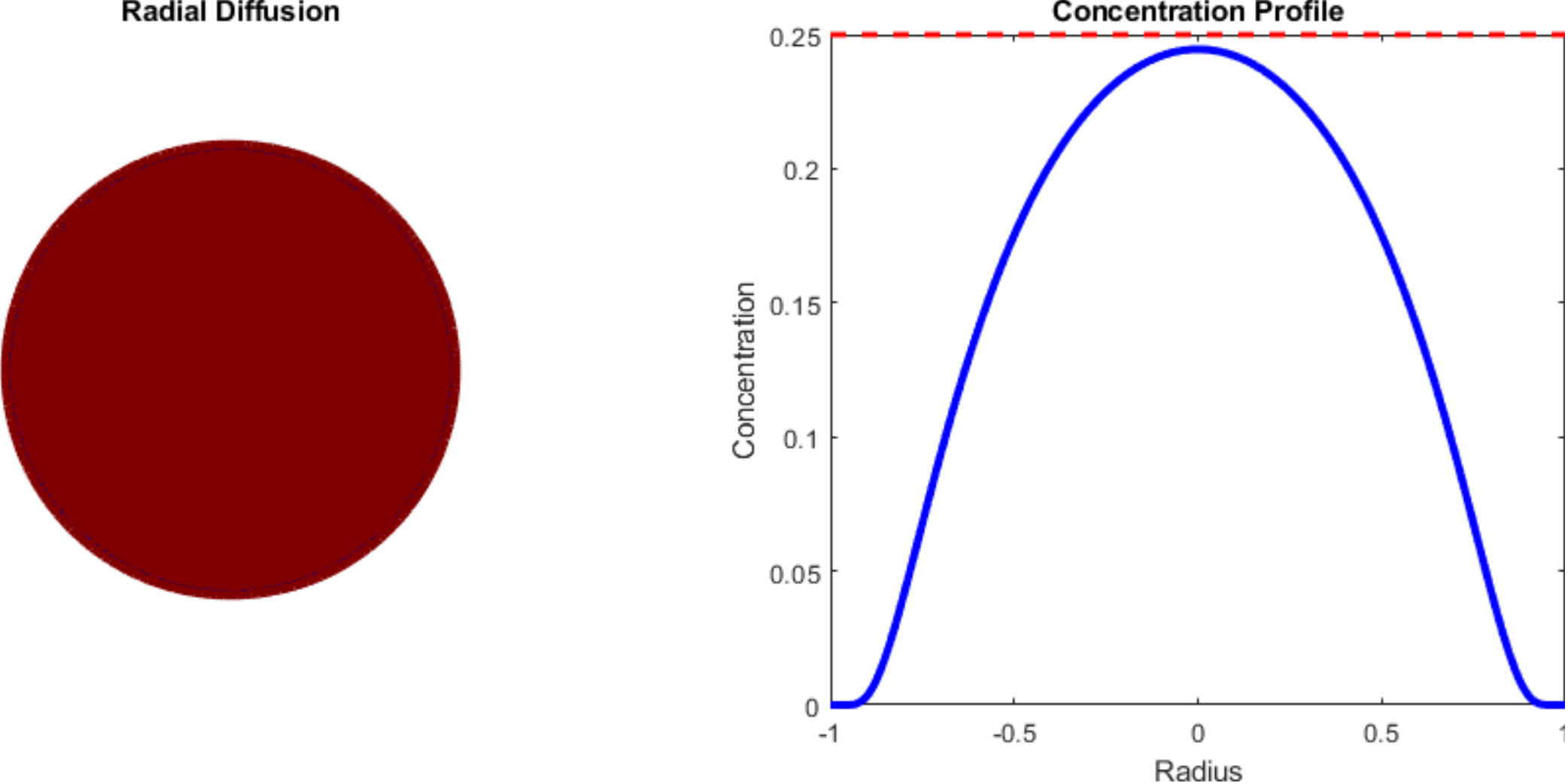}
\caption{No ablation occurs, at time $T$, when the height of the initial profile is lower than the threshold $u_T$.
This leads to an explicit formula for least effective dose shown in \eqref{Eq: LED}.}
\label{Fig: No Ablation}
\end{figure}

\begin{figure}[htbp]
\centering
\stackinset{l}{9mm}{t}{3.5mm}{\textbf{\large (a)}}{\includegraphics[width = 0.45\textwidth]{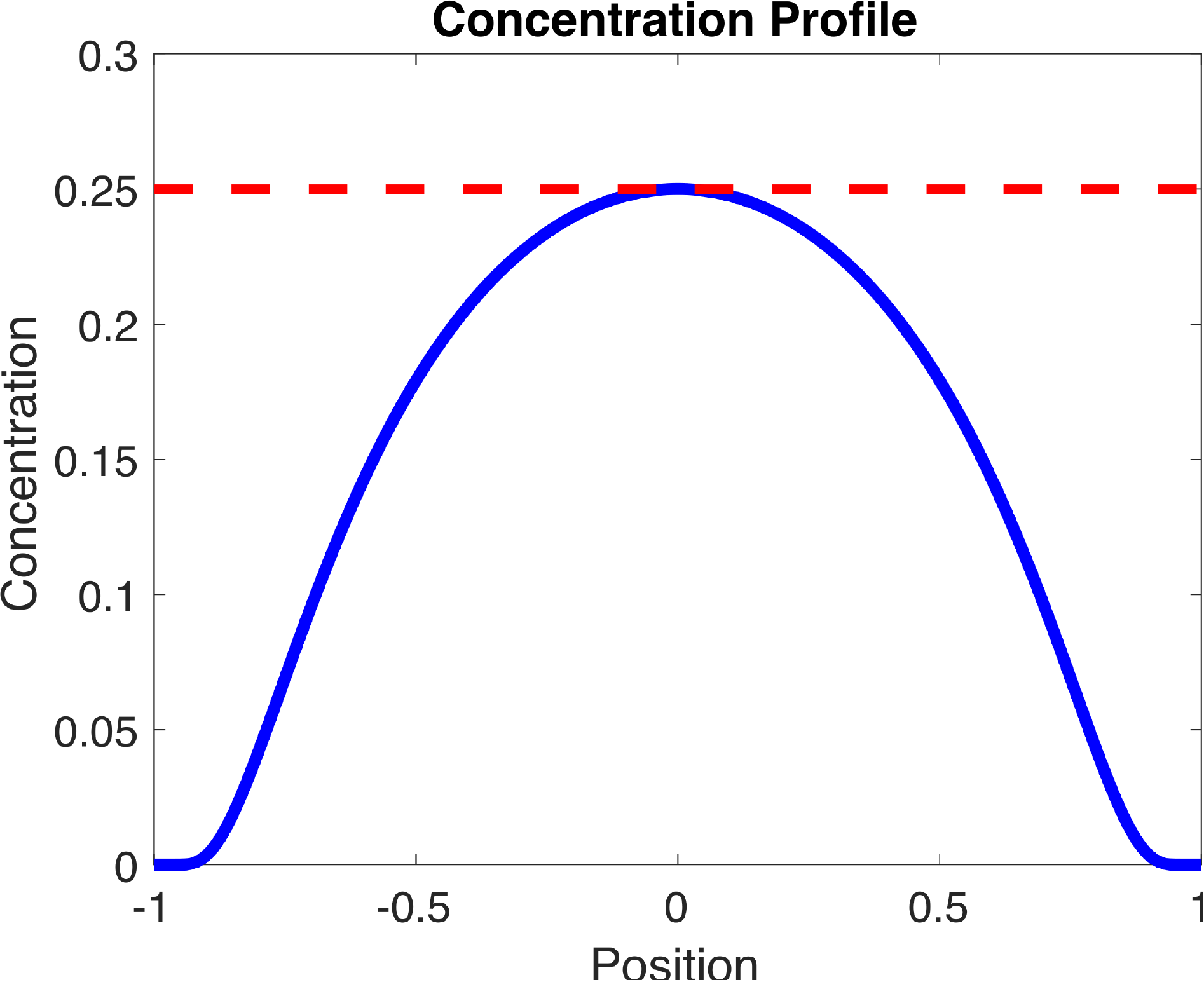}}
\refstepcounter{subfigure}\label{Fig: LED1}
\stackinset{r}{3mm}{t}{1.5mm}{\textbf{\large (b)}}{\includegraphics[width = 0.45\textwidth]{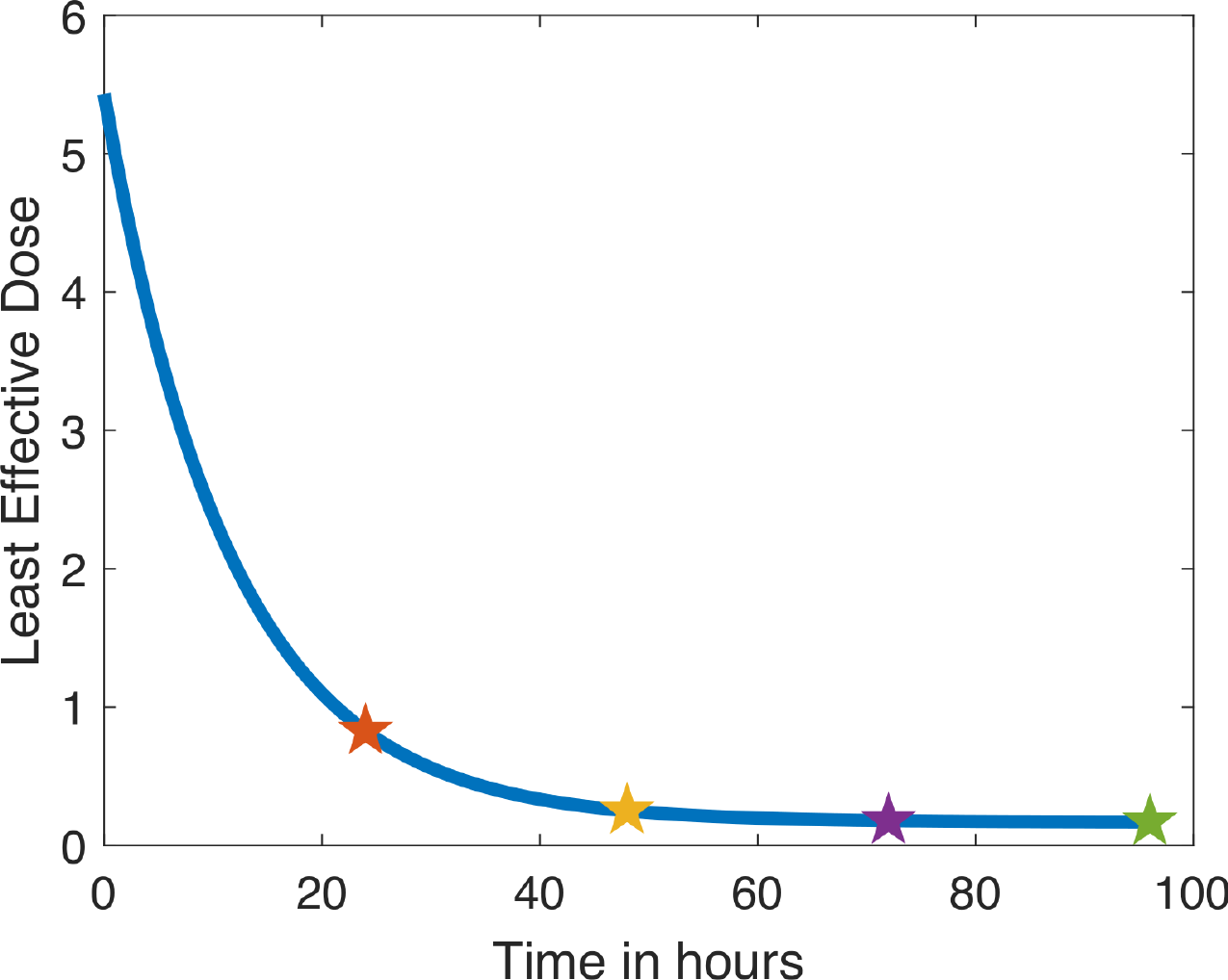}}
\refstepcounter{subfigure}\label{Fig: LED2}
\caption{Least effective dose.  \textbf{(a)}  A sample profile illustrating the least effective dose.
\textbf{(b)}  A sample relationship between the exposure time $T$ and the least effective dose $u_\text{LED}$.
The blue curve represents \eqref{Eq: LED} and the stars mark the least effective doses for
$T = 24,\, 48,\, 72,\, $ and $96$ hours.}
\label{Fig: LED}
\end{figure}

\begin{figure}[htbp]
\centering
\stackinset{l}{}{b}{3mm}{\textbf{\large (a)}}{\includegraphics[width = 0.45\textwidth]{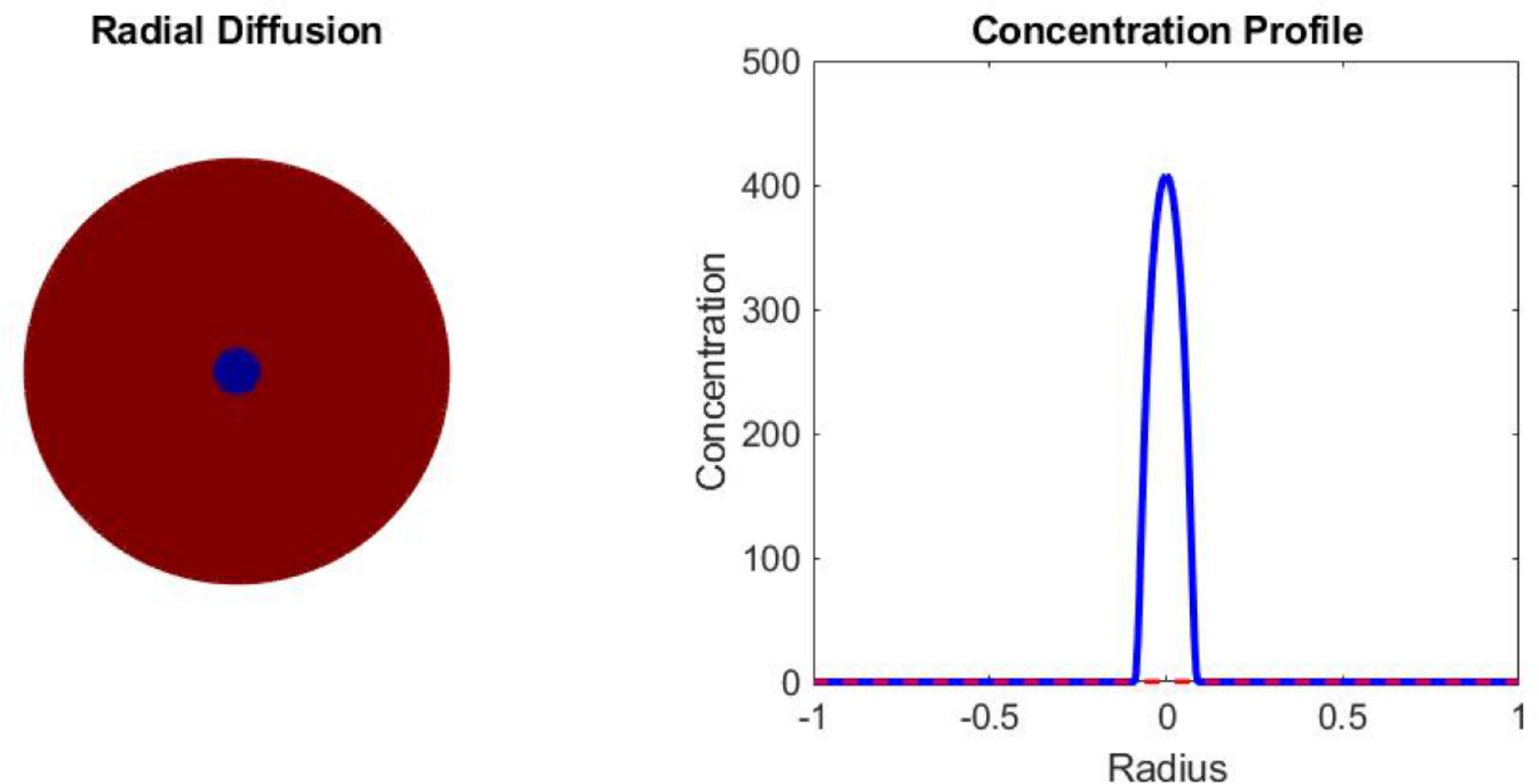}}
\refstepcounter{subfigure}\label{Fig: Partial0}\qquad
\stackinset{l}{}{b}{3mm}{\textbf{\large (b)}}{\includegraphics[width = 0.45\textwidth]{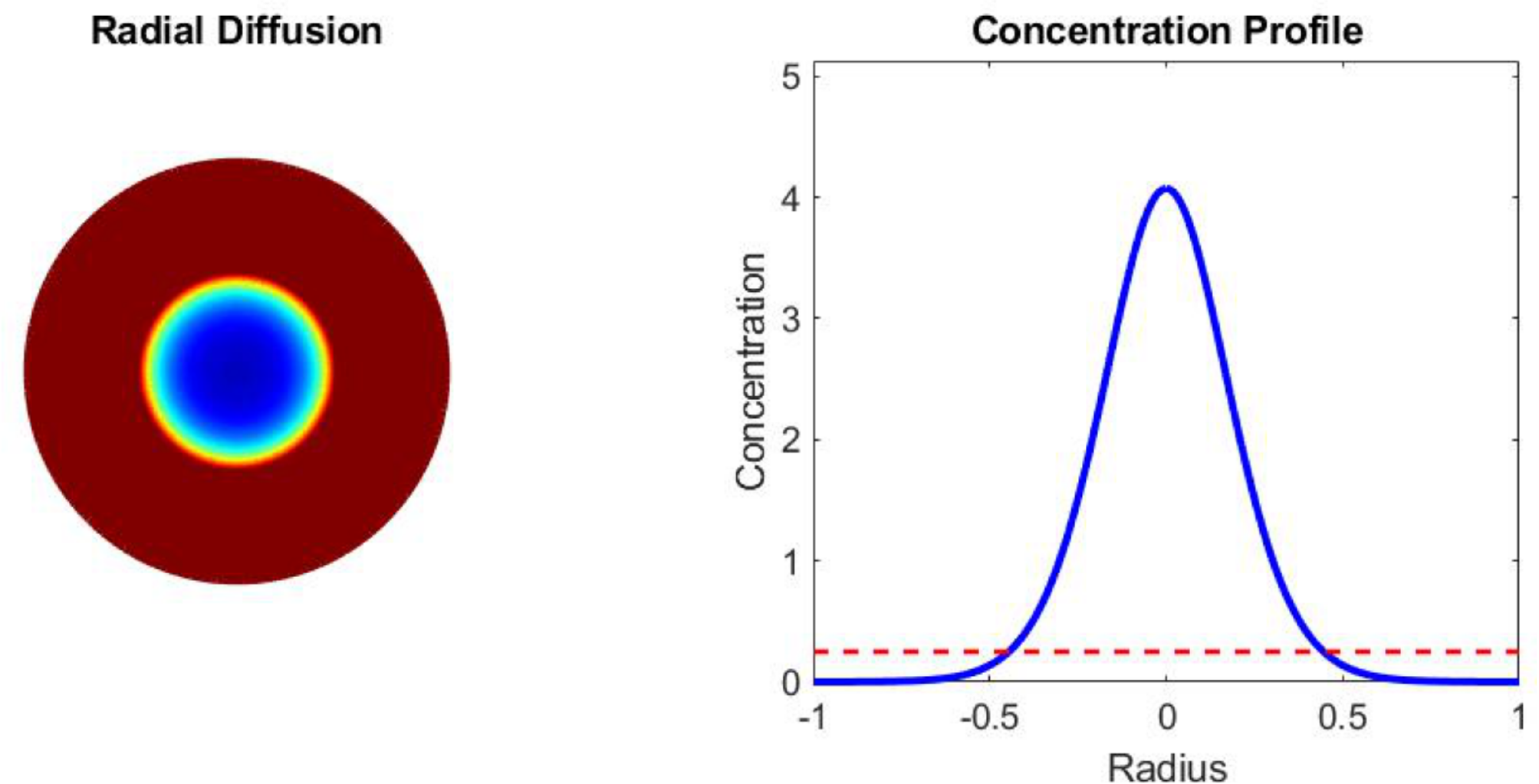}}
\refstepcounter{subfigure}\label{Fig: Partial1}
\stackinset{l}{}{b}{3mm}{\textbf{\large (c)}}{\includegraphics[width = 0.45\textwidth]{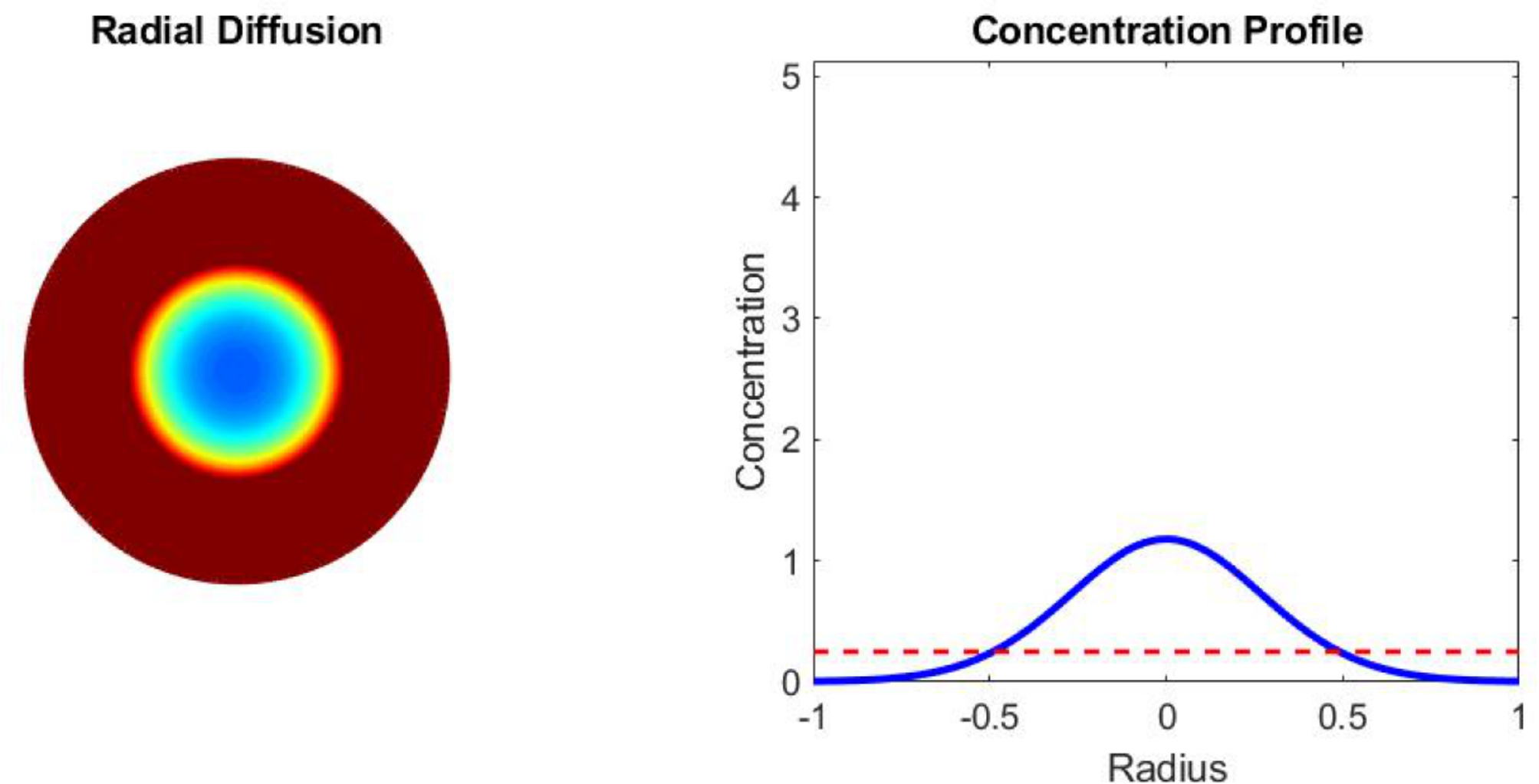}}
\refstepcounter{subfigure}\label{Fig: Partial2}\qquad
\stackinset{l}{}{b}{3mm}{\textbf{\large (d)}}{\includegraphics[width = 0.45\textwidth]{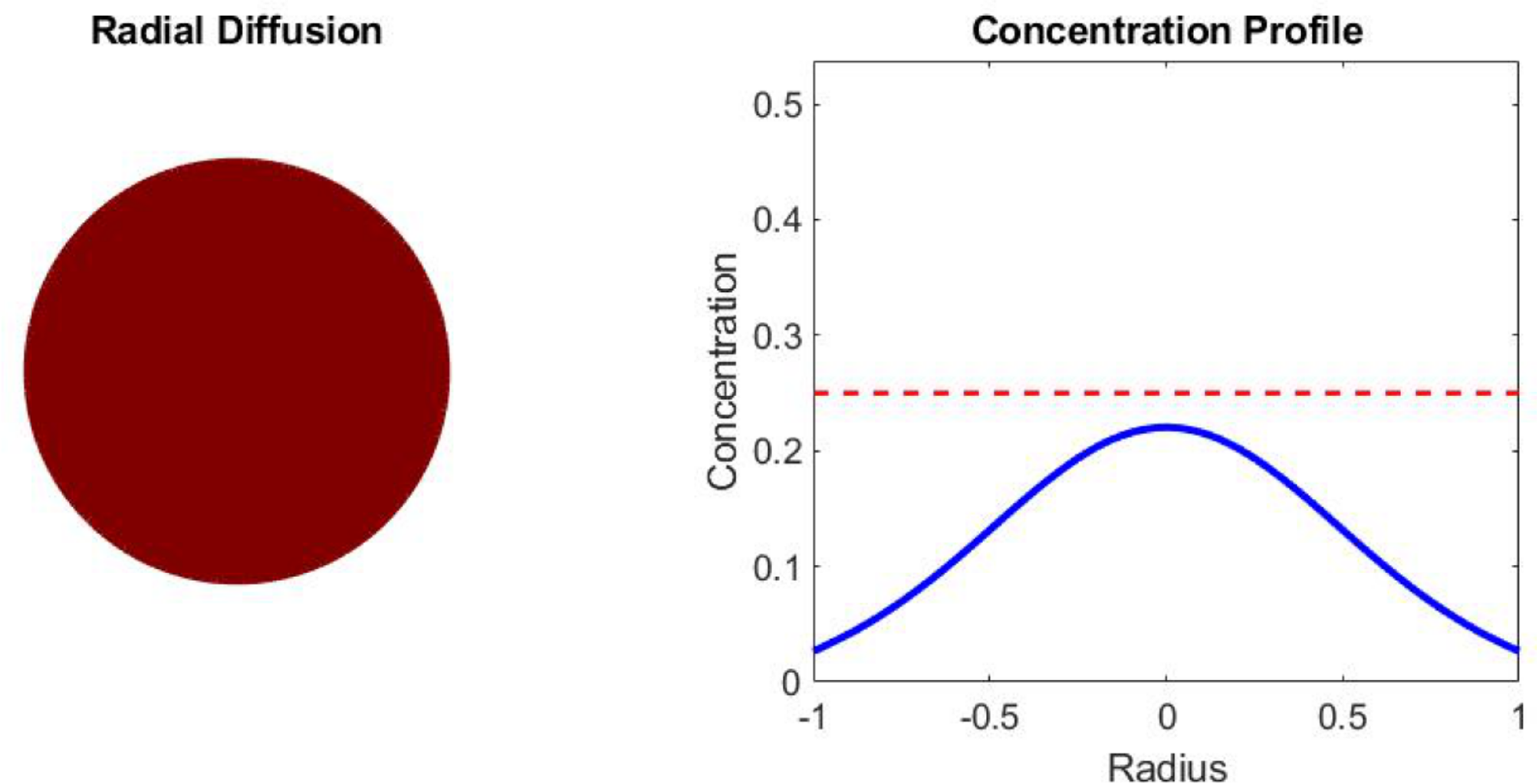}}
\refstepcounter{subfigure}\label{Fig: Partial3}
\caption{Partial ablation at time $T$.  An exaggerated initial condition was used in order to illustrate the diffusion
of the drug
in \textbf{(a)}, but the initial condition from the model \eqref{Eq: Model} is more like the profile in \textbf{(b)}.
\textbf{(a)}  Exaggerated initial condition to illustrate the
diffusion of the drug.  \textbf{(b)}  Profile similar to that used for the initial condition in \eqref{Eq: IC}.
\textbf{(c)}  Maximum radius of effective concentration is attained.  \textbf{(d)}  The drug eventually
leaks out contributing to toxicity.}
\label{Fig: Partial}
\end{figure}

\begin{figure}[htbp]
\centering
\stackinset{l}{}{b}{3mm}{\textbf{\large (a)}}{\includegraphics[width = 0.45\textwidth]{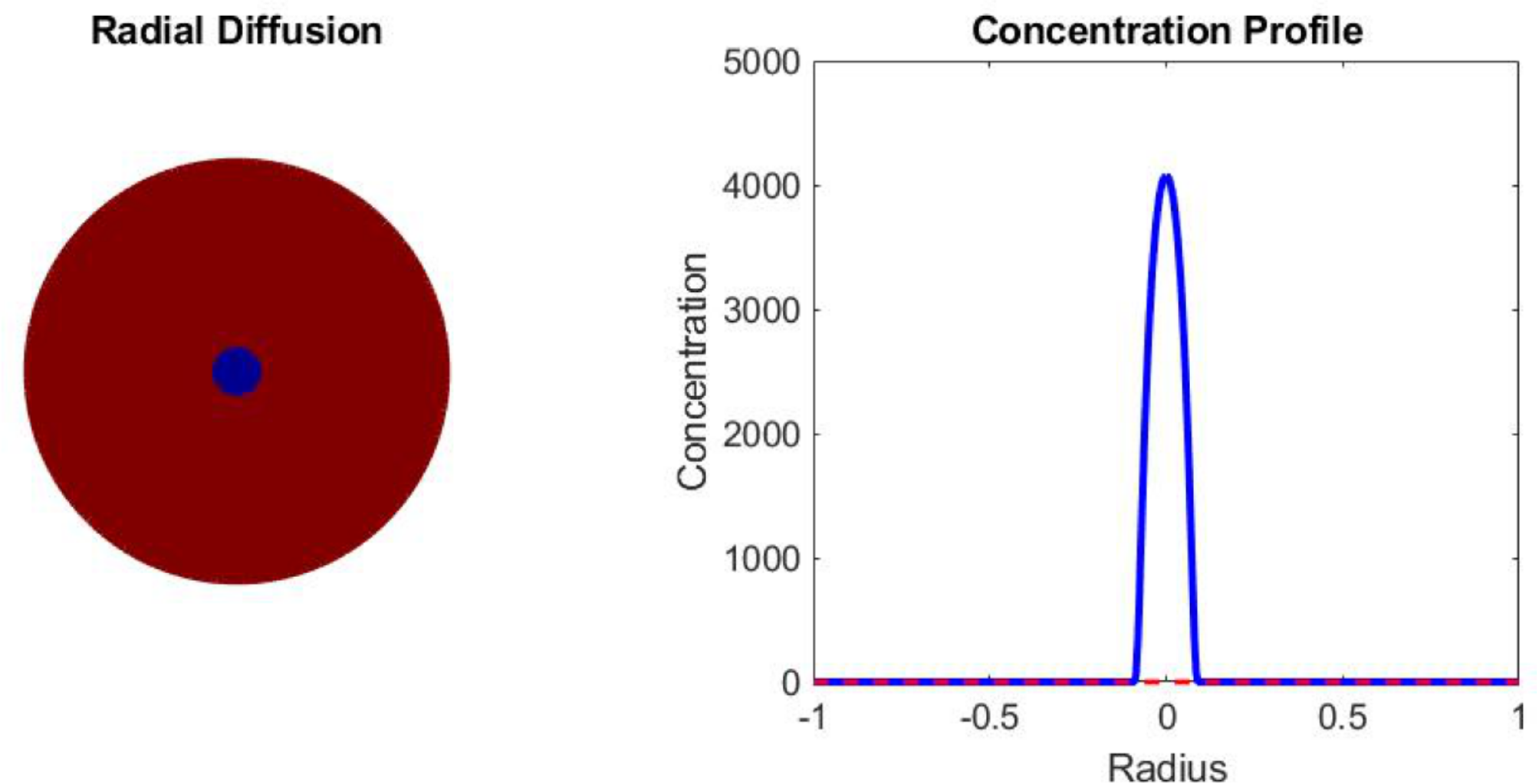}}
\refstepcounter{subfigure}\label{Fig: Full0}\qquad
\stackinset{l}{}{b}{3mm}{\textbf{\large (b)}}{\includegraphics[width = 0.45\textwidth]{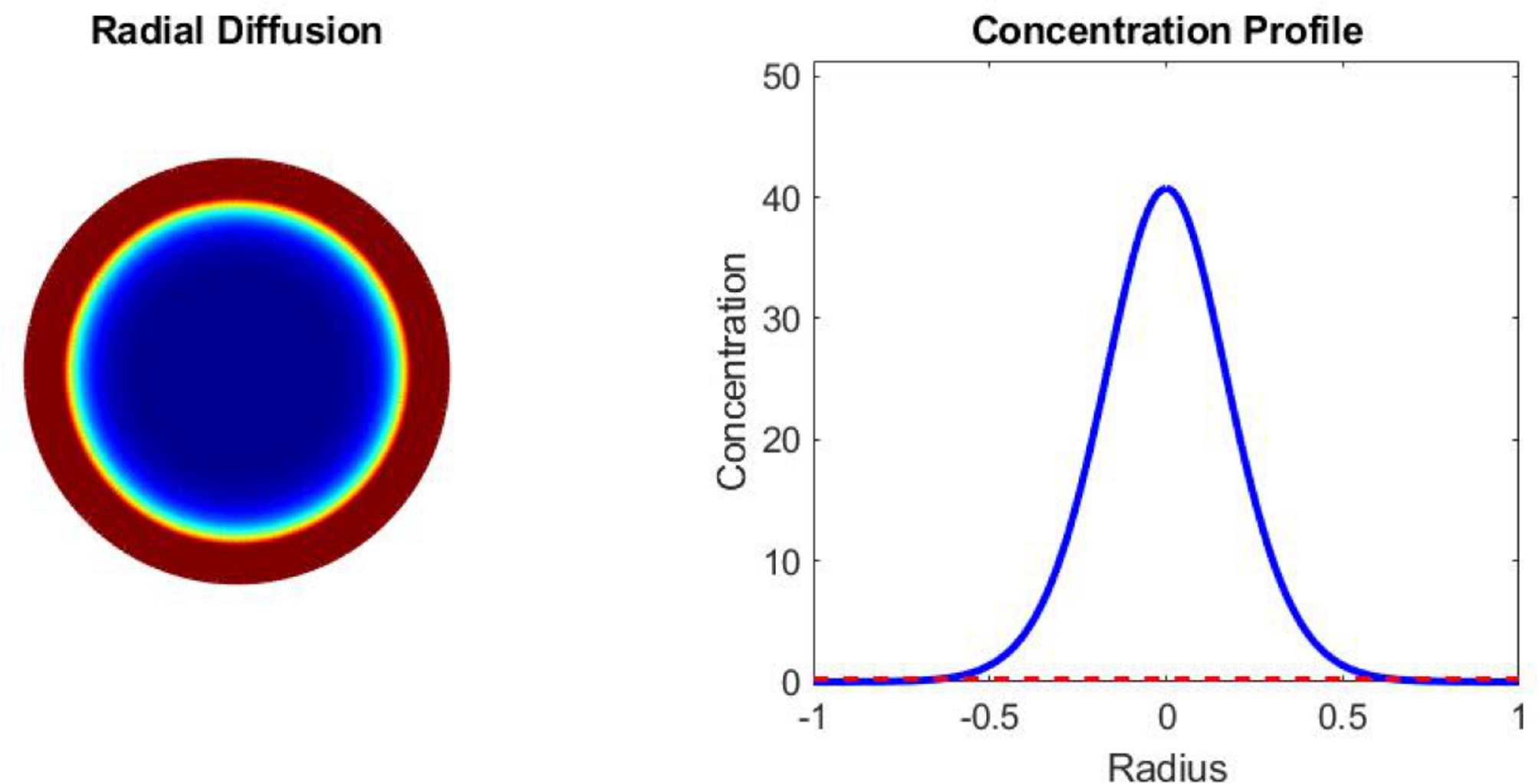}}
\refstepcounter{subfigure}\label{Fig: Full1}
\stackinset{l}{}{b}{3mm}{\textbf{\large (c)}}{\includegraphics[width = 0.45\textwidth]{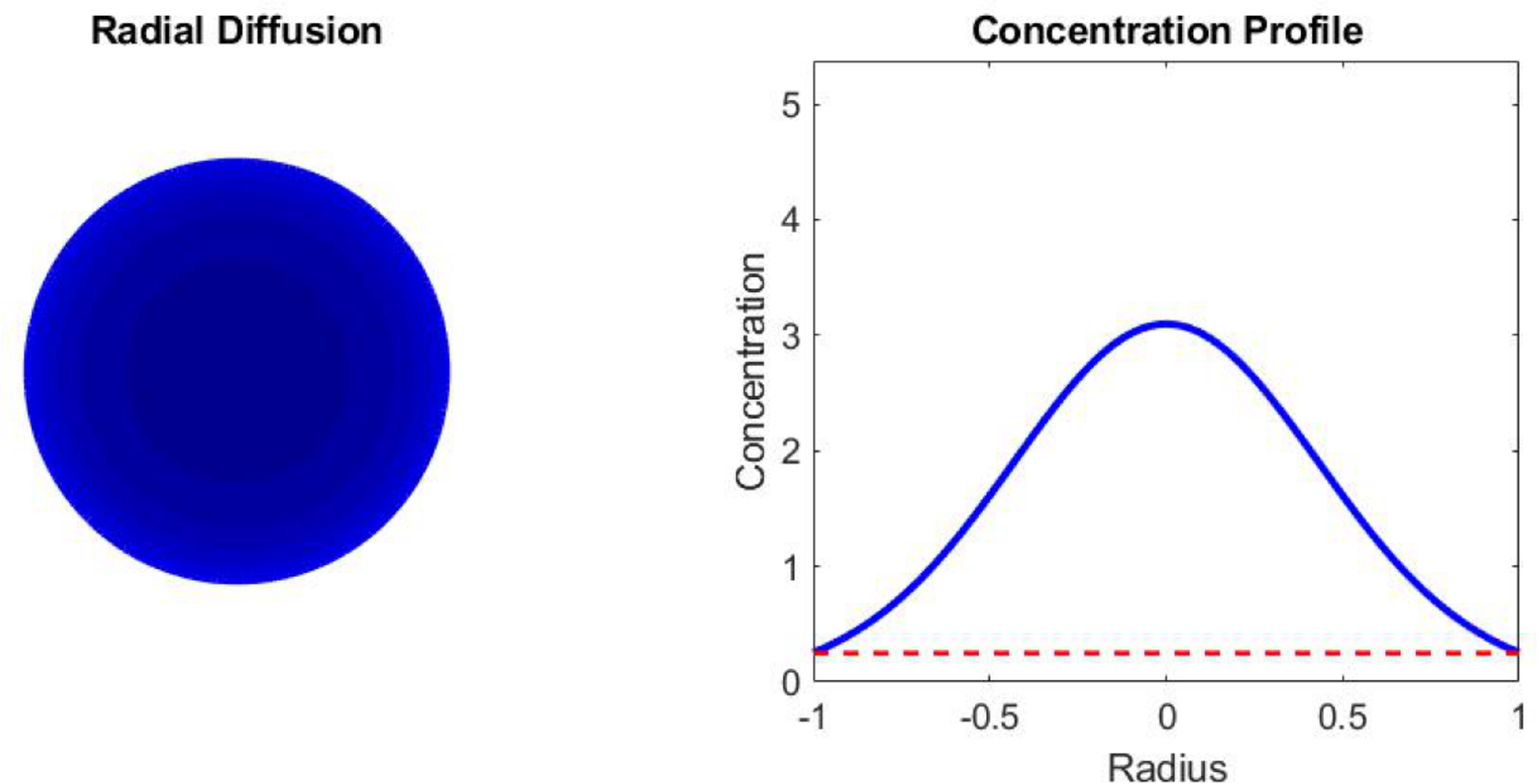}}
\refstepcounter{subfigure}\label{Fig: Full2}
\caption{Full ablation at time $T$.  \textbf{(a)} An exaggerated initial condition was used in order to 
illustrate the diffusion of the drug
in \textbf{(a)}, but the initial condition from the model \eqref{Eq: Model} is more like the profile in \textbf{(b)}.
\textbf{(a)}  Exaggerated initial condition to illustrate the
diffusion of the drug.  \textbf{(b)}  Profile similar to that used for the initial condition in \eqref{Eq: IC}.
\textbf{(c)}  Full ablation is attained, but just as in Fig. \ref{Fig: Partial}, the remainder of the drug
will leak out and contribute to toxicity.}
\label{Fig: Full}
\end{figure}

There are three possible cases for the percent of the tumor that is killed, and hence eventually ablated,
at an exposure time of $T$:
no ablation, partial ablation, and full ablation.  These are shown in Figs. \ref{Fig: No Ablation}-\ref{Fig: Full}.

In Fig. \ref{Fig: No Ablation}, the initial volume is so low that the height of the bump function does
not exceed the threshold.  This provides and opportunity to calculate the least effective dose at
a given exposure time; i.e.
when the height of the bump function is equivalent to the threshold (illustrated in Fig.
\hyperref[Fig: LED1]{\ref{Fig: LED}a}).  Notice that the height of the
bump function \eqref{Eq: IC} at the center $r=0$ is $V_0/V_b$, where $V_0$ is the initial volume
and $V_b$ is the volume of the bump function.  If we let the height be $u_T$ then the least effective
dose is
\begin{align}
\label{Eq: LED}
u_\text{LED} &= u_T\cdot V_b = u_T\frac{1}{4\pi/3}\int_0^\pi\sin\phi d\phi\int_0^{2\pi} d\theta\int_0^1
\exp\left(1 - \frac{1}{1-r^2}\right)r^2dr \nonumber\\
&= 3 u_T\int_0^1 \exp\left(1 - \frac{1}{1-r^2}\right)r^2dr =
3 \left(a - be^{-cT}\right)\int_0^1 \exp\left(1 - \frac{1}{1-r^2}\right)r^2dr.
\end{align}
From this analytical formula we may plot the least effective dose as a function of
exposure time in Fig. \hyperref[Fig: LED1]{\ref{Fig: LED}b}.  Furthermore, we observe that the steady-state
least effective does is given by
\begin{equation}
u_\text{LED}^* = u_*\cdot V_b = 3\cdot a\cdot V_b = 3\frac{u_{24}u_{72} - u_{48}^2}{u_{24} + u_{72} - 2u_{48}}
\int_0^1 \exp\left(1 - \frac{1}{1-r^2}\right)r^2dr.
\end{equation}

Figure \ref{Fig: Partial} shows the death of a fraction of the tumor cells, and hence, by our
initial assumption, partial ablation.  An exaggerated initial condition (Fig. \hyperref[Fig: Partial0]{\ref{Fig: Partial}a})
is used to illustrate the diffusion of the drug.  In order to present a profile similar to that of \eqref{Eq: IC}
(Fig. \hyperref[Fig: Partial1]{\ref{Fig: Partial}b}), we let the volume diffuse until the concentration at
the boundary is above a small, but appreciable, amount of $u = 0.01$.
This is due to the discrepancy between diffusion in real life, which happens in finite time, and mathematical
diffusion, which occurs infinitely fast.  After a maximum radius of death is attained
(Fig. \hyperref[Fig: Partial2]{\ref{Fig: Partial}c}), the percent cell death is calculated using the volume filled
at this radius.  After this maximum, the drug leaks out (Fig. \hyperref[Fig: Partial3]{\ref{Fig: Partial}d}) and eventually
contributes to toxicity, which is left for a future study.

Finally, Fig. \ref{Fig: Full} shows complete tumor death for a threshold $u_T$ at a time $T$.
As in Fig. \ref{Fig: Partial}, an exaggerated initial condition (Fig. \hyperref[Fig: Full0]{\ref{Fig: Full}a})
is used, and later a snapshot of a profile similar to \eqref{Eq: IC} (Fig. \hyperref[Fig: Full1]{\ref{Fig: Full}b})
is shown.  Since the concentration in the entire tumor exceeds the threshold, the entire tumor dies
after a time $T$.

\begin{figure}[htbp]
\centering
\includegraphics[width = 0.9\textwidth]{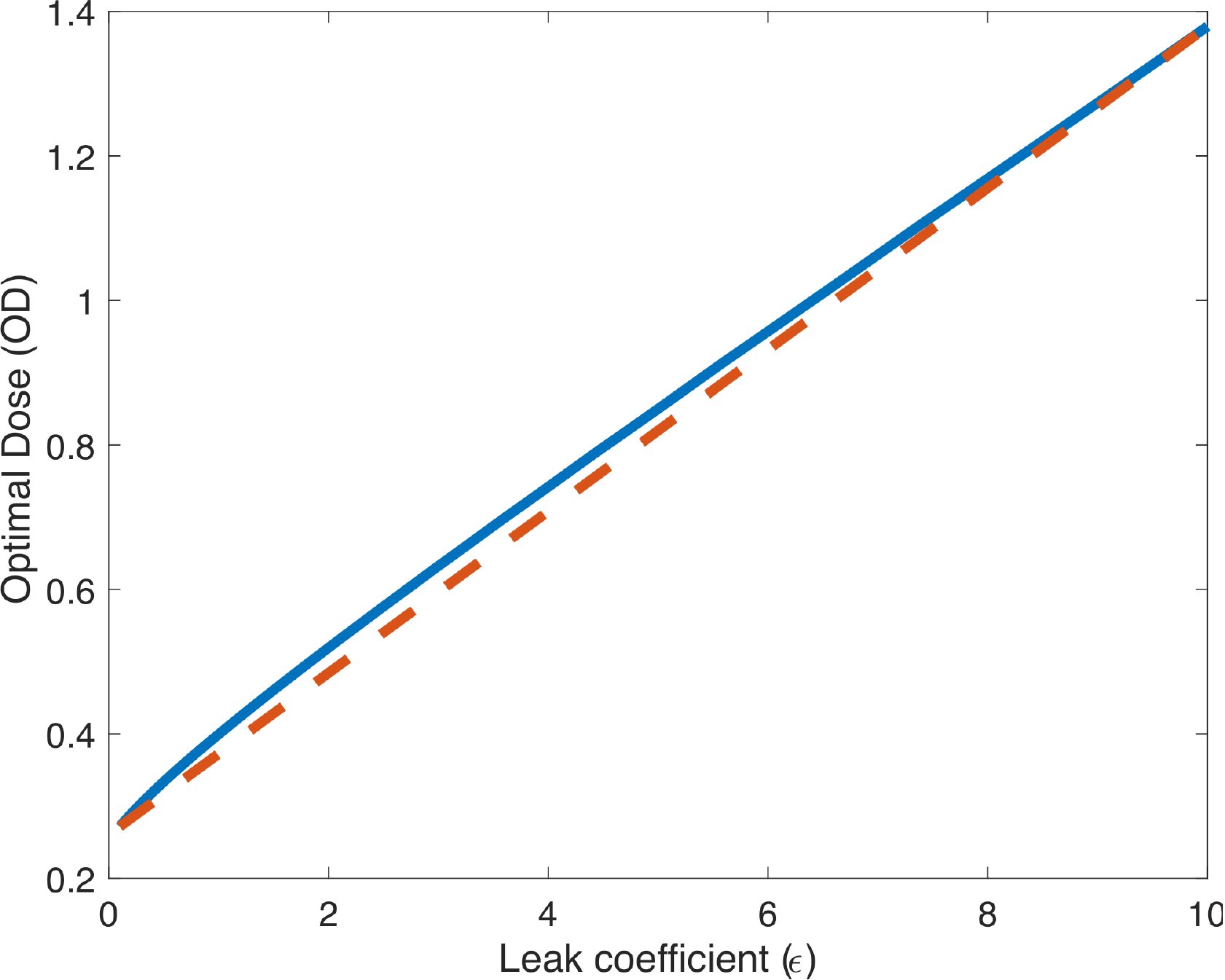}
\caption{Relationship between the optimal dose $\OD$ for a given threshold $u_T = 0.25$
and the leak coefficient $\epsilon$.  The blue curve represents the optimal dose at the given
threshold, and the red dashed curve represents a straight line from the lowest point of the blue
curve to the highest demonstrating the nonlinearity of the optimal dose curve.}
\label{Fig: Leak vs. OD}
\end{figure}

Eventually all of the drug leaks out, but for the full ablation case the toxicity would be higher than
with partial ablation.  Then we have a balancing act between efficacy and toxicity.  We may calculate
the smallest dose required to kill the entire tumor at a time $T$ using the threshold $u_T$, via the
bisection method to find the initial concentration that
allows the entire concentration profile to be just over the threshold at the boundary, similar to a shooting
method.  Since the diffusion and leak linear, we would expect the relation between the threshold
and smallest dose to be linear as well, which is precisely what we observe.  If we let $\OD_i$ be the
smallest does required to kill the tumor after a time $T_i$ we get the following relation between
the ratios of the smallest dose and the respective threshold
\begin{equation}
\frac{\OD_1}{\OD_2} = \frac{u_{T_1}}{u_{T_2}}.
\end{equation}
Since $u_T \rightarrow u_*$ as $T \rightarrow \infty$, we may define an optimal steady-state
dose as the smallest dose required to kill the tumor assuming we wait infinitely long and the tumor
growth is completely inhibited.  By this definition we notice that we may calculate $\OD$ for any
time $T$
\begin{equation}
\frac{\OD}{\OD_*} = \frac{u_T}{u_*} \Rightarrow \OD = \frac{\OD_*}{u_*}u_T
= \left[a - be^{-cT}\right]\frac{\OD_*}{u_*},
\end{equation}
which is precisely the modified Haber's rule \cite{MSJ2000} because $\OD_*/u_*$ is constant.
In addition, we may fix a threshold $u_T$, and observe how the optimal dose for that threshold
varies with the nondimensional leak coefficient $\epsilon$.  We observe in Fig.\ref{Fig: Leak vs. OD}
that this relation, represented by the blue curve, is not linear although it may seem linear without the
presence of the red dashed line from the lowest to the highest point of the blue curve in that plot.

\section{Comparison with clinical data}\label{Sec: Comparison}

\begin{figure}[htbp]
\includegraphics[width = 0.9\textwidth]{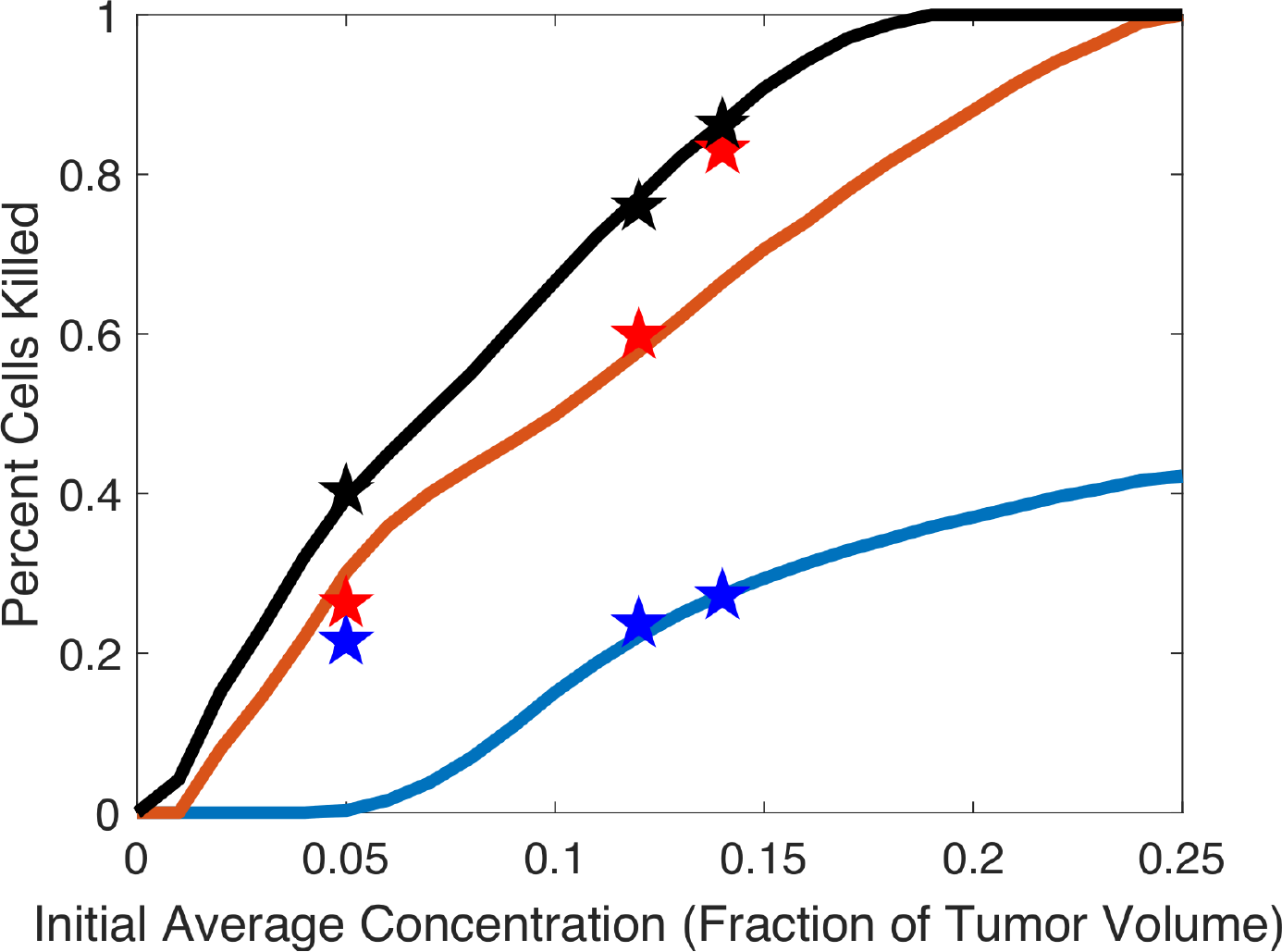}
\caption{Simulated dose-response curves for cell line ``C32'' and drug ``Selumetinib''
compared to experimental data points (stars).  The
blue curves and markers are at 24 hours, red is 48 hours, and black is 72 hours.}
\label{Fig: Dose-Response}
\end{figure}

While there was no data on the rate of cell death in the study by Morhard \ea \cite{Morhard2017}
because it focused on tumor shrinkage, we may use other clinical/experimental data sets to show
qualitative agreement.  This would indicate the modeling framework is reasonable, and then the
models may be modified for different applications.  In this light, a ``proof of concept'' comparison is
made between the model and a sample data set
obtained from LINCS database (http://lincs.hms.harvard.edu/db/) which, among other things, offers cellular responses  to 
chemical and genetic pertubations. In this article we consider apoptosis fraction as the cellular response observed under 
 different dosage of different compounds. So, our experimental data consists of a set of dose-response pairs observed for several cell-
 lines and compounds at 24, 48 and 72 hours.
  The dose-response curves from the model and the observed 
data points are shown in Fig. \ref{Fig: Dose-Response}.  

By employing \eqref{Eq: time-threshold} we find the concentration thresholds at the three time points
for a given initial drug concentration.  For the dose-response curves in this article we match an initial
dose of $1\mu \text{M}$ to an initial concentration of $1/12$ tumor volume because
the experiments of Morhard \ea \cite{Morhard2017} showed complete ablation for all of their hamsters
at $1/4$ tumor volume, which is approximately what would be achieved by most of the cell line - drug
pairs if the 72 hour time point response is
extrapolated to $1/4$ tumor volume.  Furthermore, we match two other doses by varying the initial volume
until the distance between two of the three time points is minimized.  We choose to match
two time points since it is evident that the data set is quite noisy and it has outliers that do not obey the
Hill equation \cite{Hill1910}.  One concern may be the nonuniformity of the initial volume - dose relation,
but this is to be expected since the relation is dependent on the tumor size and concentration of the drug
in the solution.

In this section, using a leak coefficient of $\epsilon = 5$,
we illustrate our  performance in modeing the apoptosis fraction observed for cell line ``C32'' and drug
``Selumetinib''(we offer a few more illustrative performances in Appendix \ref{Appendix: Dose-response}.   
 Since the data is quite noisy for low doses a truncated data set, shown in 
Table \ref{Tab: C32S}, is much more reliable for comparisons with the model.  The responses for a dose of
$1\mu \text{M}$ from the pair is matched with $1/12$ tumor volume from the model.  Equation 
\ref{Eq: time-threshold} gives us the relation between the three time points (24, 48, and 72 hours) and the
thresholds ($u_{24}$, $u_{48}$, and $u_{72}$).  These thresholds are then used to produce the
simulated dose-response curves in Fig. \ref{Fig: Dose-Response}.  The data from the $1\mu \text{M}$
dose will match up exactly with the curves.  In order to match the data from the $0.316 \mu \text{M}$
and $3.16 \mu \text{M}$ doses with the respective initial concentration in the model we identify the concentration
for which the difference between the response in the data and the model of two time points is minimized.
For example, in Fig. \ref{Fig: Dose-Response}, the $0.316 \mu \text{M}$ dose corresponds to an initial
concentration of approximately $1/20$ of the tumor volume since the 72 hour and 48 hour time points are closest
to the curve at that dose, and similarly the $3.16 \mu \text{M}$ corresponds to approximately $7/50$ of the
tumor volume since the 72 hour and 24 hour time points are closest.  Since 72 hours is near the steady state
of this process, the curves often match the data at that time point.  For the 24 and 48 hour time points, the
data becomes quite noisy with outliers and responses that would not be possible to achieve, indicating possibly
large errors.  If the distance between all three points were minimized, these outliers would have a stronger effect.

\begin{table}[htbp]
\caption{Truncated data set for cell line ``C32'' with drug ``Selumetinib''}

\centering
\setlength{\tabcolsep}{14pt} 
\renewcommand{\arraystretch}{1}
\begin{tabular}{ccc}
Dose (micromole) & Time (hours) & Response (apoptosis fraction)\\
\hline
\multirow{3}{*}{0.316} & 24 & 0.2148\\
	& 48 &  0.262\\
	& 72 &  0.4026\\
\hline
\multirow{3}{*}{1.000} & 24 &  0.2365\\
	& 48 &  0.5982\\
	& 72 &  0.7591\\
\hline
\multirow{3}{*}{3.160} & 24 &  0.2716\\
	& 48 &  0.8303\\
	& 72 &  0.8621\\
\hline
\end{tabular}
\label{Tab: C32S}
\end{table}

\begin{figure}[htbp]
\centering
\includegraphics[width = 0.9\textwidth]{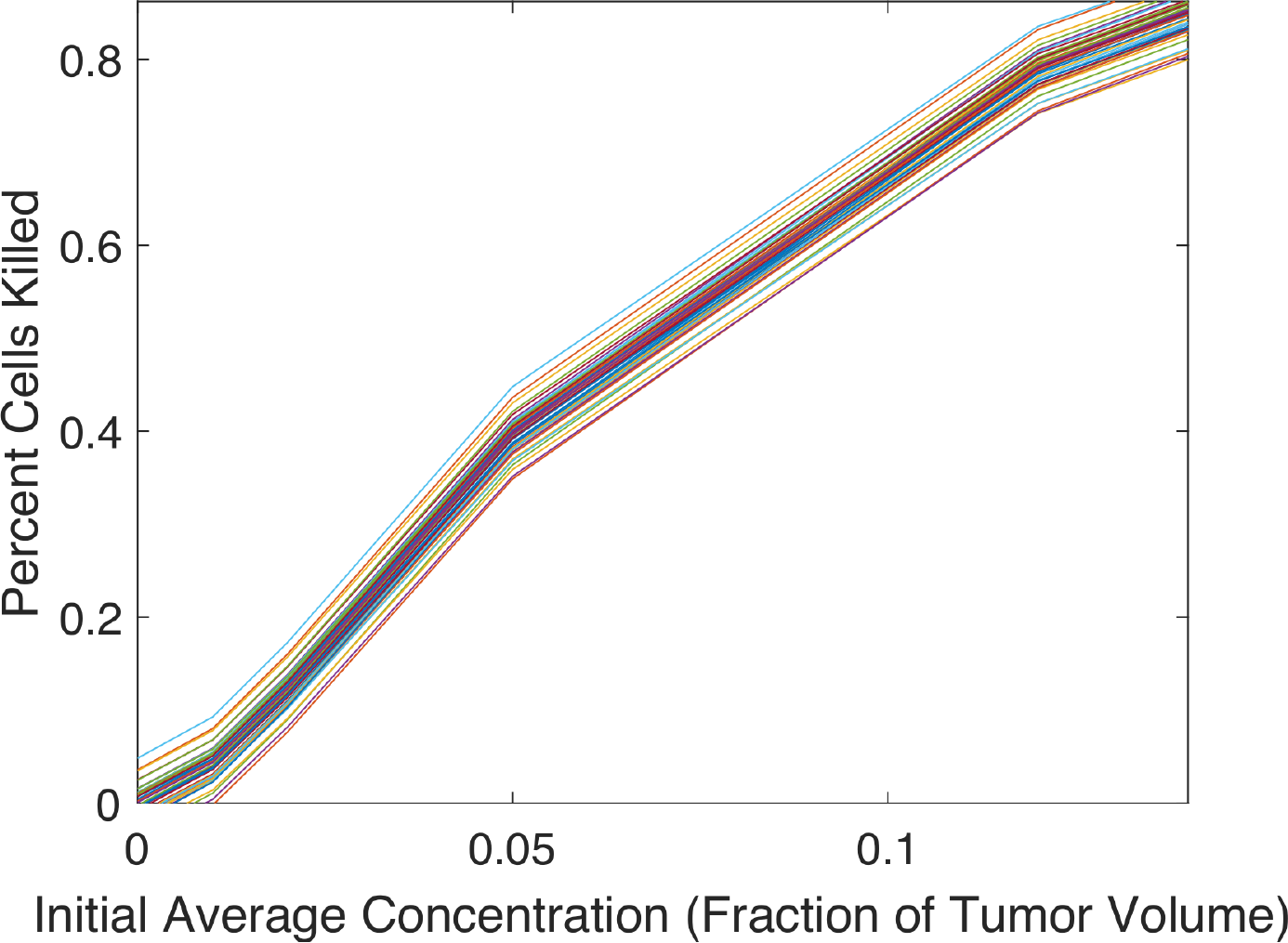}
\caption{Fifty sigmoidal curves fitting the empirical data, which is given a Gaussian spread to simulate error
at each dose point, for cell line ``C32'' and drug ``Selumetinib'' at 72 hours of exposure time.}
\label{Fig: ConfBand_Illustration}
\end{figure}

Besides the foregoing three dose points we assume that at a dose of
zero the response should be around zero. Once we match the original concentration levels with the tumor volumes, we empirically fit a sigmoidal curve 
to the experimental dose-response data observed at 72 hours. The empirical model is given by 
\begin{equation}
\mbox{Apoptosis fraction} = a + \frac{d}{1+\left(\frac{\delta}{concentration}\right)^\theta}
\label{sigmoid}
\end{equation}
where $a$ is the lower asymptote of the response curve, $d$ is the range of response, $\delta$ is typically interpreted as $EC_{50}$ and $\theta$ is the Hill coefficient. The parameters of the Hill equation (\ref{sigmoid}) are estimated using non-linear least squares fit.  We then generate 50 replicates from the fitted sigmoid curve using parametric bootstrap procedure with errors drawn from a Gaussian process with exponential correlation function. The noise variance is comparable to the residual variance obtained from the least square fit (\ref{sigmoid}) and the range parameter is chosen to ensure smoothness and monotonicity of each replicate. Fig. \ref{Fig: ConfBand_Illustration} shows 50 bootstrap replicates for cell line ``C32'' and drug ``Selumetinib''.

\begin{figure}[htbp]
\centering
\includegraphics[width = 0.9\textwidth]{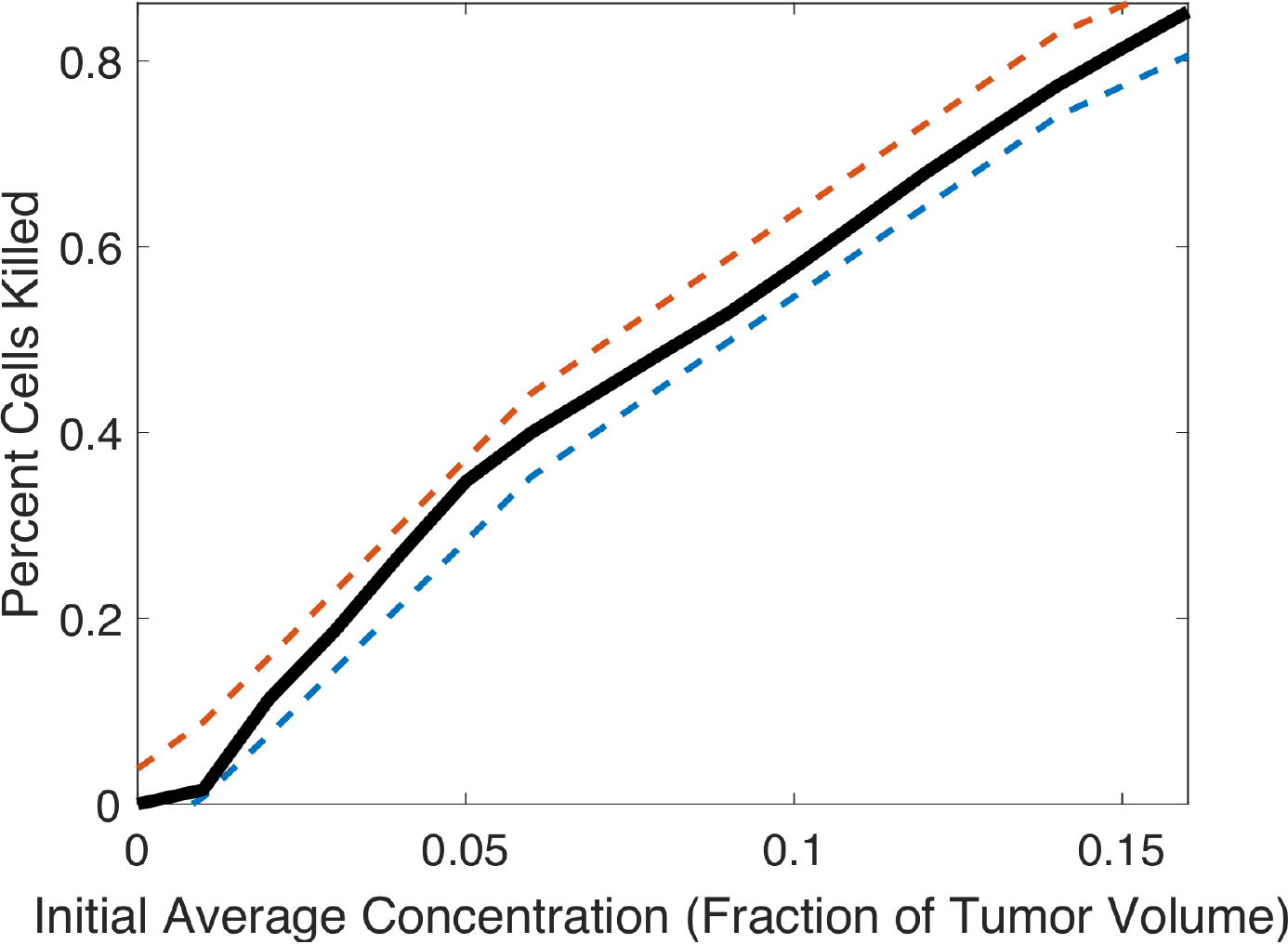}
\caption{Theoretical dose-response curve for cell line ``C32'' and drug ``Selumetinib'' within a 95\%
confidence band at 72 hours of exposure time.  The black curve represents the dose-response from the
model.  The blue and red dashed lines represent the piece-wise linear curve connecting the lower and upper
quantiles of each of the 95\% confidence intervals.}
\label{Fig: ConfBand}
\end{figure}

From these bootstrap replicates of sigmoidal curves we compute the point-wise $95\%$ confidence intervals at each matched dose, then join them into a
piece-wise linear approximate confidence band.  An example of this 
is shown in Fig. \ref{Fig: ConfBand} for the foregoing illustrative case. The dose-response curve produced by our model is represented as the black curve, and the point-wise empirical quantiles are connected by the piece-wise linear dashed curves. Observe that the curve produced by our model is contained within the 95\% pointwise confidence interval, indicating statistical adequacy of the the deterministic model. The remainder of the data pairs are relegated
to Appendix \ref{Appendix: Confidence bands}. 

We would like to point out that since we are computing pointwise confidence intervals, there could be situations where for some dose levels our model output lies outside the 95\% confidence interval computed at those doses. Rare occurrences (less than 5\% of the times) of such instances do not necessarily indicate statistical inadequacy of the posited model. However,  frequent occurrences of such events (more than 10\% of the times) do indicate statistical inadequacy. 

Given the empirical evidence that our model is statistically adequate to generate dose-response curves from experimental data, we proceed to compute the theoretical EC50s for this dataset. The model is cut off at the maximum empirical response, and the EC50s, by  definition, are computed as the dose to achieve half that response. We use a bisection scheme similar to that of the optimal dose. EC50s are then approximated within
a maximum response error of $|u_\text{approx} - u(\text{EC50})| < 0.0025$.  The spatial resolution of
the diffusion scheme will limit the error bound, so this must be chosen to be equivalent or less than the desired error. Table \ref{Tab: EC50s} shows the  theoretical EC50s and the associated errors. 

\begin{table}[htbp]
\caption{Theoretical EC50s with response errors}

\centering
\setlength{\tabcolsep}{14pt} 
\renewcommand{\arraystretch}{1}
\begin{tabular}{cccc}
Cell line & Drug & EC50 & Response error\\
\hline
\multirow{5}{*}{C32} & SB590885 & 0.0562 & 0.0006 \\
	& PLX-4720 & 0.0684 & 0.0015 \\
    & AZ-628 & 0.0596 & 0.0007 \\
    & Selumetinib & 0.0596 & 0.0007 \\
    & Vemurafenib & 0.0840 & 0.0017 \\
\hline
\multirow{4}{*}{COLO 858} & SB590885 & 0.1055 & 0.0015 \\
	& PLX-4720 & 0.1260 & 0.0005 \\
    & AZ-628 & 0.0830 & 0.0019 \\
    & Selumetinib & 0.1035 & 0.0014 \\
\hline
\multirow{2}{*}{RVH-421} & AZ-628 & 0.1191 & 0.0013\\
    & Selumetinib & 0.1221 & 0.0002 \\
\hline
\multirow{2}{*}{WM-115} & SB590885 & 0.1074 & 0.0021 \\
    & AZ-628 & 0.0664 & 0.0008 \\
\hline
\multirow{5}{*}{M27-mel} & SB590885 & 0.1191 & 0.0024 \\
	& PLX-4720 & 0.0664 & 0.0008 \\
    & AZ-628 & 0.0908 & 0.0003 \\
    & Selumetinib & 0.1006 & 0.0007 \\
    & Vemurafenib & 0.1230 & 0.0023 \\
\hline
\end{tabular}
\label{Tab: EC50s}
\end{table}

\begin{figure}[htbp]
\includegraphics[width = 0.9\textwidth]{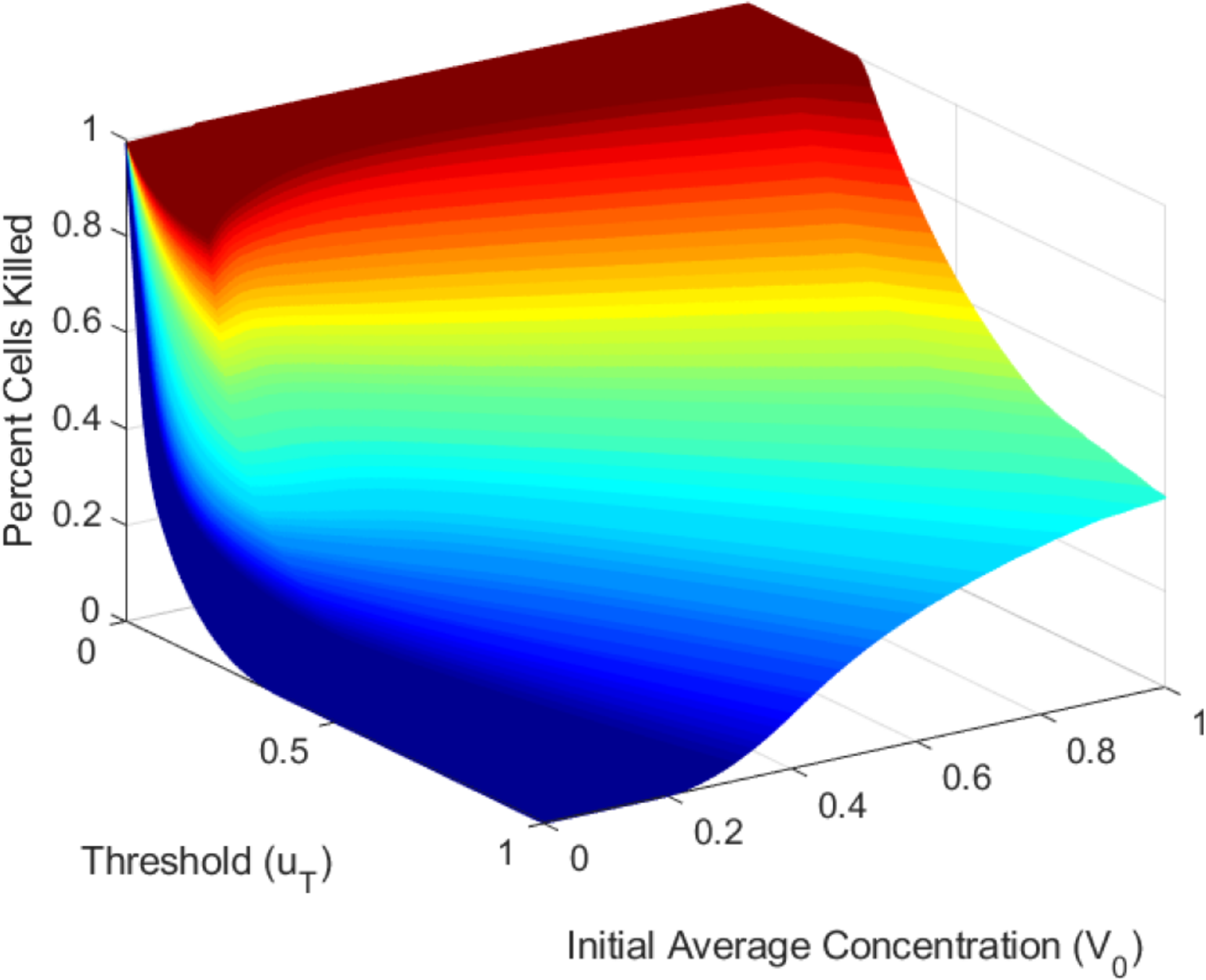}
\caption{Simulated dose-threshold-response surface.  This can be converted into a dose-time-response
surface by calibrating \eqref{Eq: time-threshold} in Sec. \ref{Sec: Cell Death} the time-threshold relation
for each cell line and drug.}
\label{Fig: DoseSurf}
\end{figure}

While a dose-response curve may be easily fit to the empirical data, a reliable dose-time-response
surface is much more difficult due to how noisy the data is.  With a mechanistic model we get a
full dose-threshold-response surface (Fig. \ref{Fig: DoseSurf}), which can be turned into an individual's
dose-time-response surface by using \eqref{Eq: time-threshold} in Sec. \ref{Sec: Cell Death} to relate
the threshold to time.  We observe that the surface in Fig. \ref{Fig: DoseSurf} is qualitatively similar to
that of Miller \ea \cite{MSJ2000}, further solidifying the agreement between this new model and
previous studies.

\section{Conclusion}
\label{Sec: Conclusion}

Cancer treatments have come a long way in terms of their increased efficacy and reduced toxicity.
However, it is still very much a ``trial and error'' process.  Since individuals contain such a variety
of biological properties, and in fact separate tumors in the same individual are also quite different,
we need to develop better individualized treatments.  While there is an abundance of statistical
models \cite{HRGP2015}, there are few mechanistic models detailing the distribution of drugs inside
a tumor leading to tumor cell death.  There has been recent work on transport models for drugs
into solid tumors \cite{WaiteRoth12, KGR2013, SoltaniChen2011, SoltaniChen2012, SSRSBBM15},
but there are no models in the literature that describe the diffusion from an injection into a solid tumor
causing it to ablate.

In this investigation we developed a radially symmetric concentration diffusion model in a solid spherical
tumor with leaky boundaries in Sec. \ref{Sec: Diffusion}, which is then nondimensionalized to absorb
the constants (radius, diffusivity, leak coefficient) into the nondimensional leak coefficient, effectively
making it a one parameter model.  Then in Sec. \ref{Sec: Cell Death}, the concentration from the
diffusion model is related to cell death assuming a fixed constant concentration required to kill a
cell at a time $T$, which we call the \emph{concentration threshold}, $u_T$.  Three time points,
at a specific initial dose, are used to relate the threshold to time using \eqref{Eq: time-threshold}.
Numerical simulations used to produce dose-response curves are described in Sec. \ref{Sec: Numerics},
and the curves themselves are illustrated in Sec. \ref{Sec: Comparison}, where they are tested against
empirical data.  While \cite{Morhard2017} did not include dose-response curves in their study, we
use a different empirical data set to show qualitative agreement and use it as a ``proof of concept'' comparison.
The comparison with a completely different data set solidifies the argument that this is a modeling
framework that can be applied to other experiments and not just a model for the Morhard \ea experiments
\cite{Morhard2017}.

It is true that, due to its simplicity, the model presented here misses some of the behavior associated
with the diffusion of a drug leading to cell death.  Critics may even suggest that it is na\"{i}ve.  It is
completely legitimate to call into question the assumption that the tumor is a uniformly dense sphere (i.e.,
constant diffusivity and radial symmetry).  One may also point to the use of a constant (in space, but not in time)
concentration threshold for the death of a tumor cell since oxygen availability varies in the tumor \cite{Pappas2016, 
GAP16, KHP14, IRP13}.  However, scrutiny is precisely what we seek because this will lead to more sophisticated models,
but these will still be in the framework of drug diffusion inside a solid tumor leading to tumor cell death, and eventually
ablation.  We shall endeavor to develop more models in this framework, and we hope that other researchers will
as well.

\section*{Acknowledgment}

The authors would like to express their gratitude to NIH (Grant \#: 1R01GM122084-01) for supporting this work.
A.R. and S.G. are also grateful to D. Pappas and A. Ibragimov for fruitful discussions.
A.R. and S.G. appreciate the support of the Department of Mathematics and Statistics at TTU, and R.P.
appreciates the support of the Department of Electrical and Computer Engineering at TTU.

\bibliographystyle{unsrt}
\bibliography{Cancer}

\appendix
\section{Dose-response curves}\label{Appendix: Dose-response}

\begin{figure}[htbp]
\stackinset{l}{-10mm}{t}{16mm}{\textbf{\small \rotatebox{90}{C32}}}{\stackinset{l}{7mm}{t}{-10mm}{\textbf{\small SB590885}}{\includegraphics[width = 0.19\textwidth]{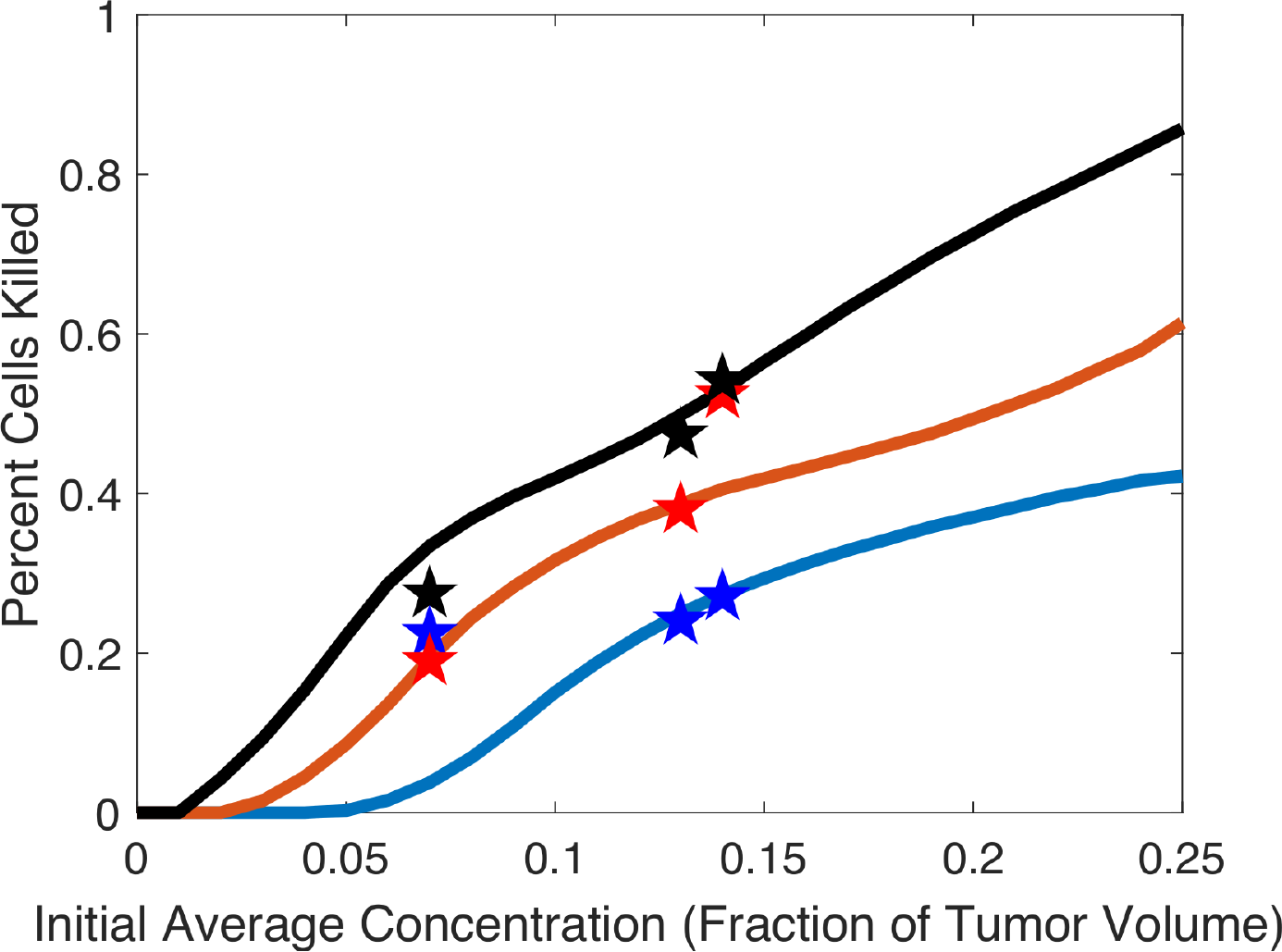}}}
\stackinset{l}{7mm}{t}{-10mm}{\textbf{\small PLX-4720}}{\includegraphics[width = 0.19\textwidth]{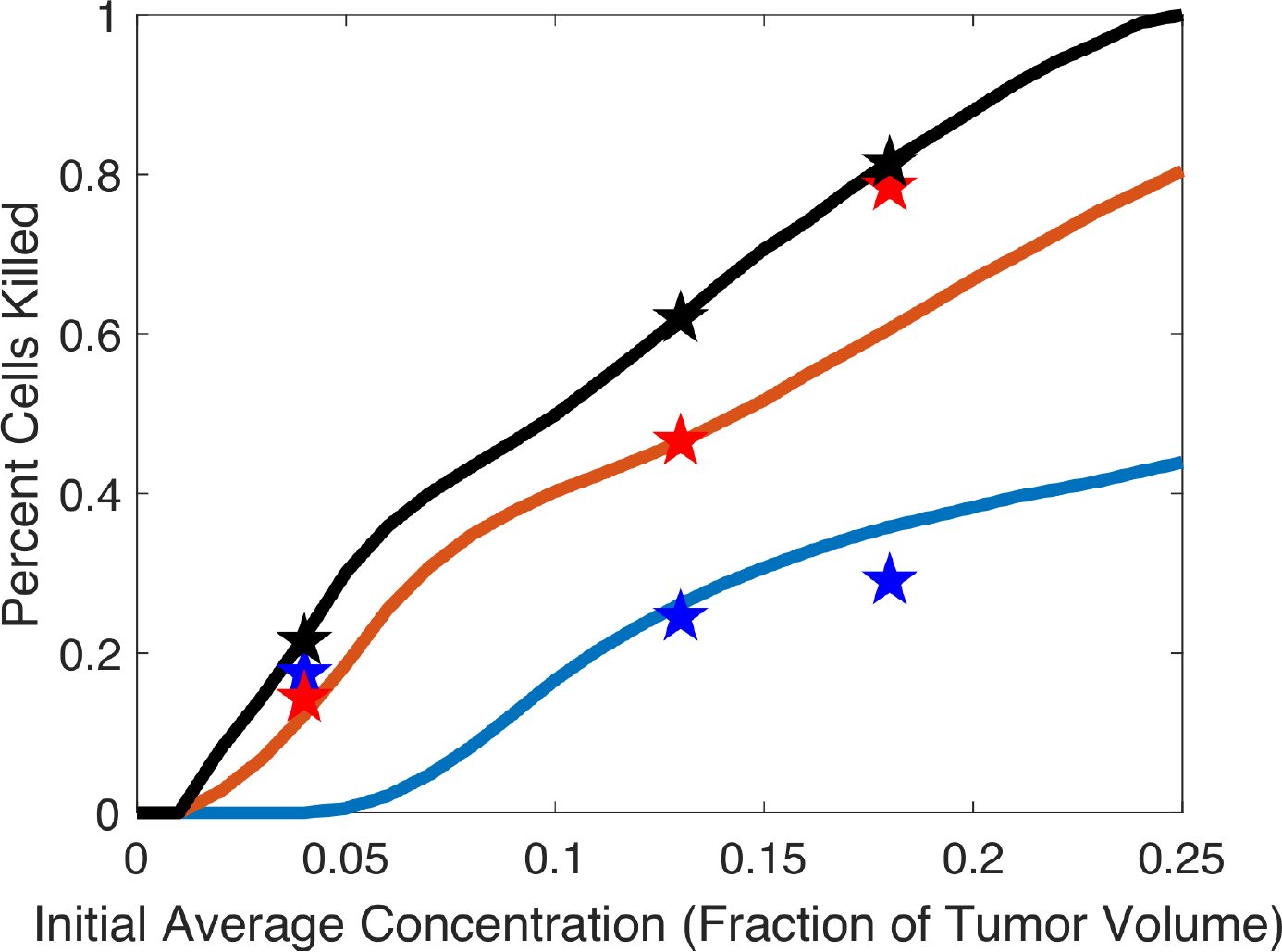}}
\stackinset{l}{9mm}{t}{-10mm}{\textbf{\small AZ-628}}{\includegraphics[width = 0.19\textwidth]{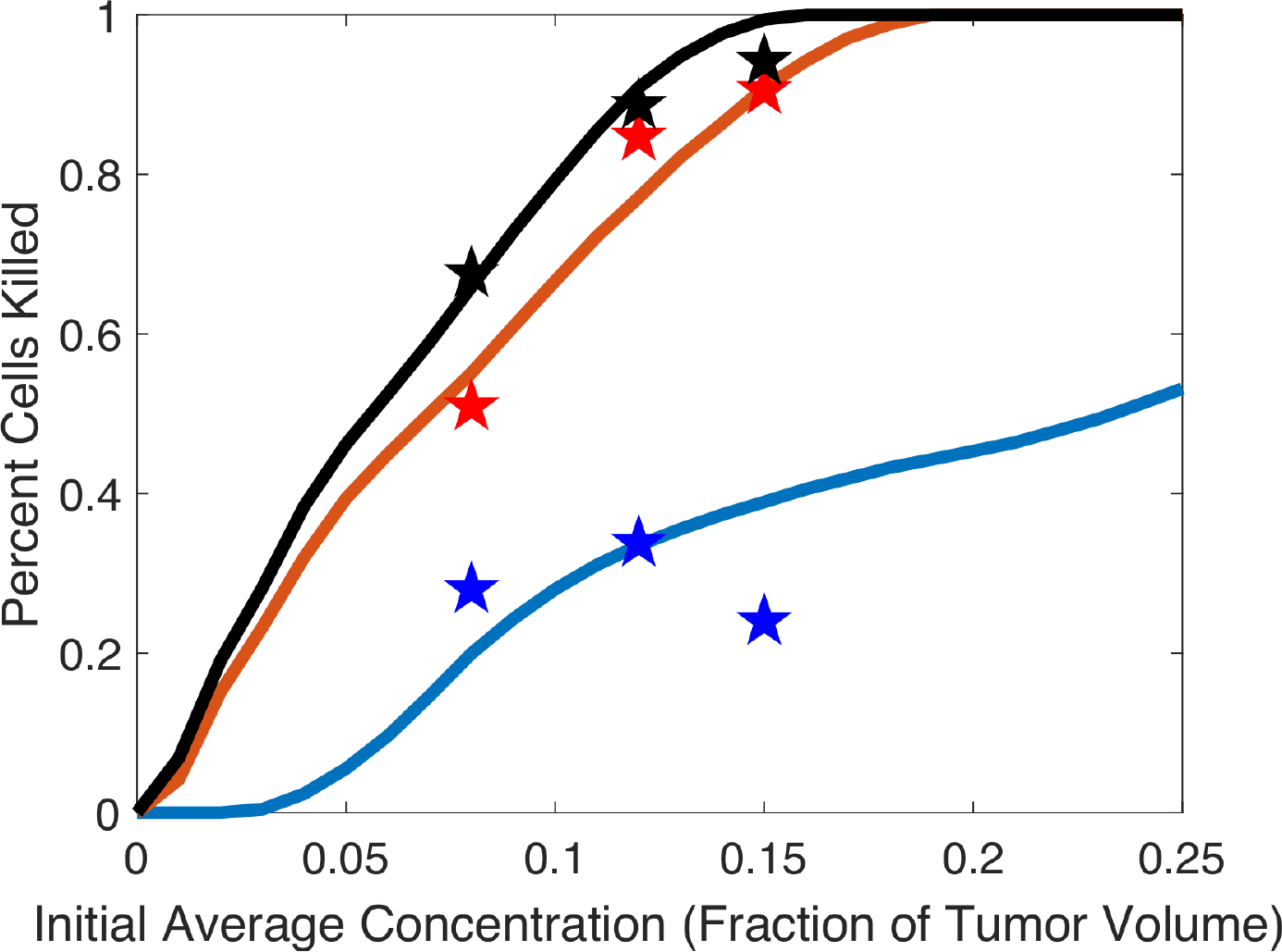}}
\stackinset{l}{4mm}{t}{-10mm}{\textbf{\small Selumetinib}}{\includegraphics[width = 0.19\textwidth]{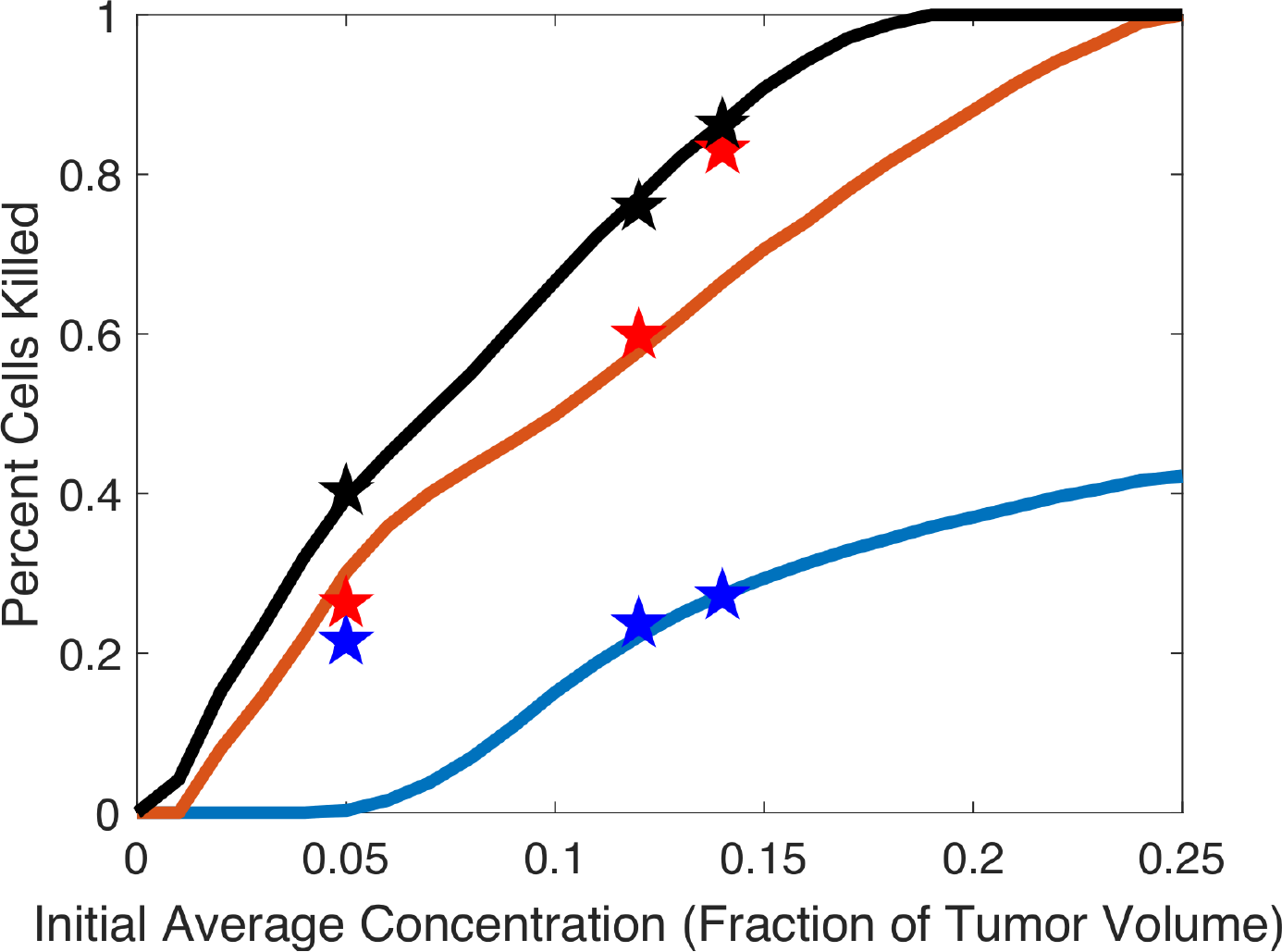}}
\stackinset{l}{4mm}{t}{-10mm}{\textbf{\small Vemurafenib}}{\includegraphics[width = 0.19\textwidth]{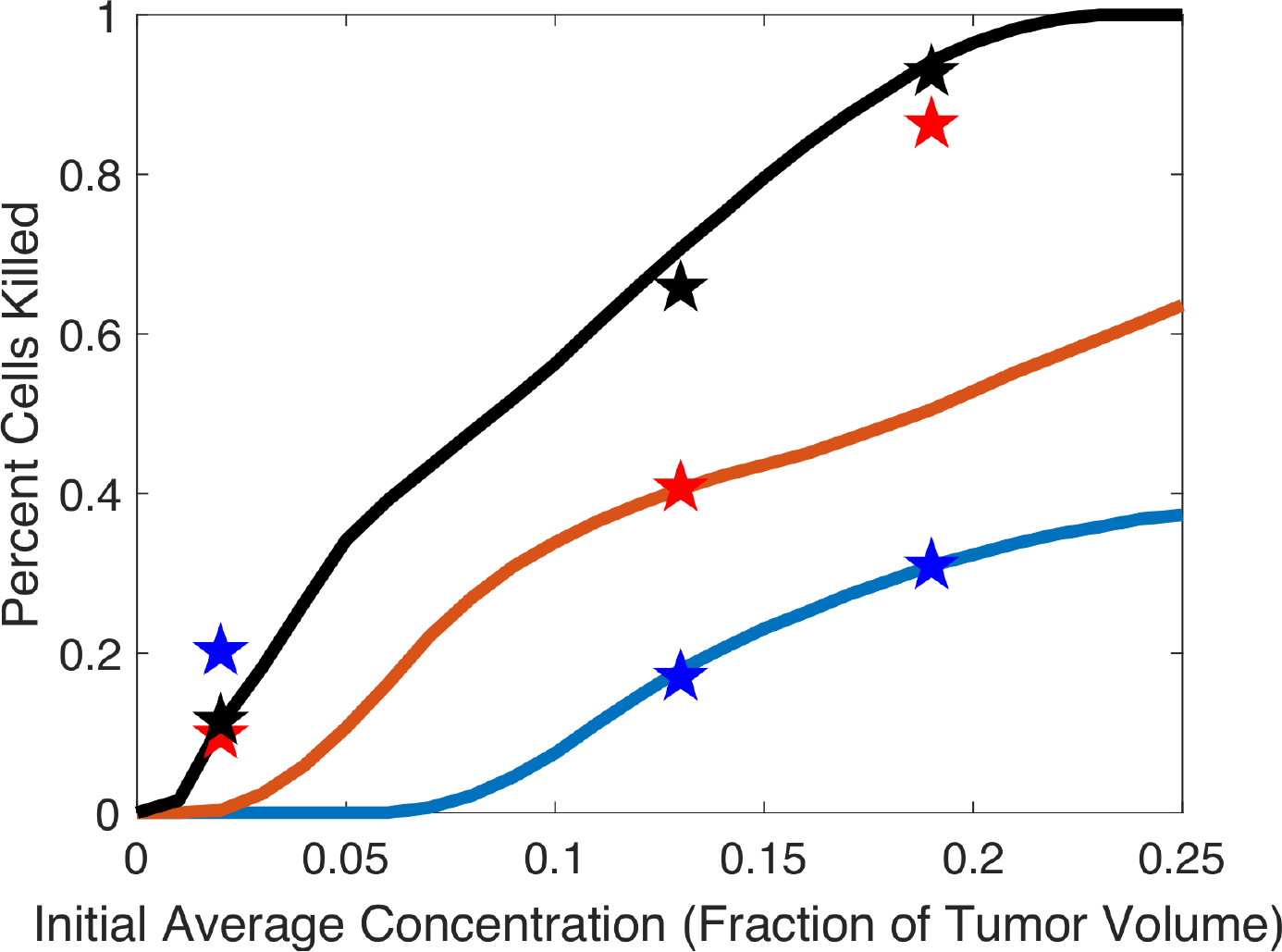}}

\bigskip
\stackinset{l}{-10mm}{t}{}{\textbf{\small \rotatebox{90}{COLO 858}}}{\includegraphics[width = 0.19\textwidth]{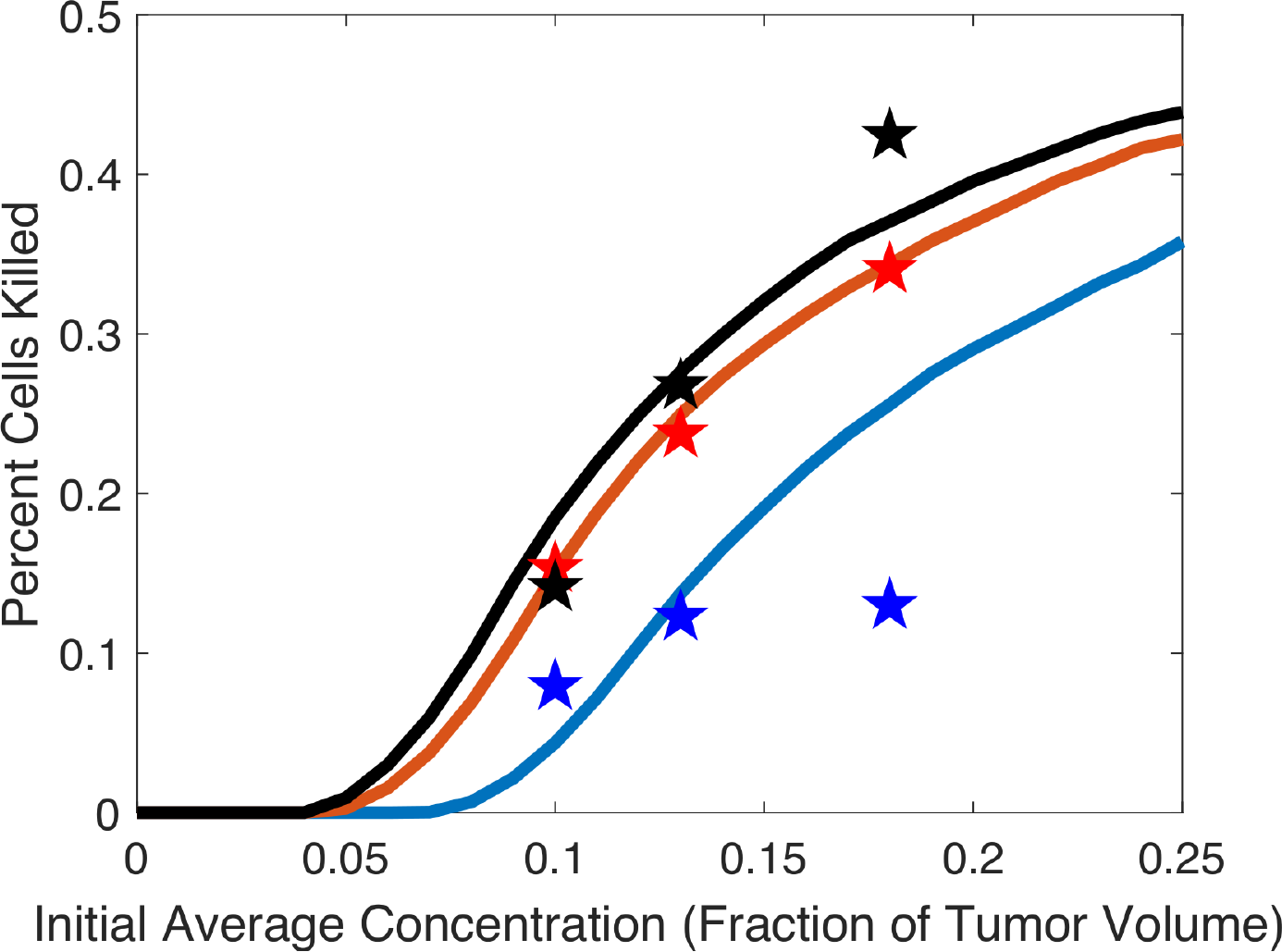}}
\includegraphics[width = 0.19\textwidth]{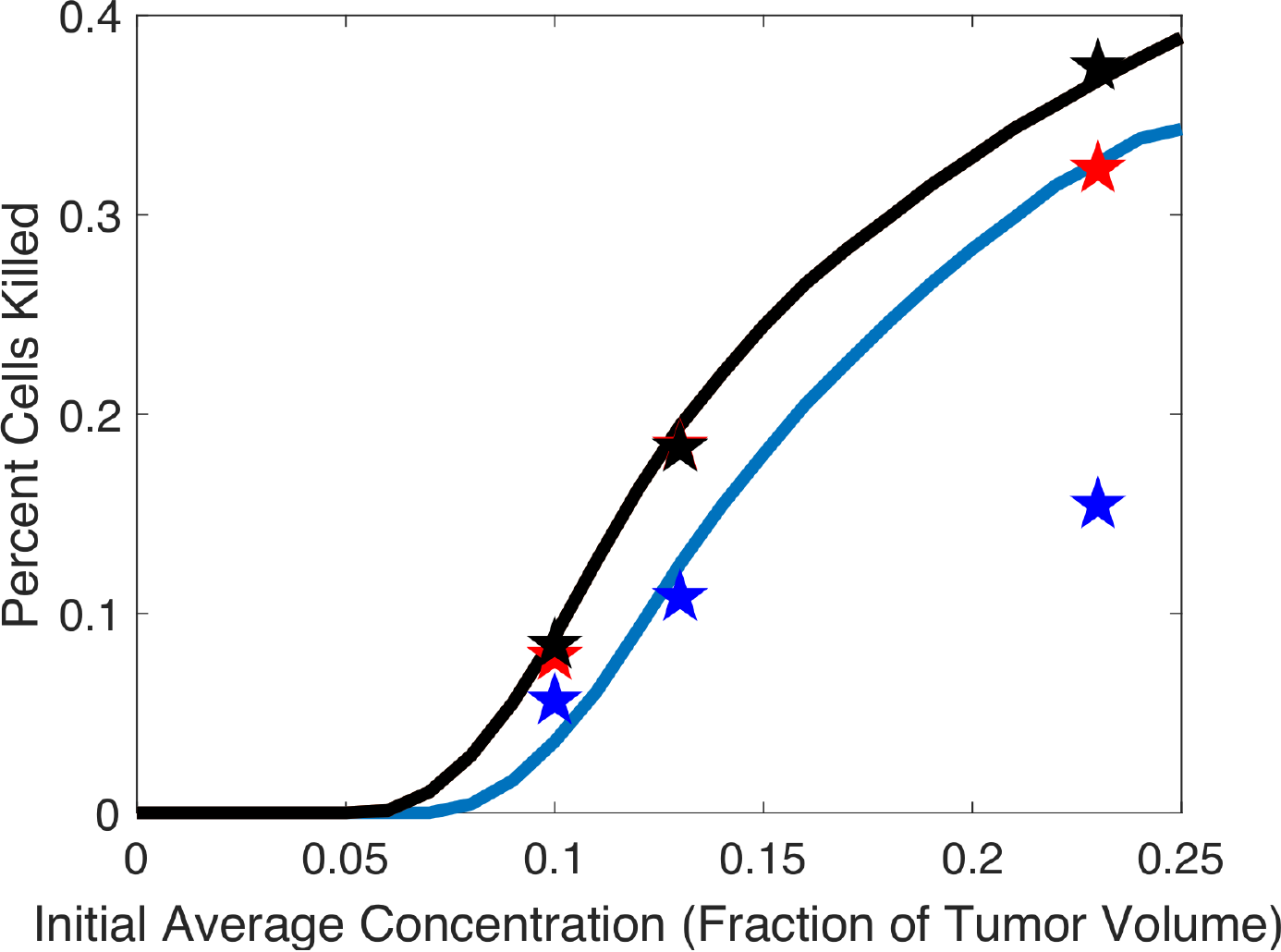}
\includegraphics[width = 0.19\textwidth]{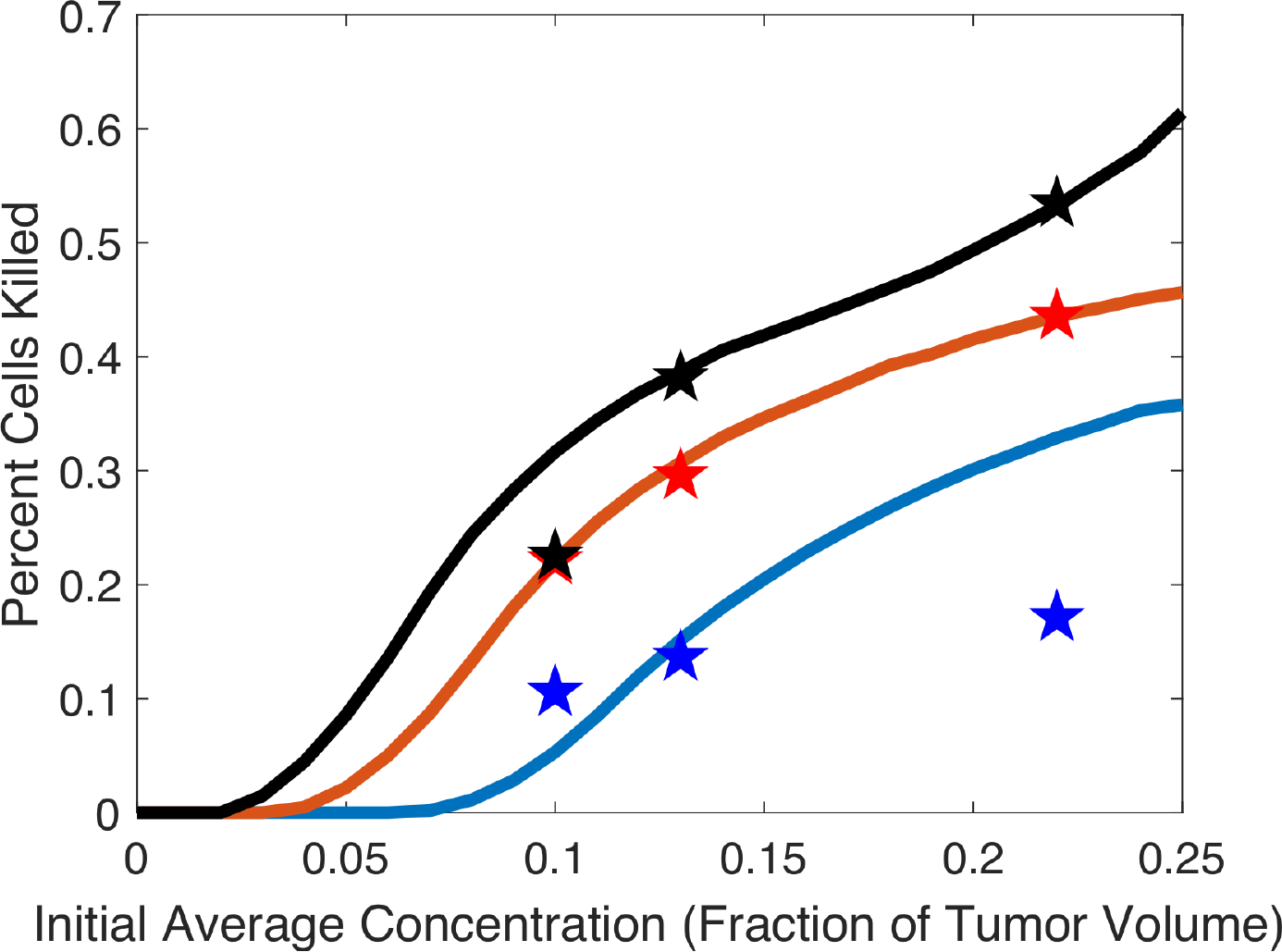}
\includegraphics[width = 0.19\textwidth]{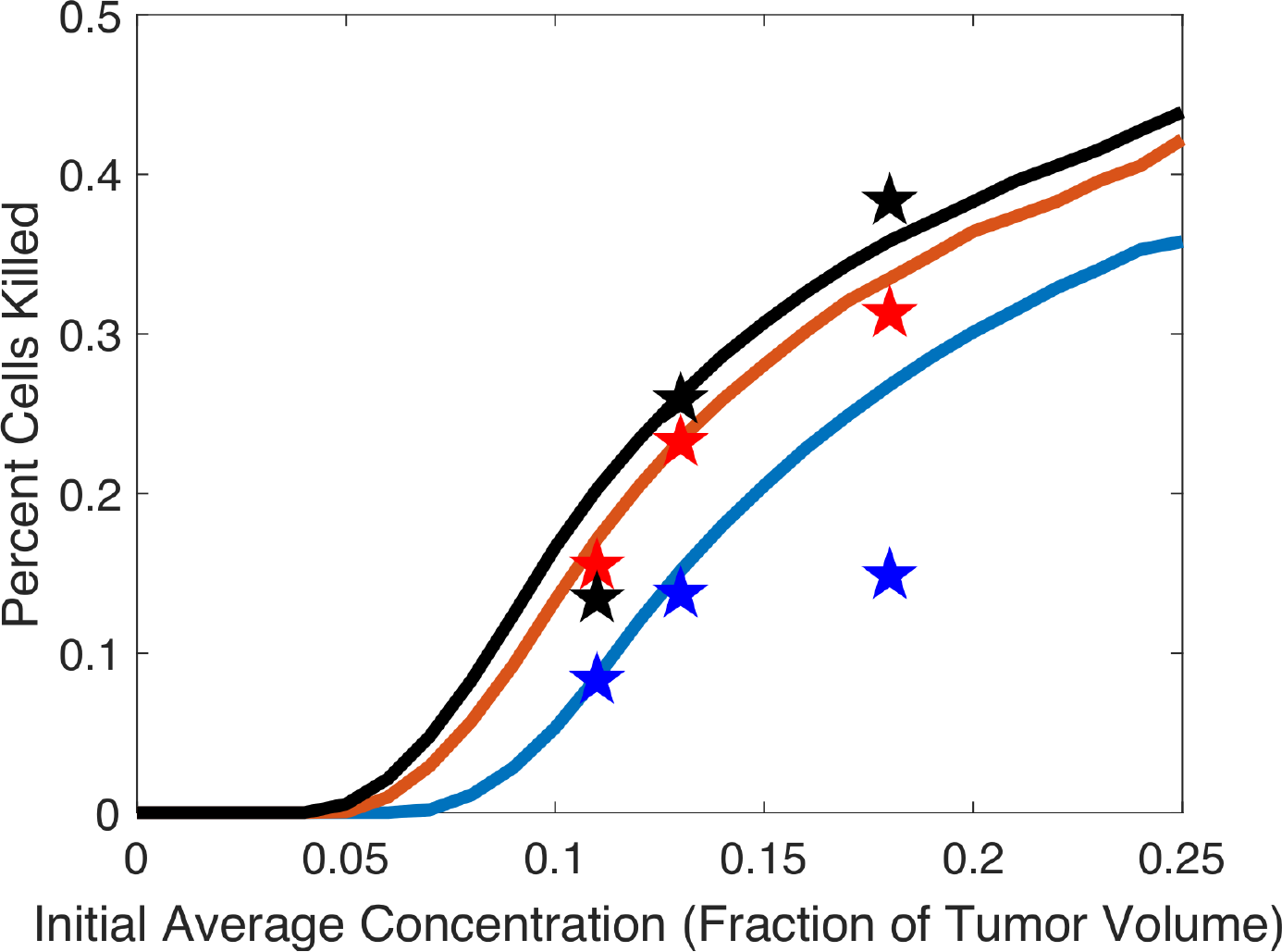}
\includegraphics[width = 0.19\textwidth]{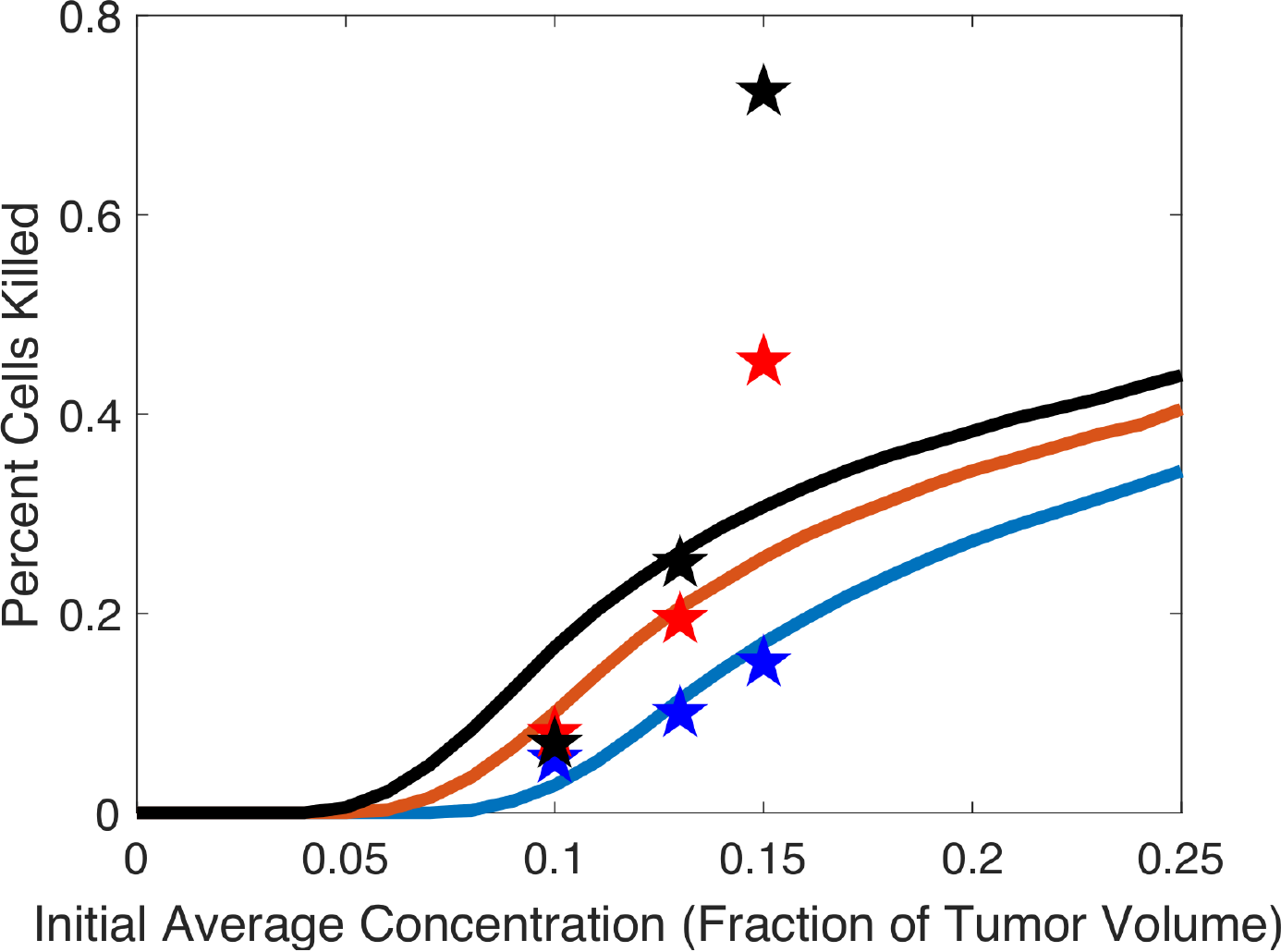}

\bigskip
\stackinset{l}{-10mm}{t}{1mm}{\textbf{\small \rotatebox{90}{RVH-421}}}{\includegraphics[width = 0.19\textwidth]{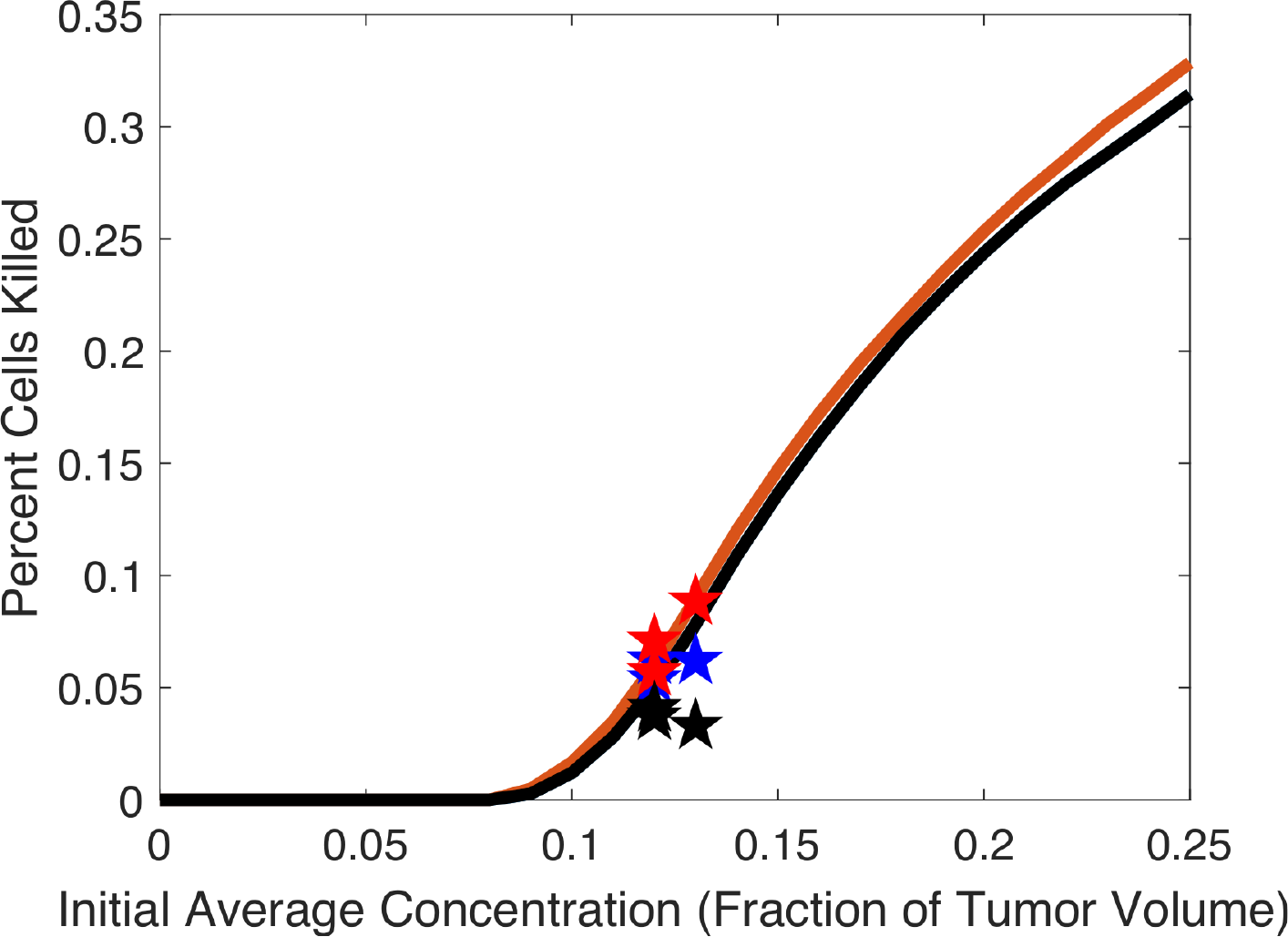}}
\includegraphics[width = 0.19\textwidth]{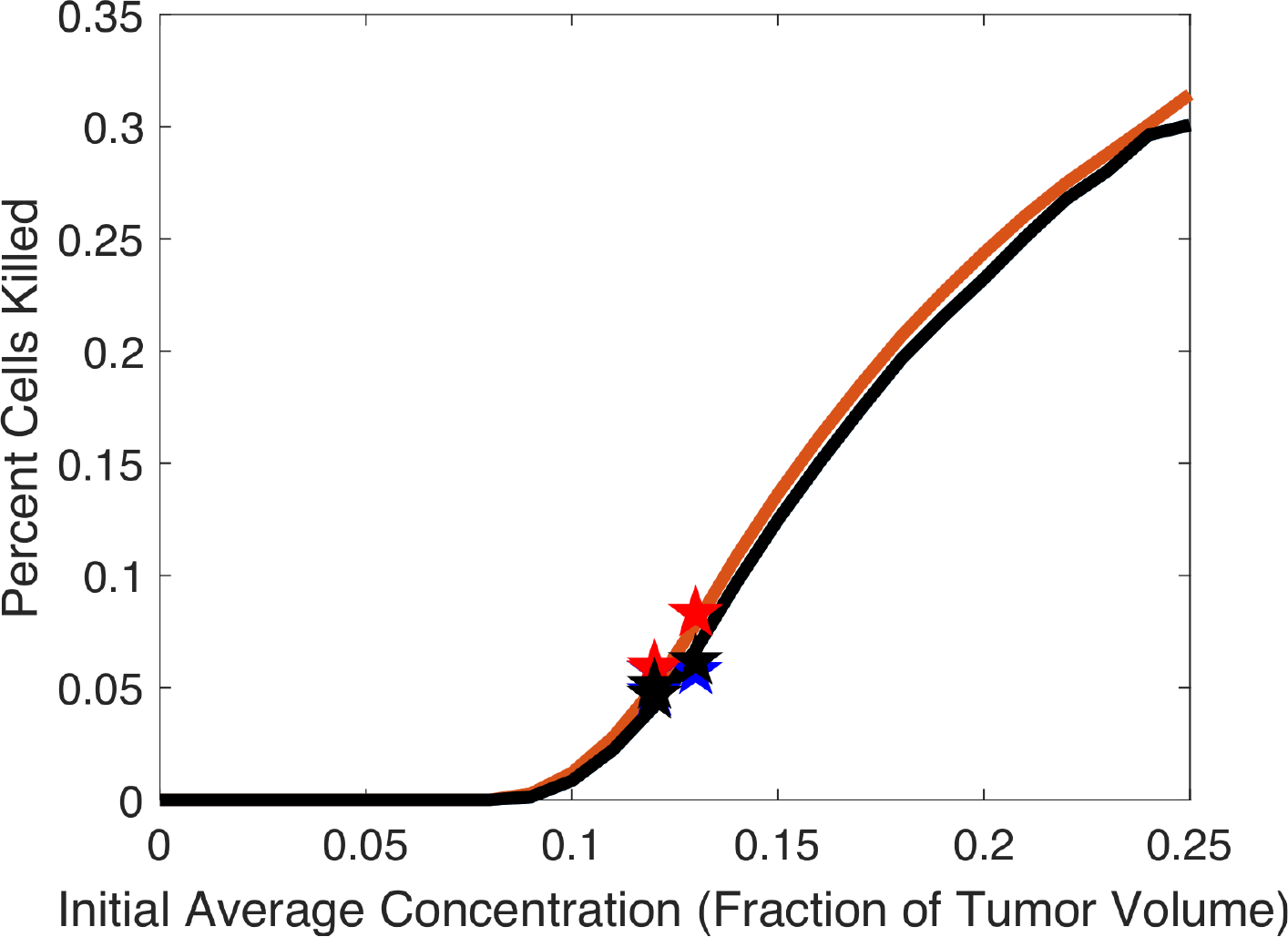}
\includegraphics[width = 0.19\textwidth]{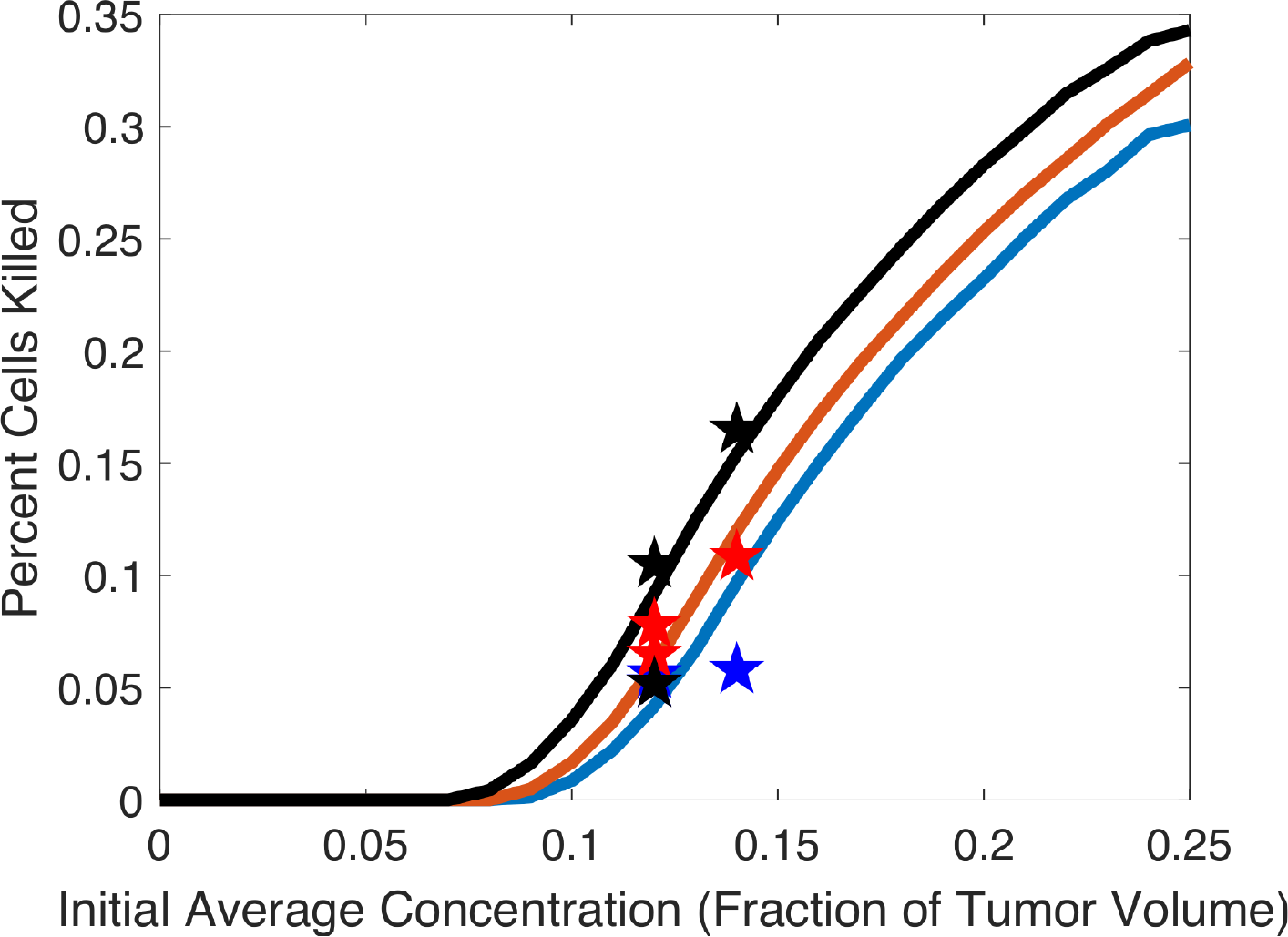}
\includegraphics[width = 0.19\textwidth]{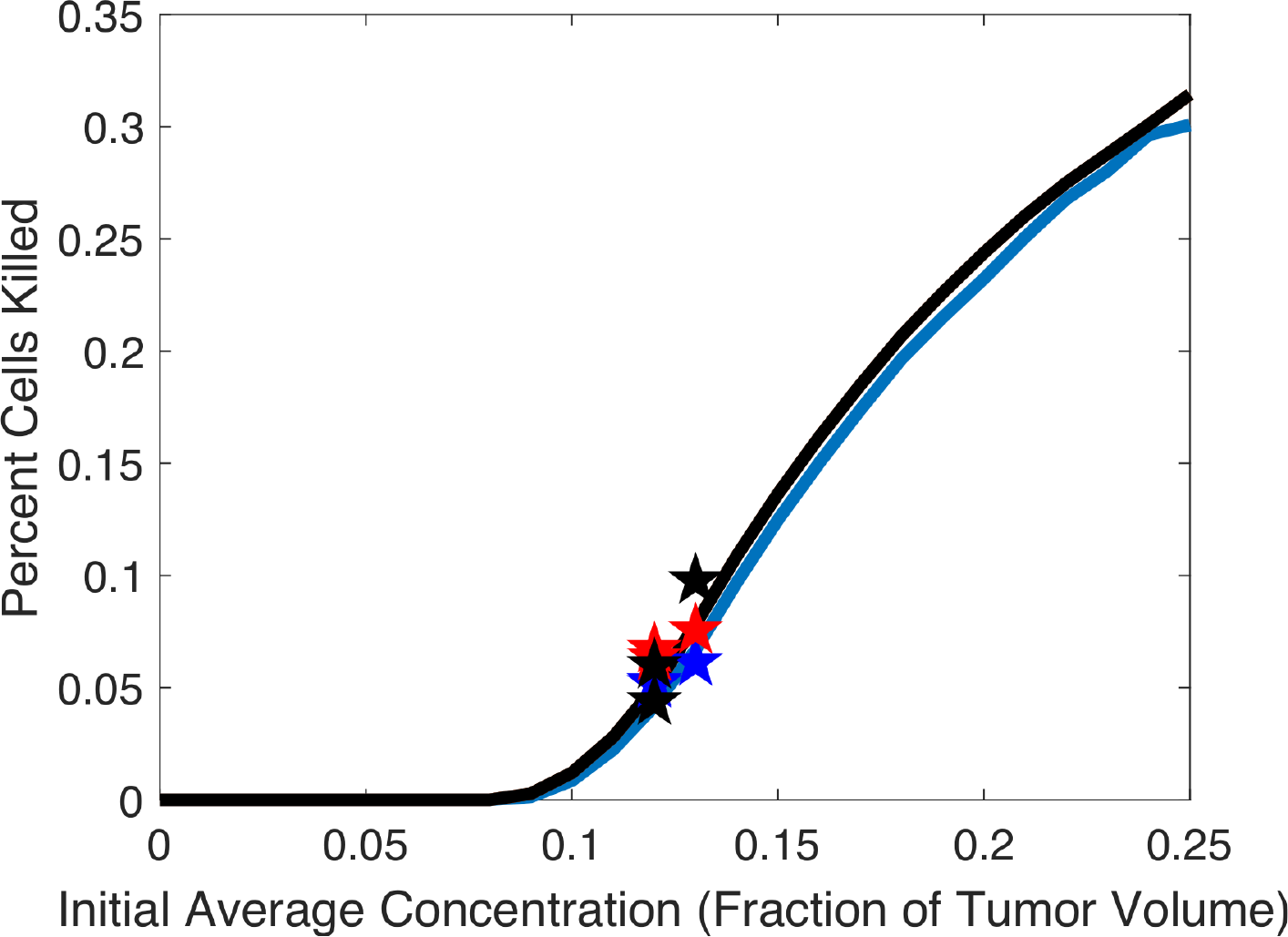}
\includegraphics[width = 0.19\textwidth]{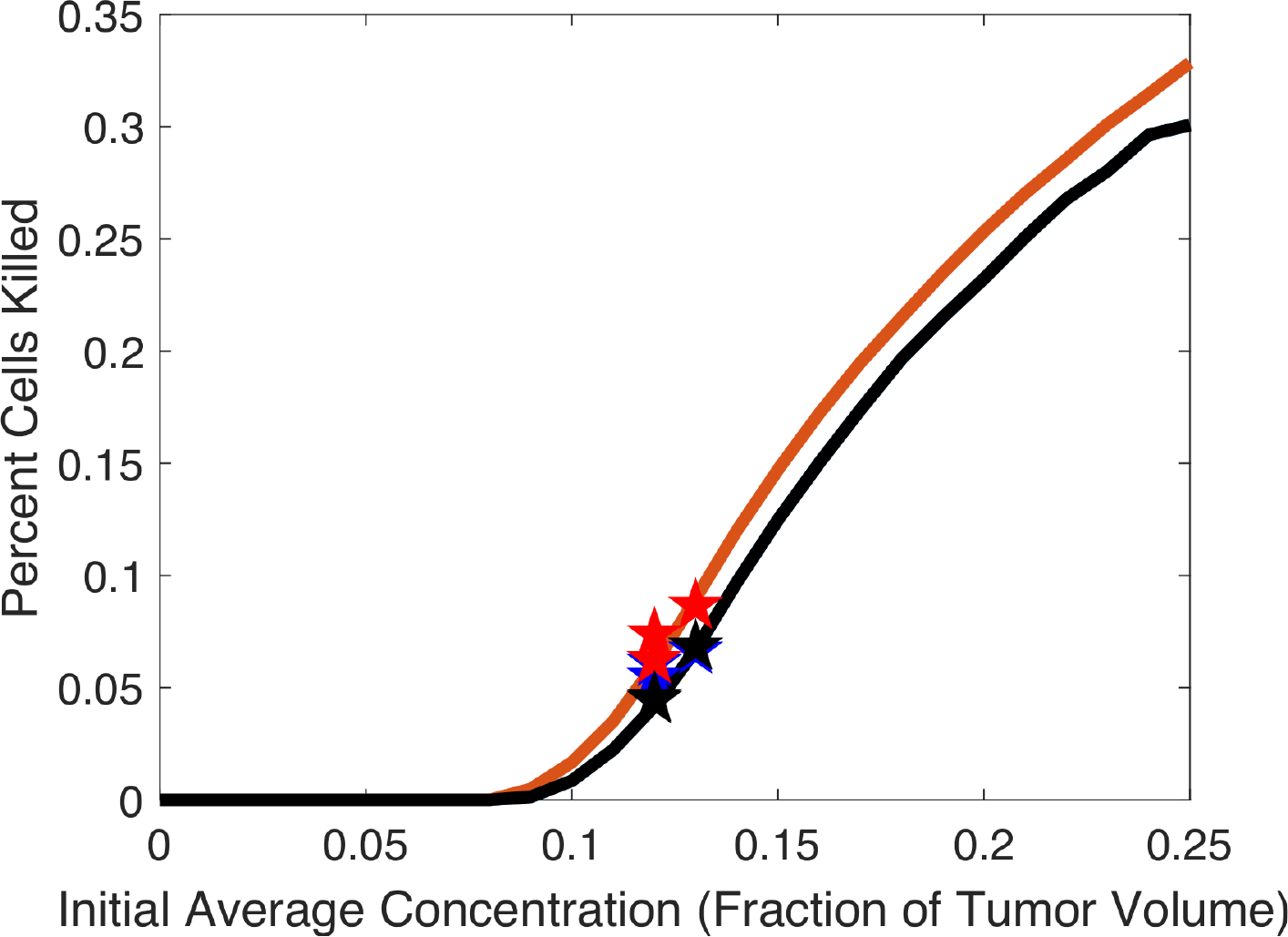}

\bigskip
\stackinset{l}{-10mm}{t}{1mm}{\textbf{\small \rotatebox{90}{WM-115}}}{\includegraphics[width = 0.19\textwidth]{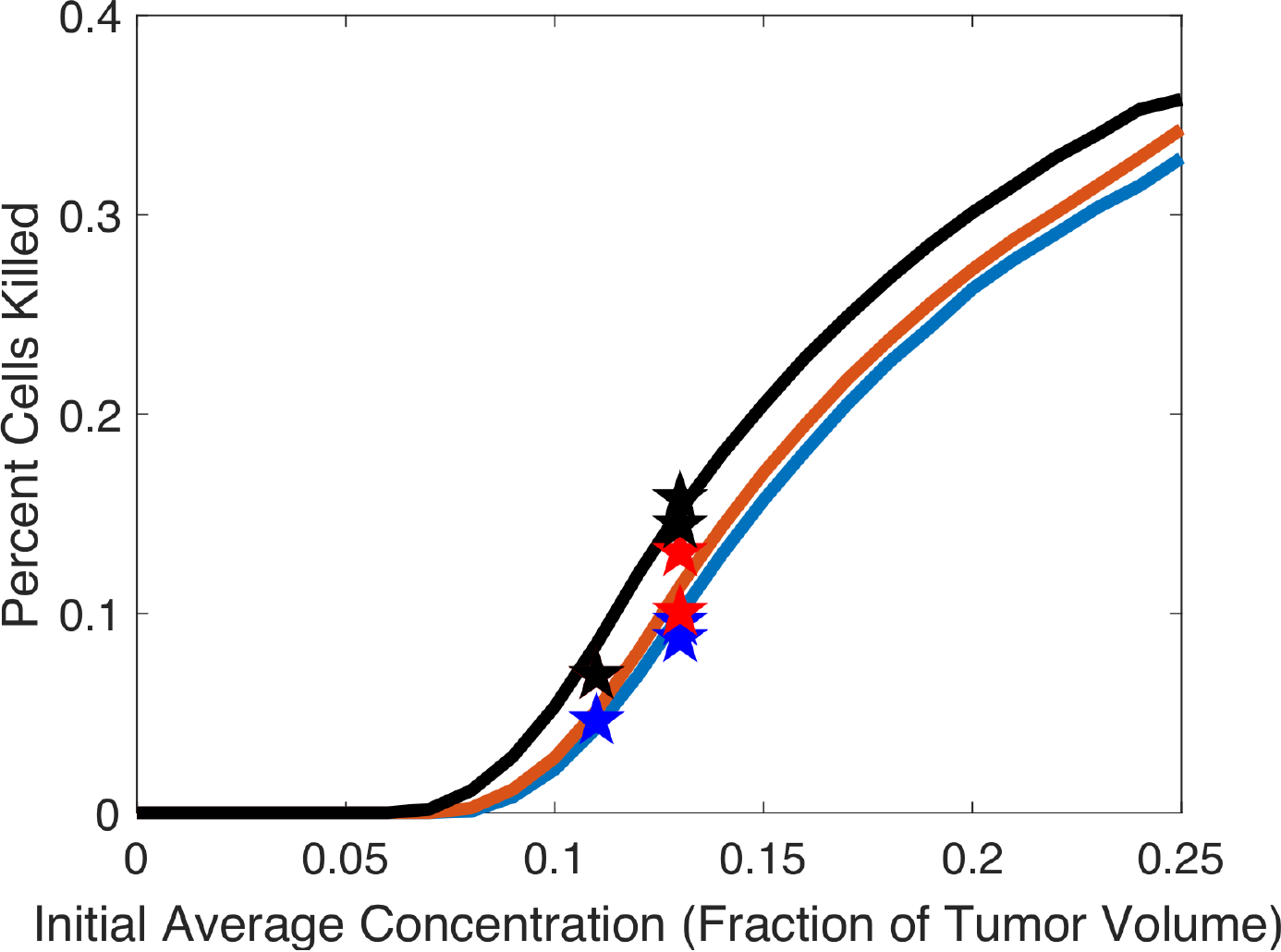}}
\includegraphics[width = 0.19\textwidth]{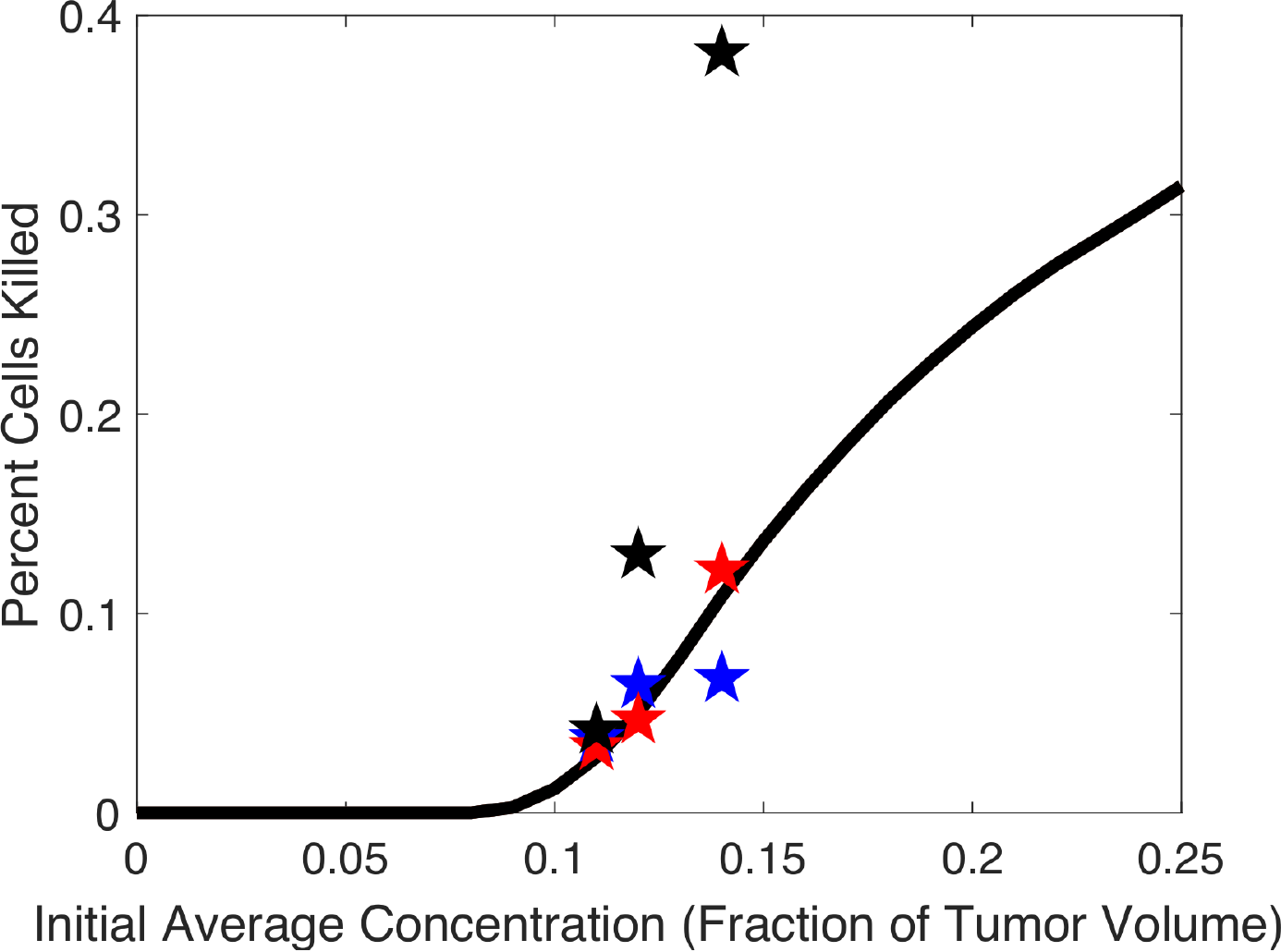}
\includegraphics[width = 0.19\textwidth]{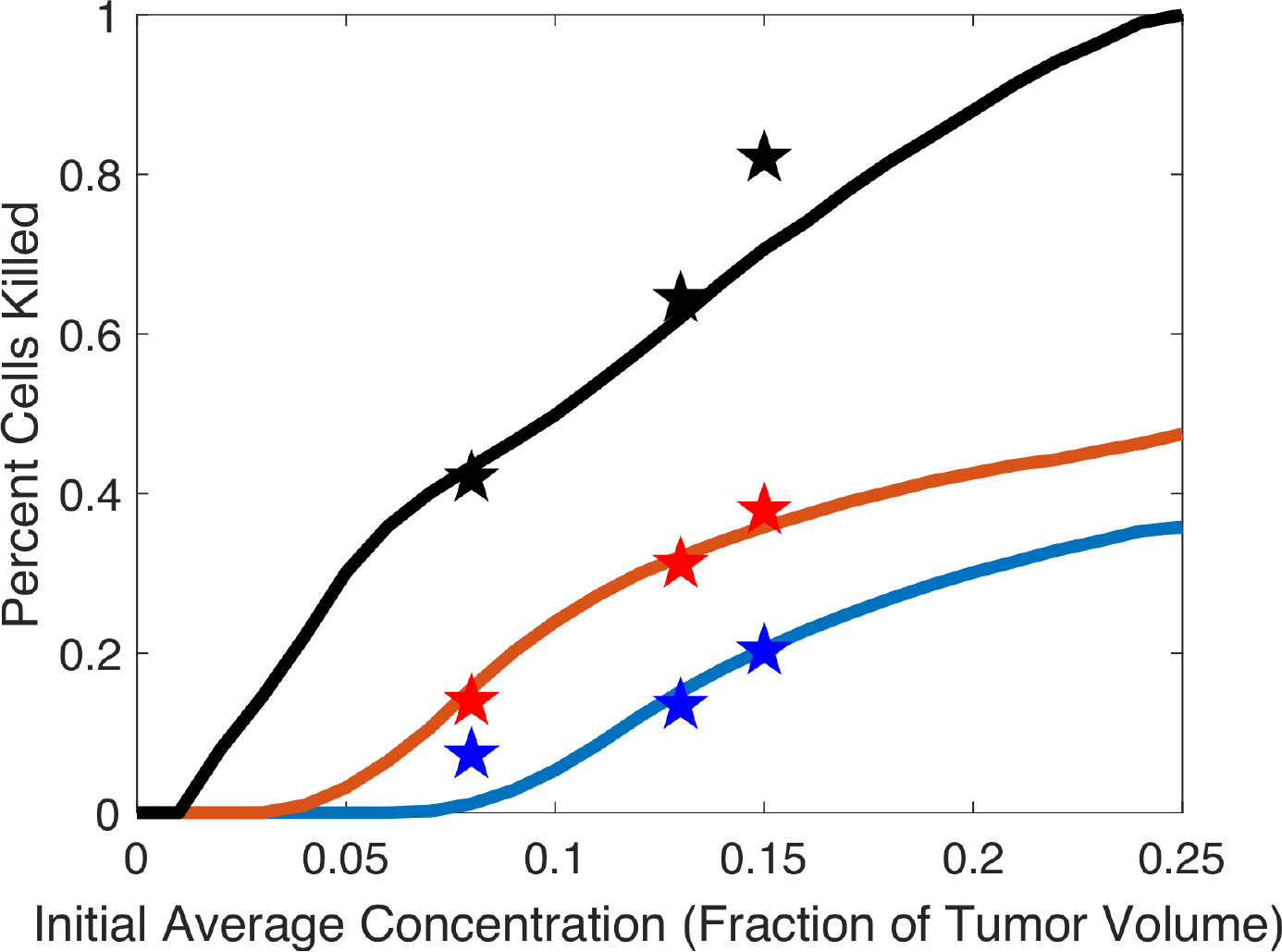}
\includegraphics[width = 0.19\textwidth]{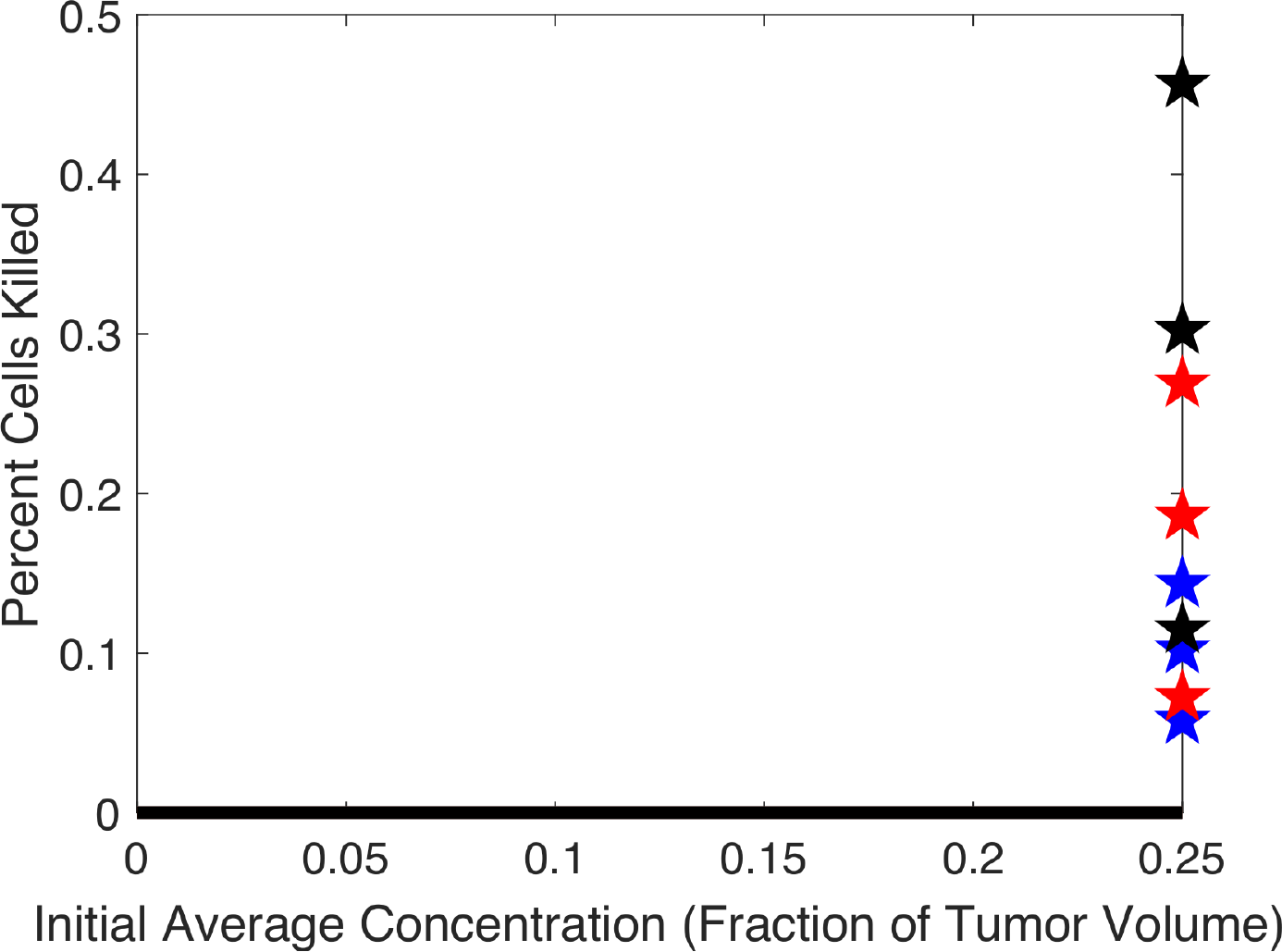}
\includegraphics[width = 0.19\textwidth]{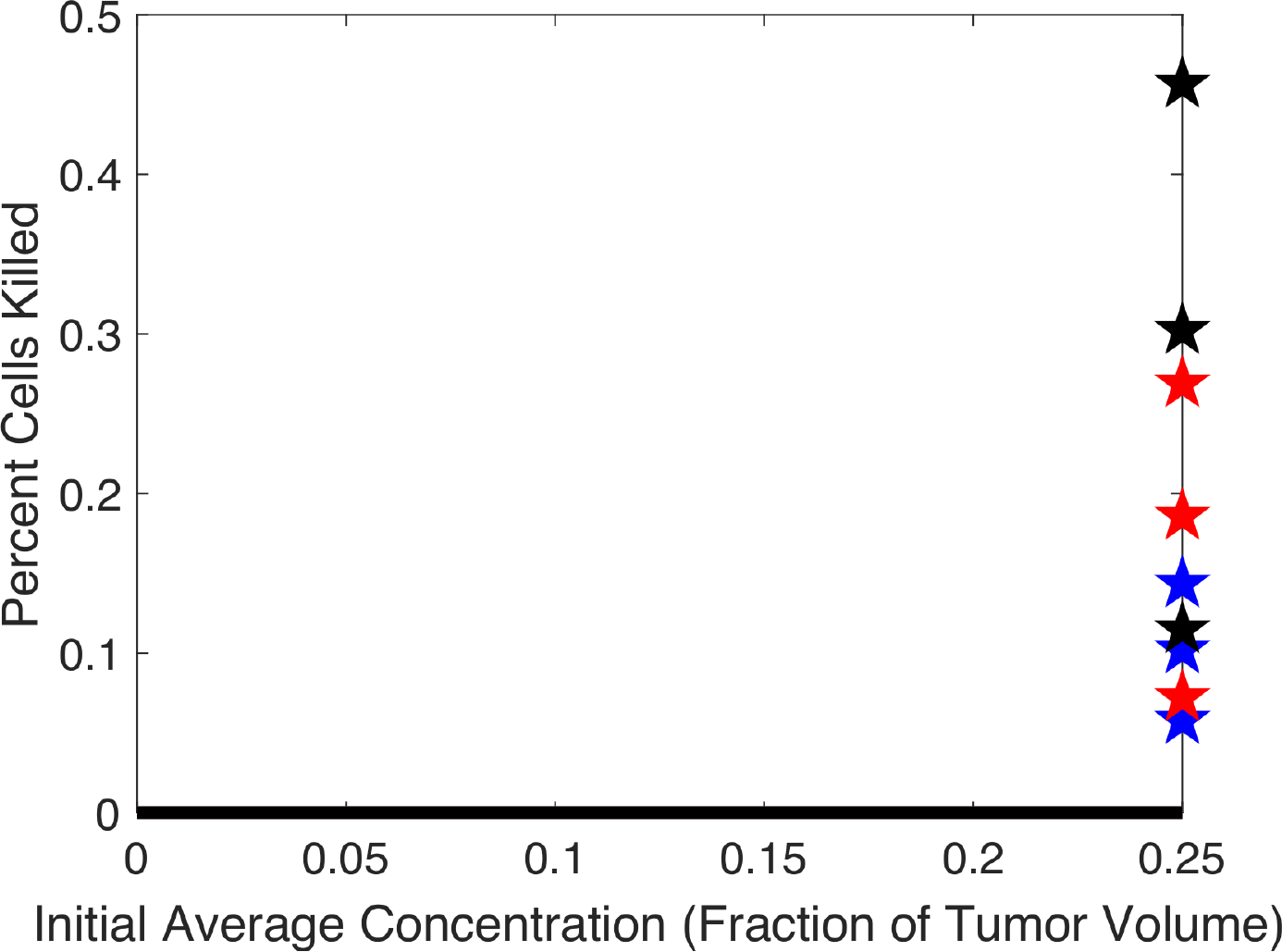}

\bigskip
\stackinset{l}{-10mm}{t}{}{\textbf{\small \rotatebox{90}{WM1152C}}}{\includegraphics[width = 0.19\textwidth]{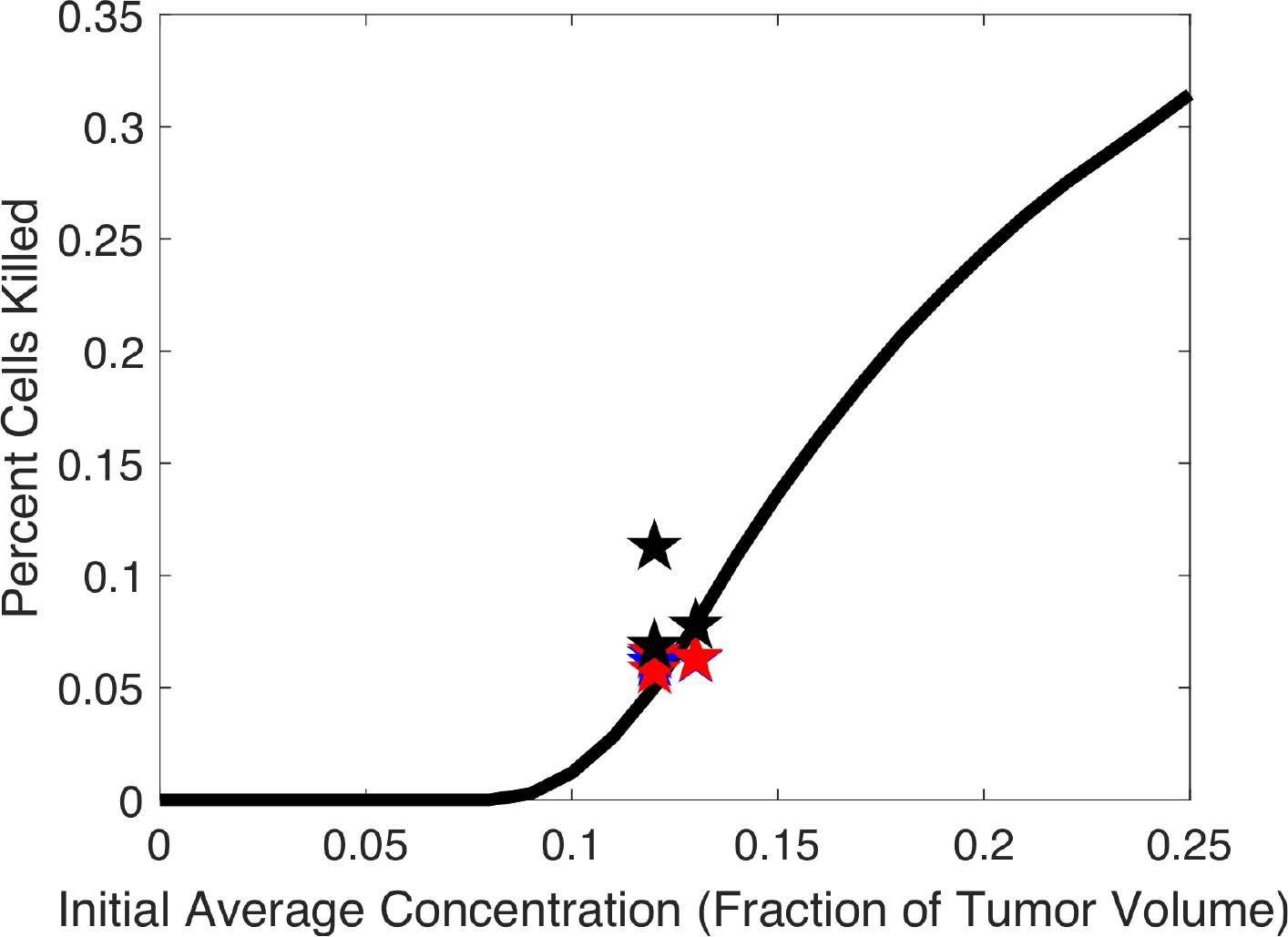}}
\includegraphics[width = 0.19\textwidth]{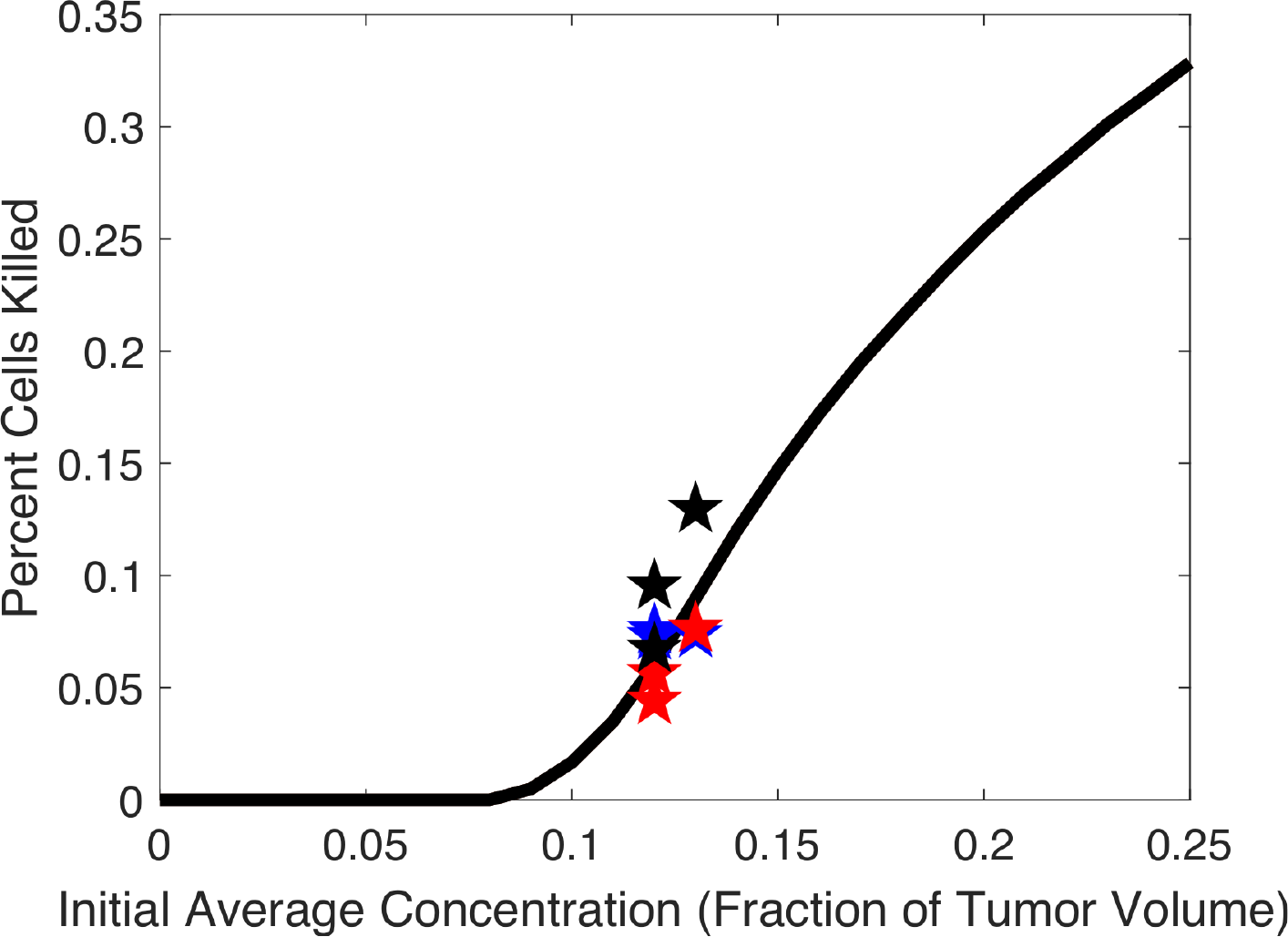}
\includegraphics[width = 0.19\textwidth]{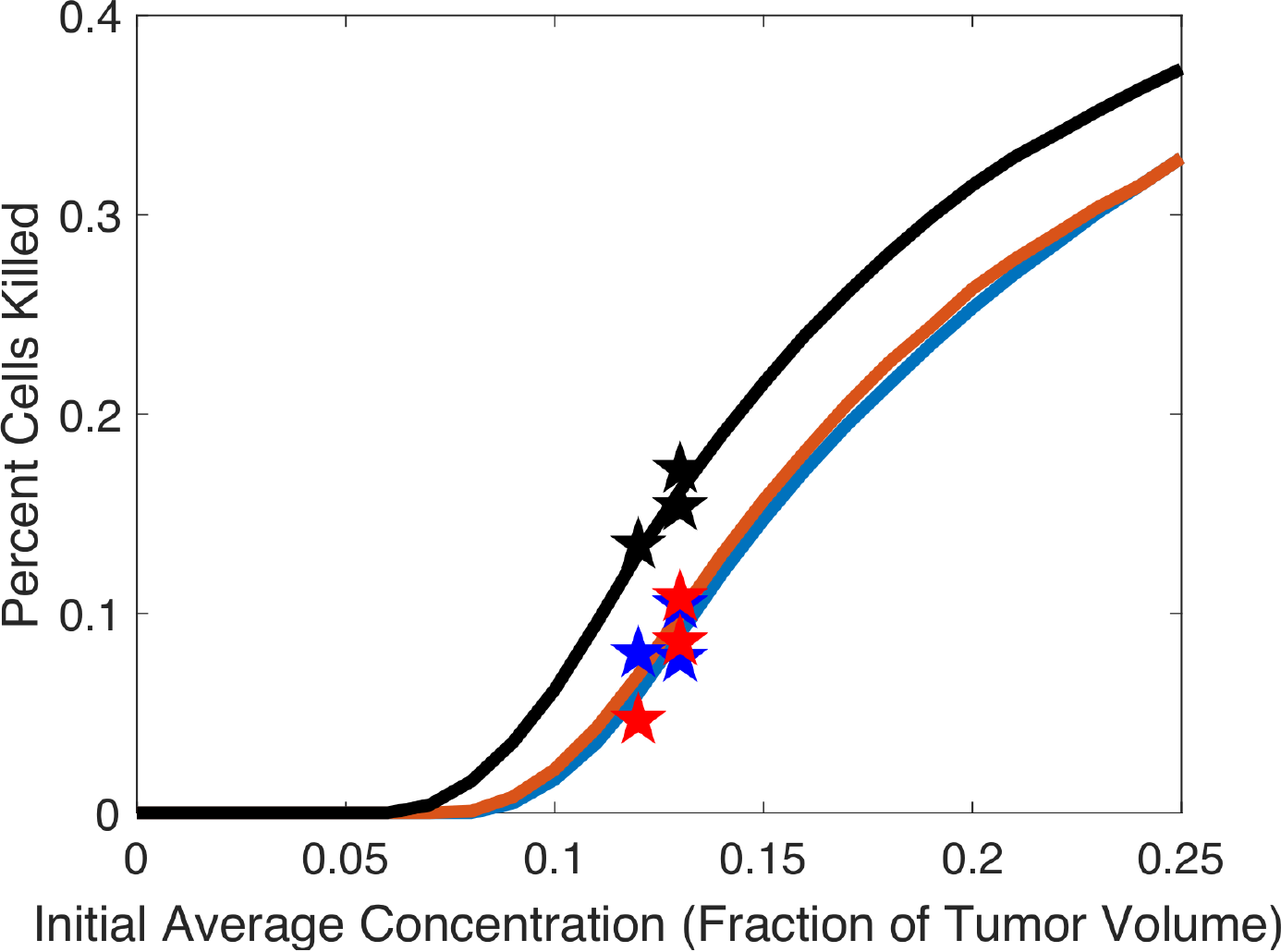}
\includegraphics[width = 0.19\textwidth]{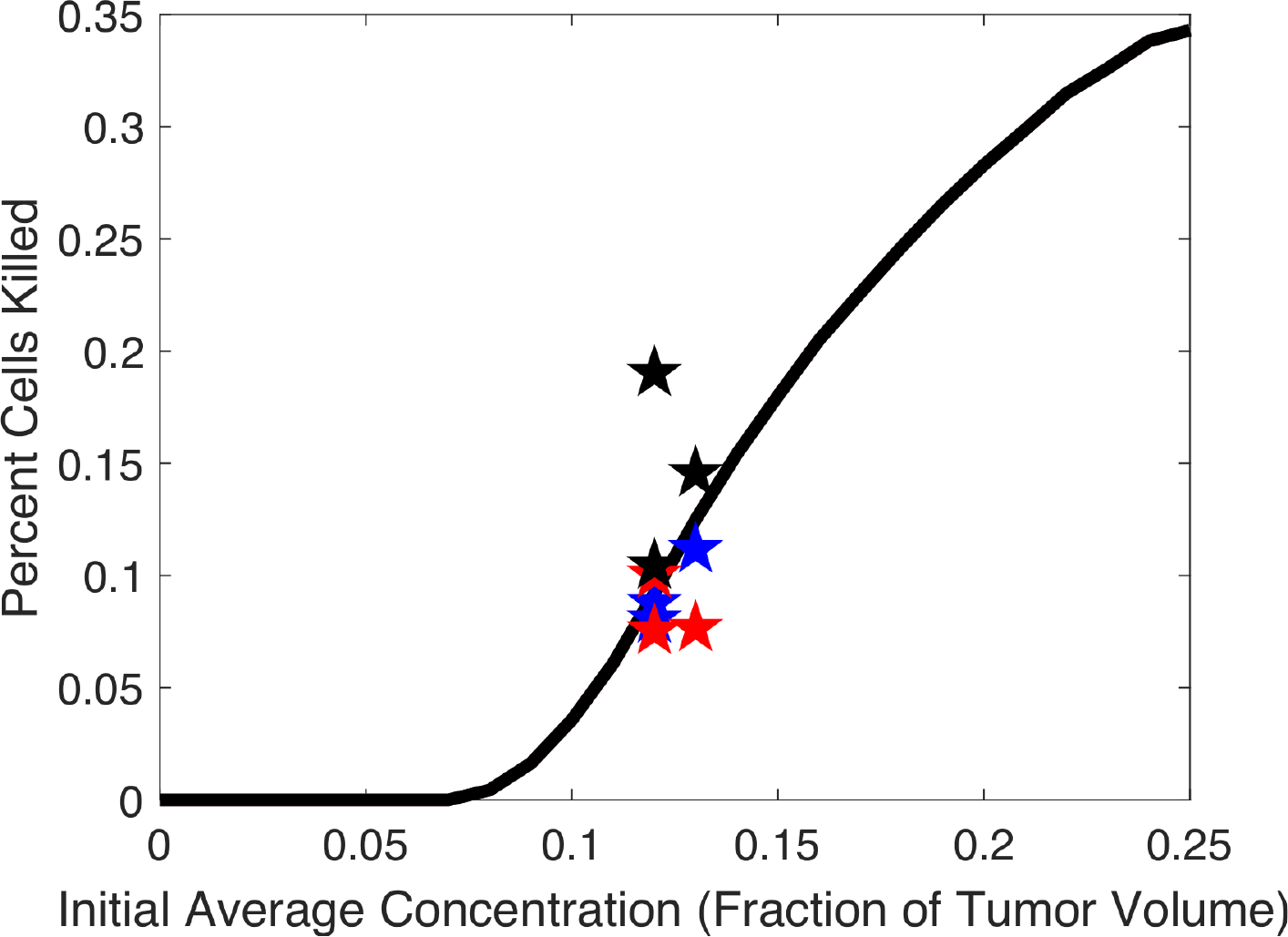}
\includegraphics[width = 0.19\textwidth]{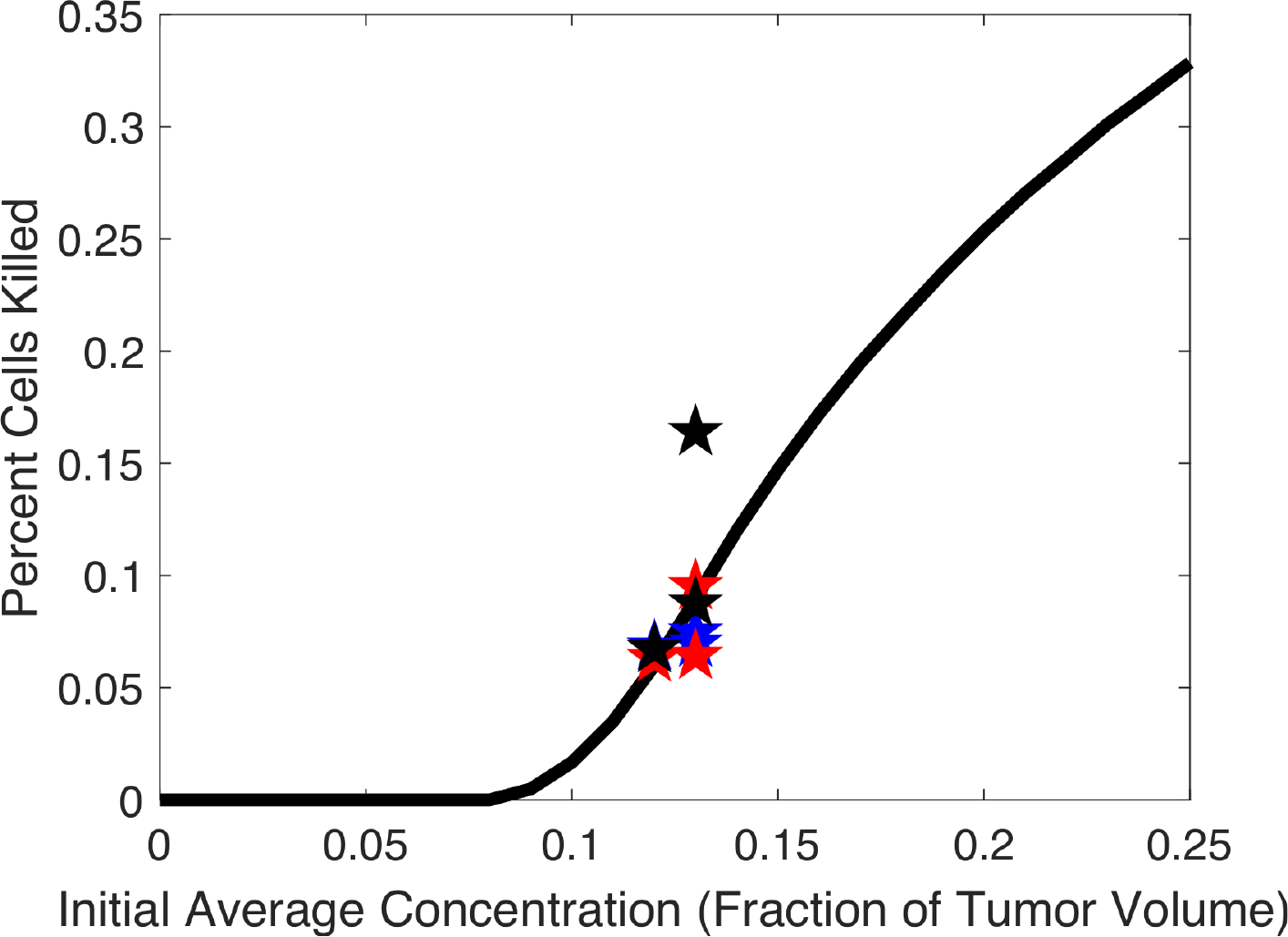}

\bigskip
\stackinset{l}{-10mm}{t}{1mm}{\textbf{\small \rotatebox{90}{LOXIMVI}}}{\includegraphics[width = 0.19\textwidth]{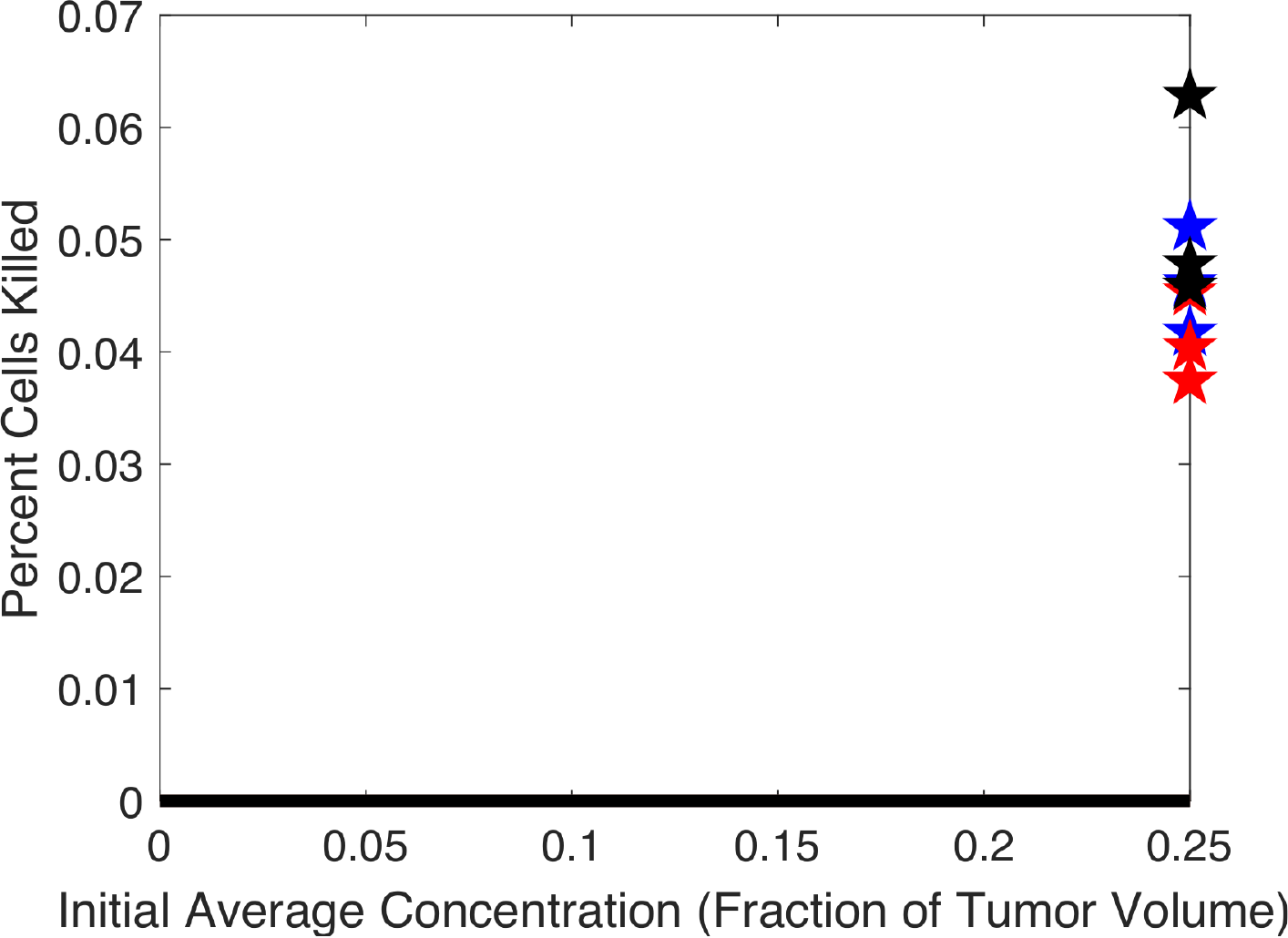}}
\includegraphics[width = 0.19\textwidth]{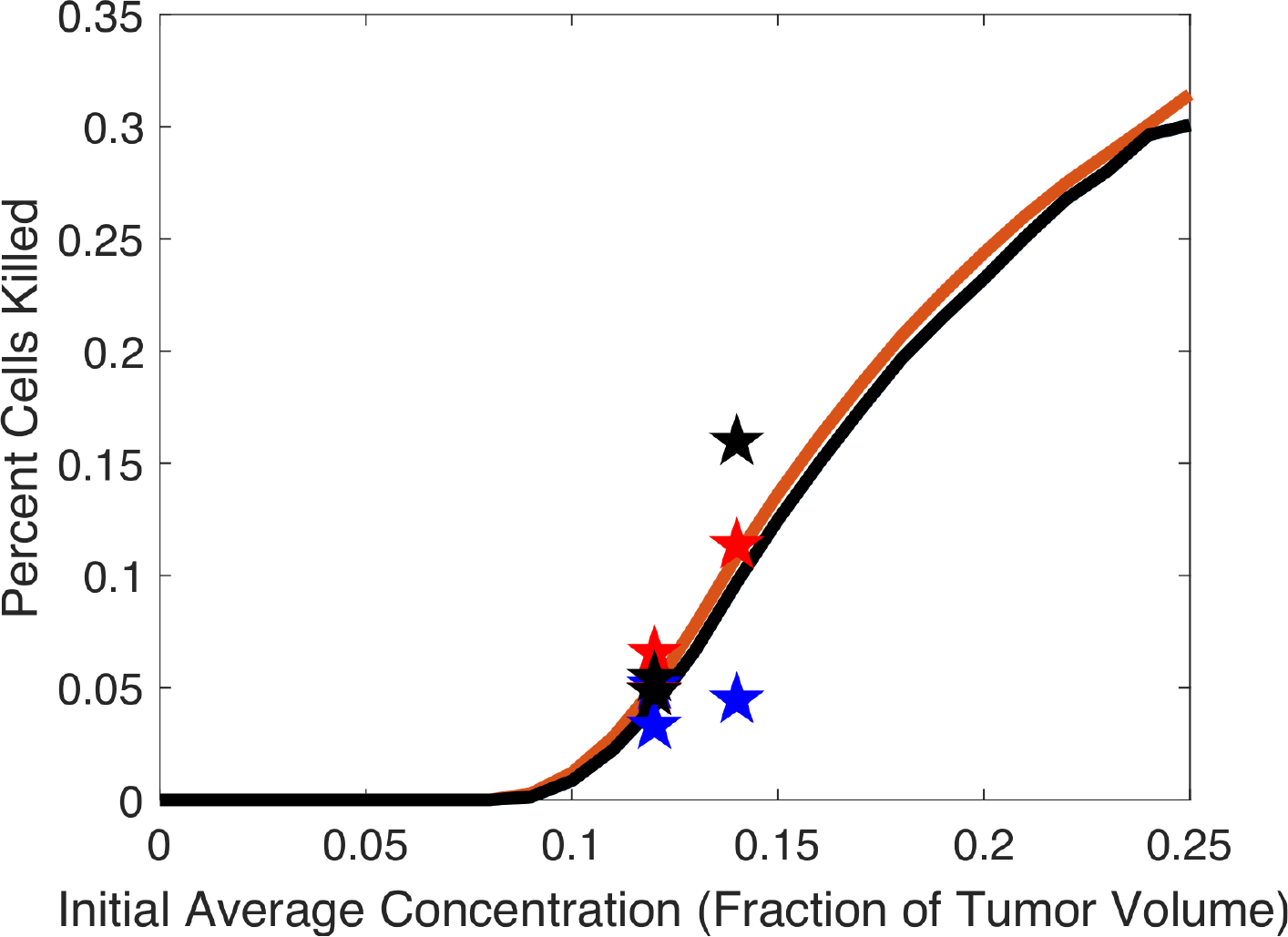}
\includegraphics[width = 0.19\textwidth]{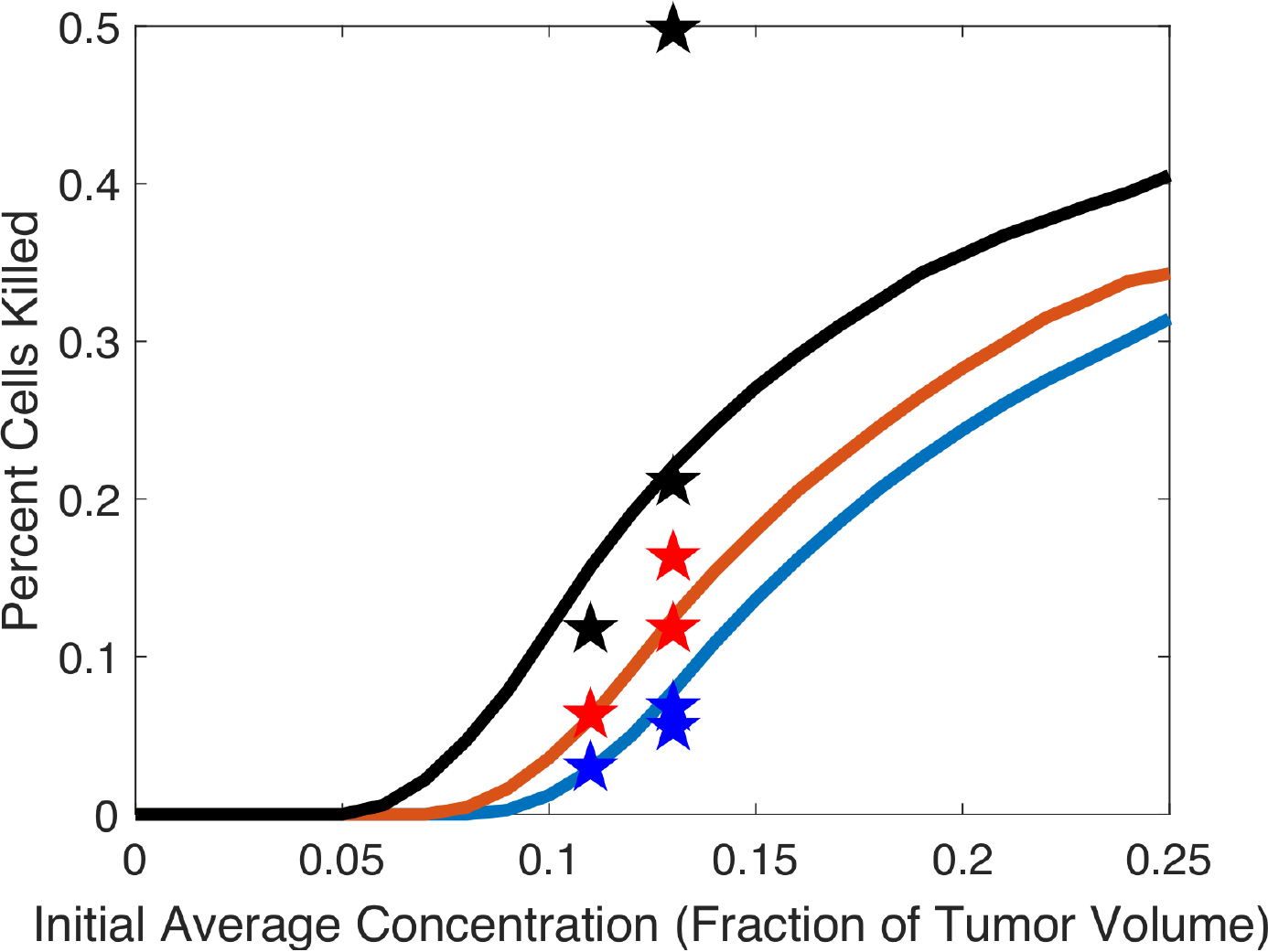}
\includegraphics[width = 0.19\textwidth]{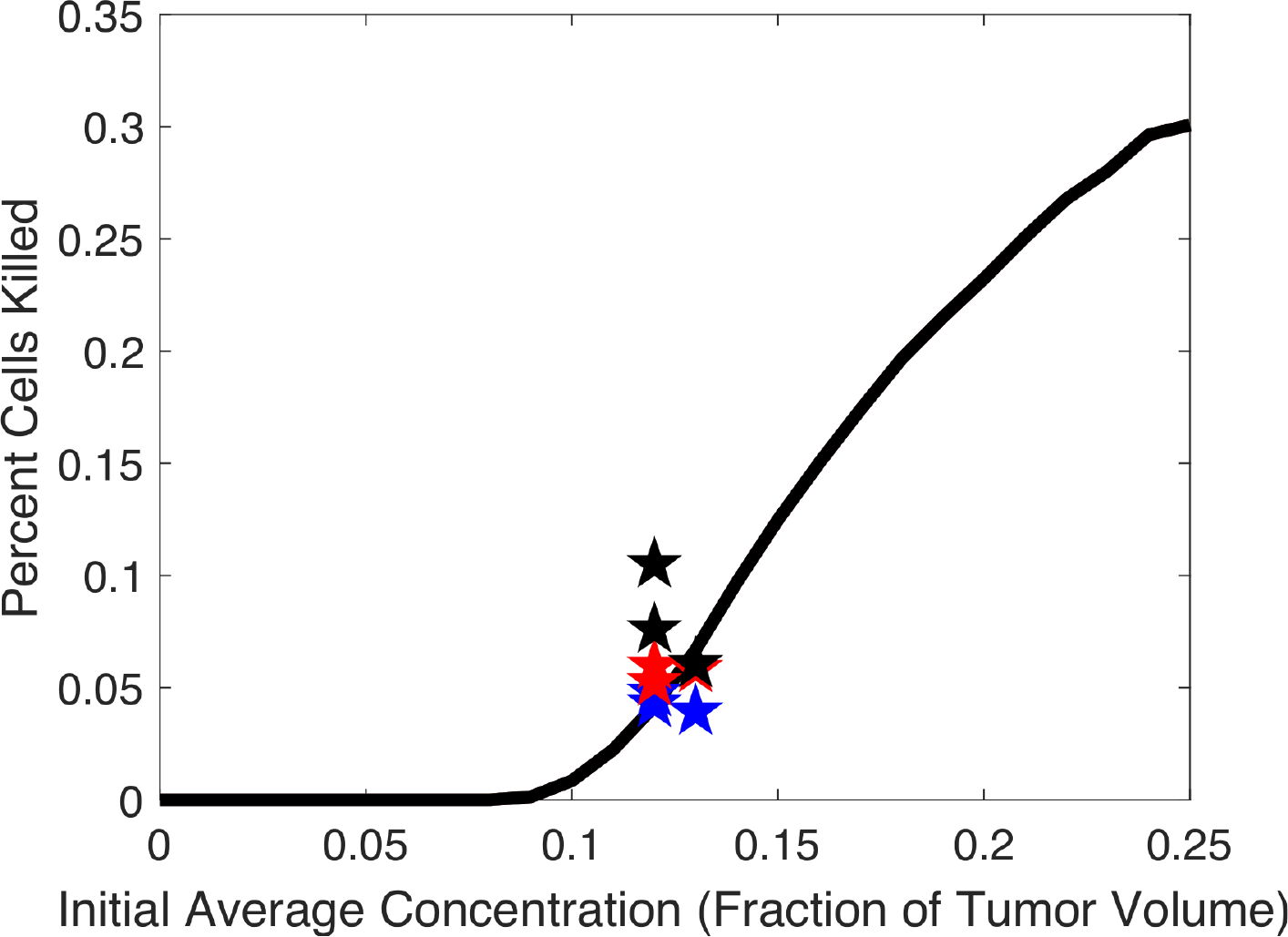}
\includegraphics[width = 0.19\textwidth]{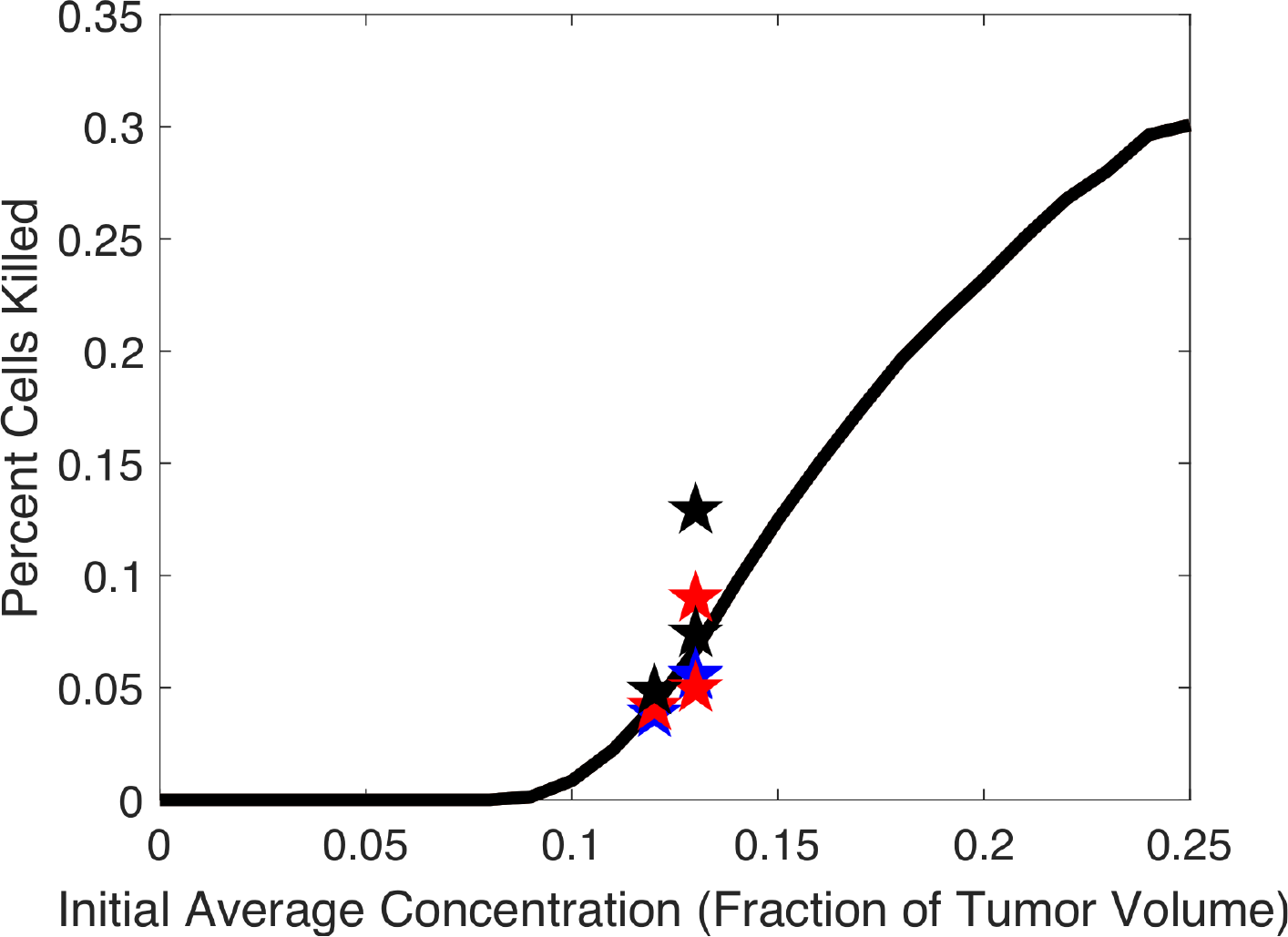}

\bigskip
\stackinset{l}{-10mm}{t}{1mm}{\textbf{\small \rotatebox{90}{M27-mel}}}{\includegraphics[width = 0.19\textwidth]{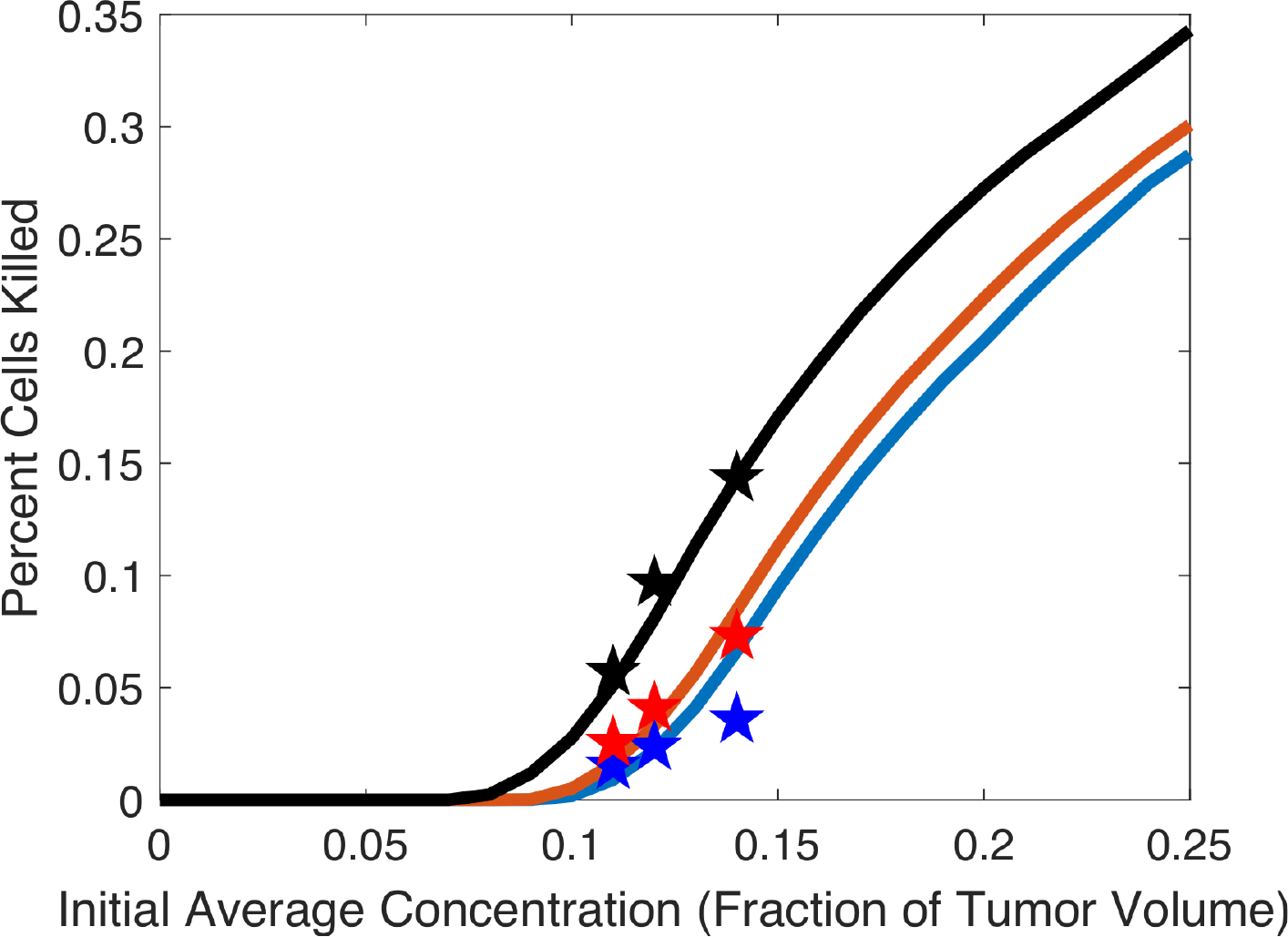}}
\includegraphics[width = 0.19\textwidth]{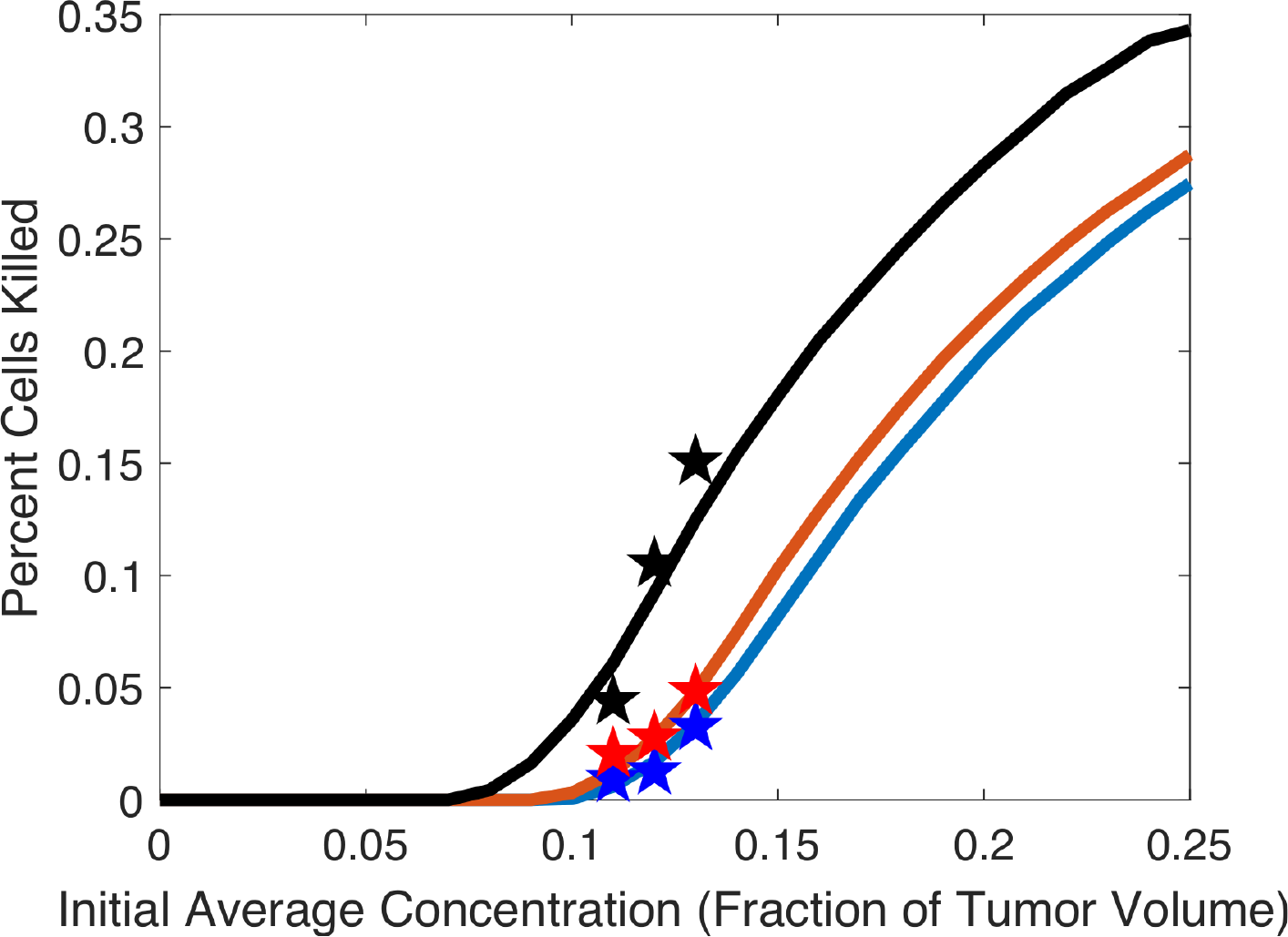}
\includegraphics[width = 0.19\textwidth]{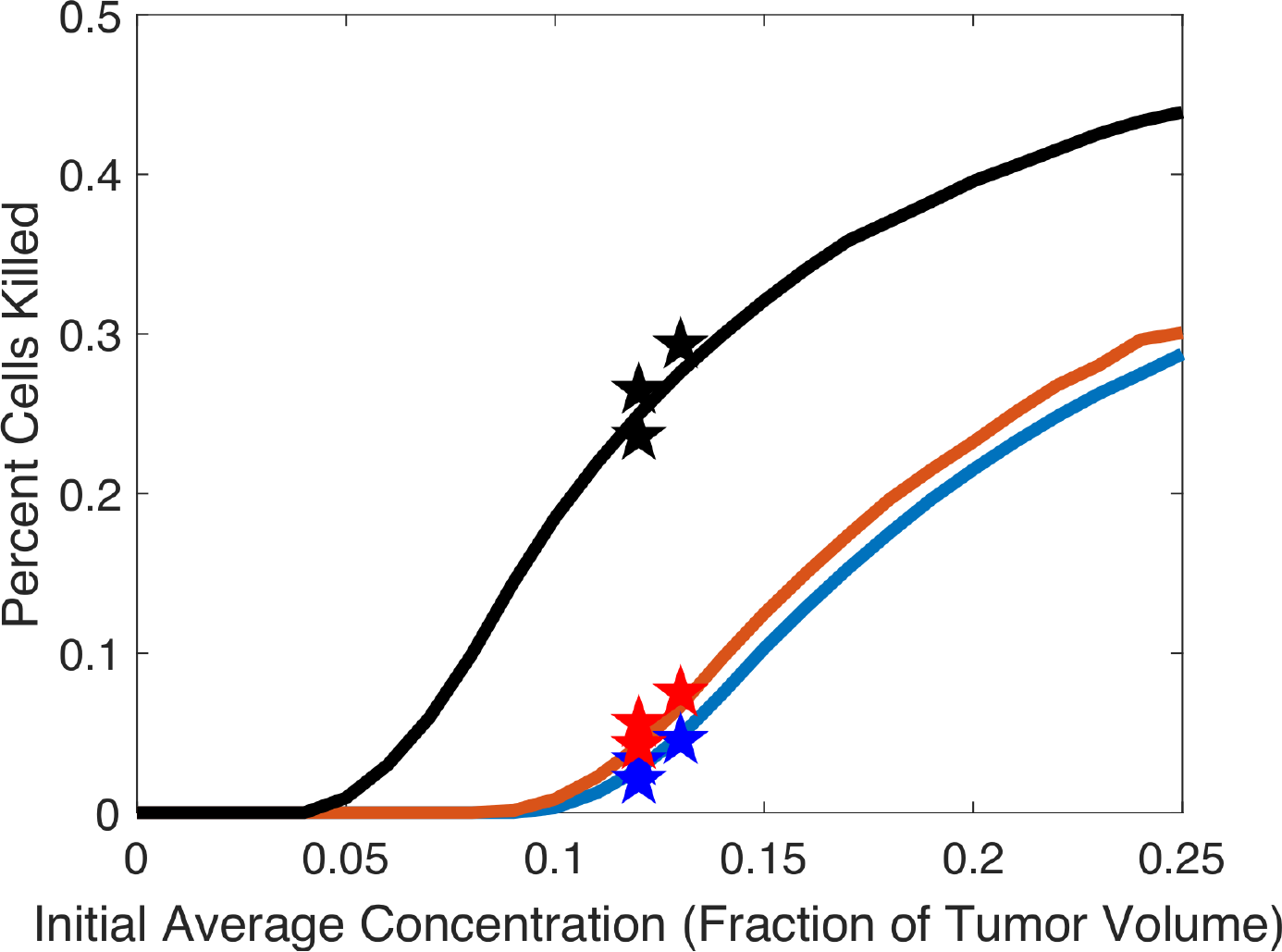}
\includegraphics[width = 0.19\textwidth]{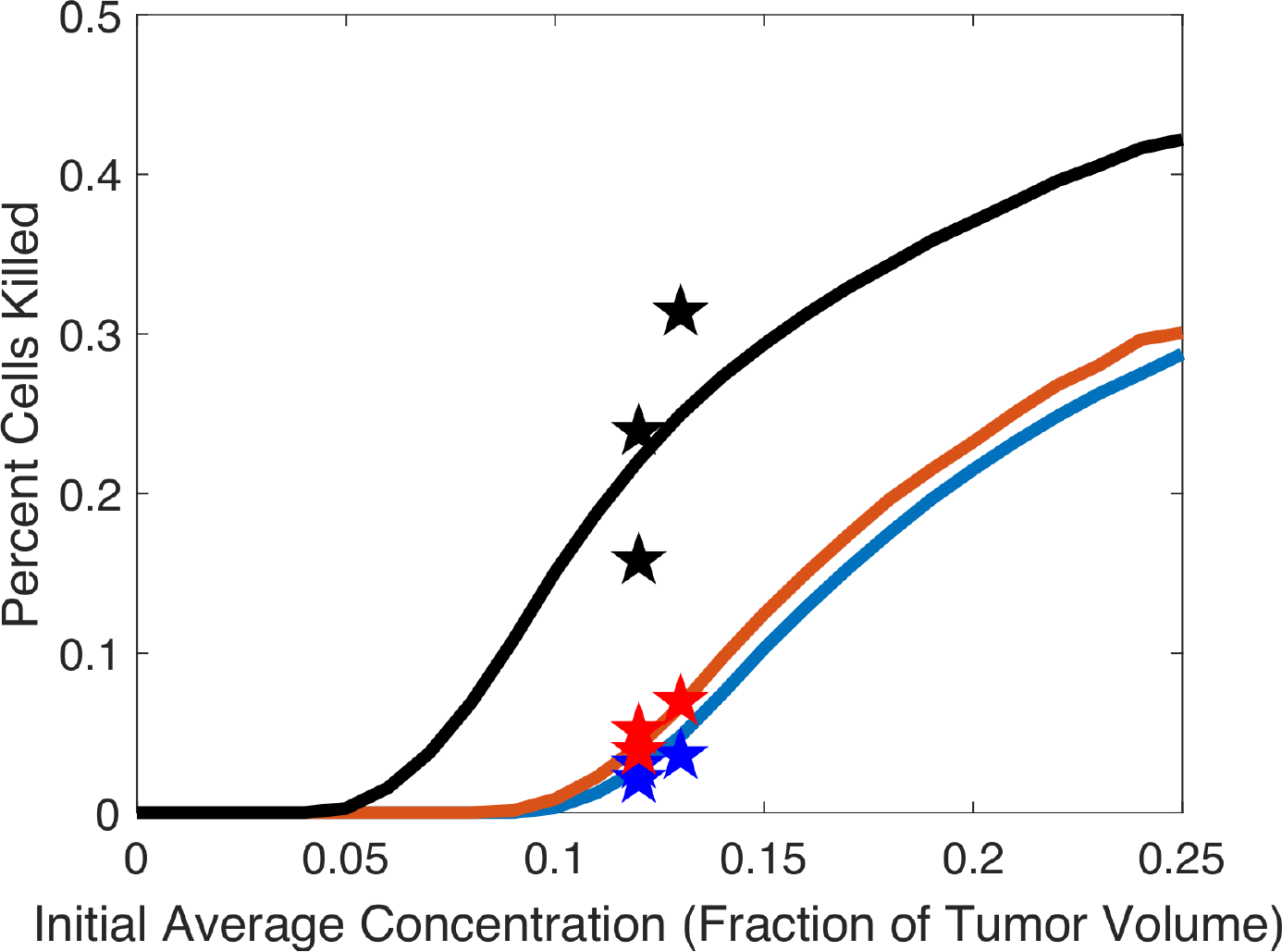}
\includegraphics[width = 0.19\textwidth]{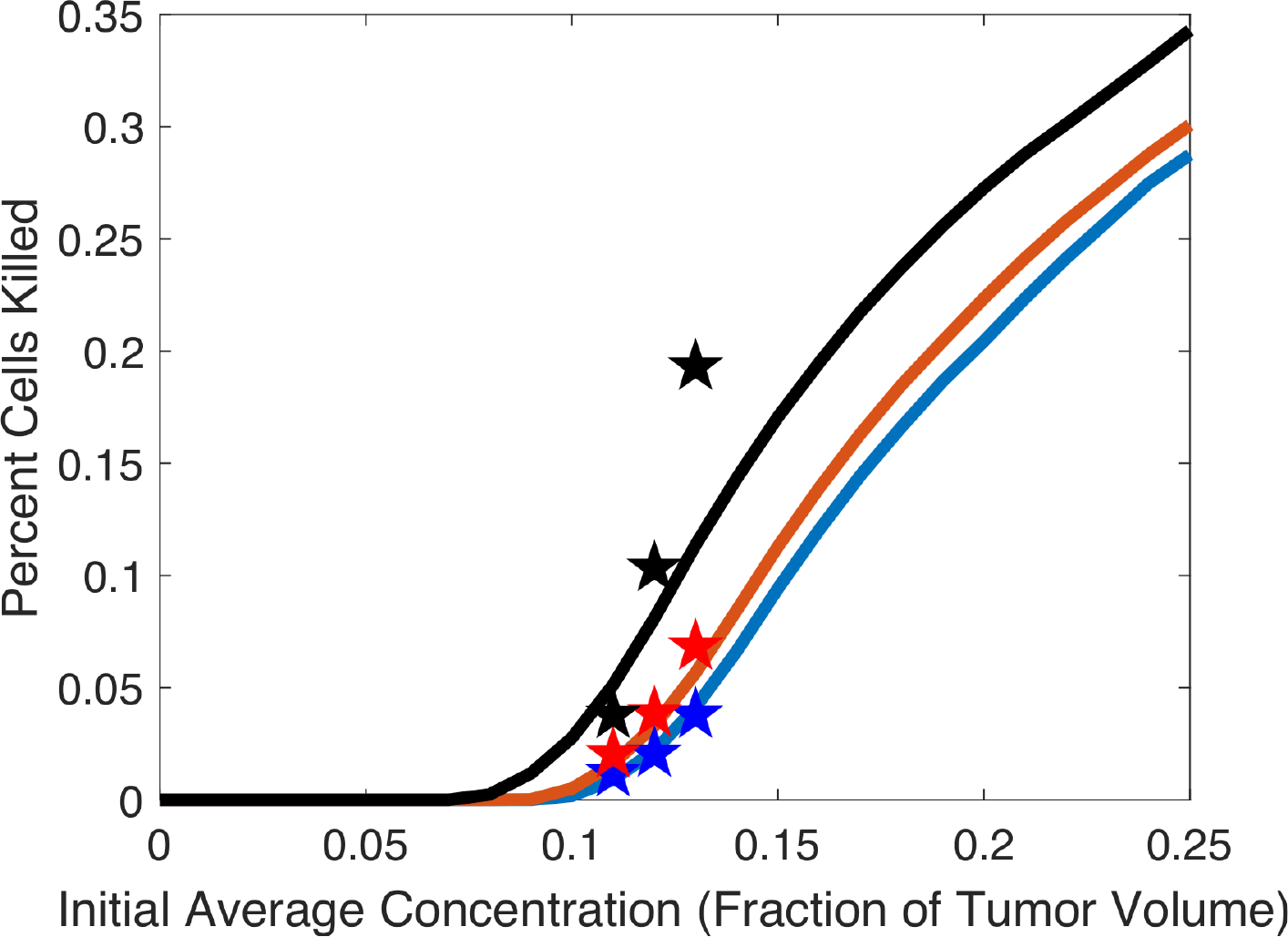}

\caption{Matrix of dose-response curves from the entire data set.  The columns denote the drugs
and the rows denote the cell lines for the data pairs.  Dose-response curves are produced in accordance
with the techniques delineated in Sec. \ref{Sec: Comparison}.}
\label{Fig: DoseResponseMatrix}
\end{figure}

Here we present the remainder of the dose-response plots from Sec. \ref{Sec: Comparison}.
For each of the curves, an initial dose of $1 \mu\text{M}$ is mapped to an initial concentration
of $1/12$ tumor volume.  Equation 
\ref{Eq: time-threshold} gives us the relation between the three time points (24, 48, and 72 hours) and the
thresholds ($u_{24}$, $u_{48}$, and $u_{72}$).  These thresholds are then used to produce the
simulated dose-response curves in Fig. \ref{Fig: DoseResponseMatrix}.  Then
the two other initial doses are matched by varying the dose
until a minimum distance between two of the time points and the curves is achieved.  We choose to match
two time points since it is evident that the data set is quite noisy and it has outliers that do not obey the
Hill equation \cite{Hill1910}.

\section{Confidence bands}\label{Appendix: Confidence bands}

\begin{figure}[htbp]
\stackinset{l}{-10mm}{t}{16mm}{\textbf{\small \rotatebox{90}{C32}}}{\stackinset{l}{7mm}{t}{-10mm}{\textbf{\small SB590885}}{\includegraphics[width = 0.19\textwidth]{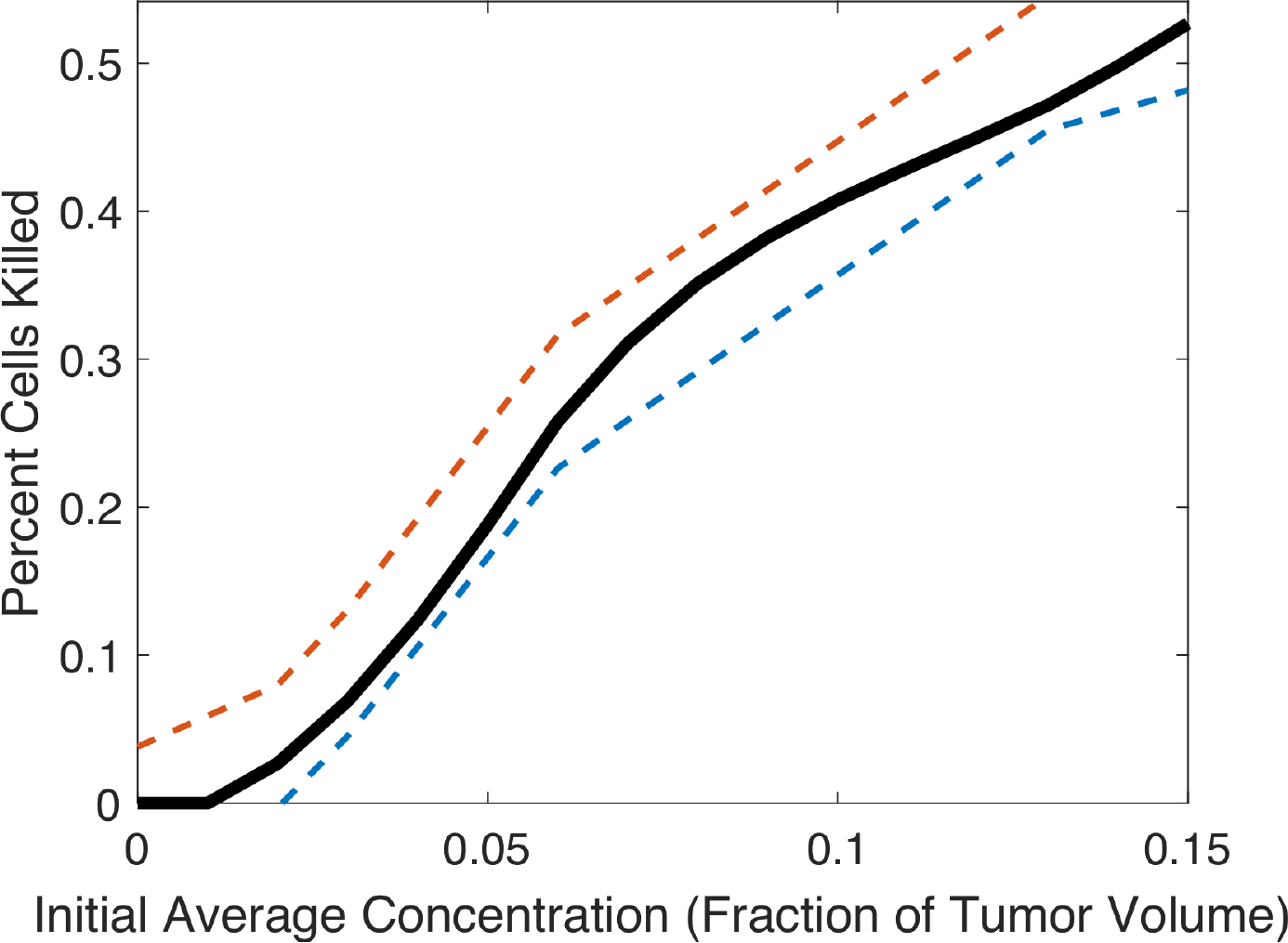}}}
\stackinset{l}{7mm}{t}{-10mm}{\textbf{\small PLX-4720}}{\includegraphics[width = 0.19\textwidth]{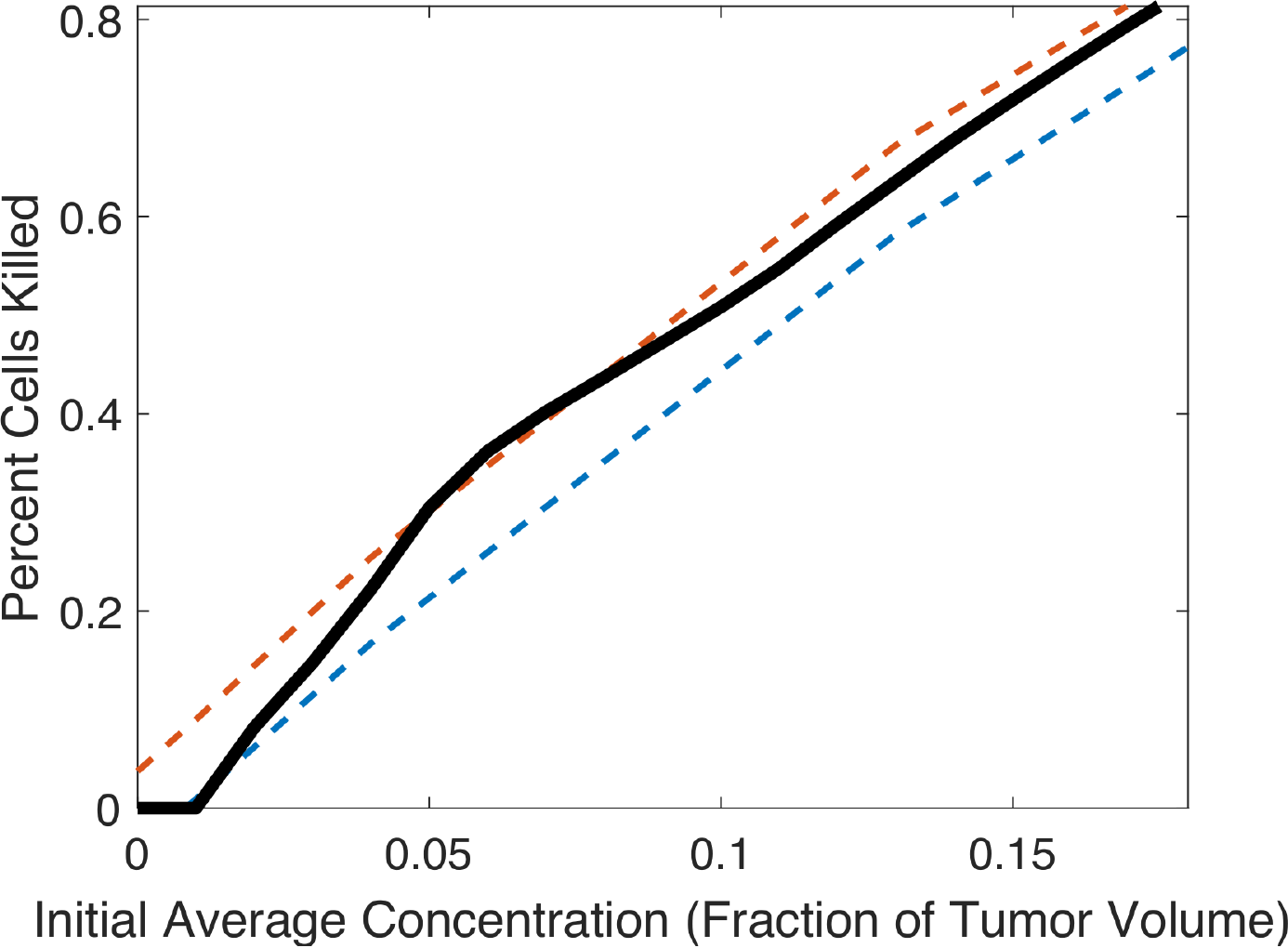}}
\stackinset{l}{9mm}{t}{-10mm}{\textbf{\small AZ-628}}{\includegraphics[width = 0.19\textwidth]{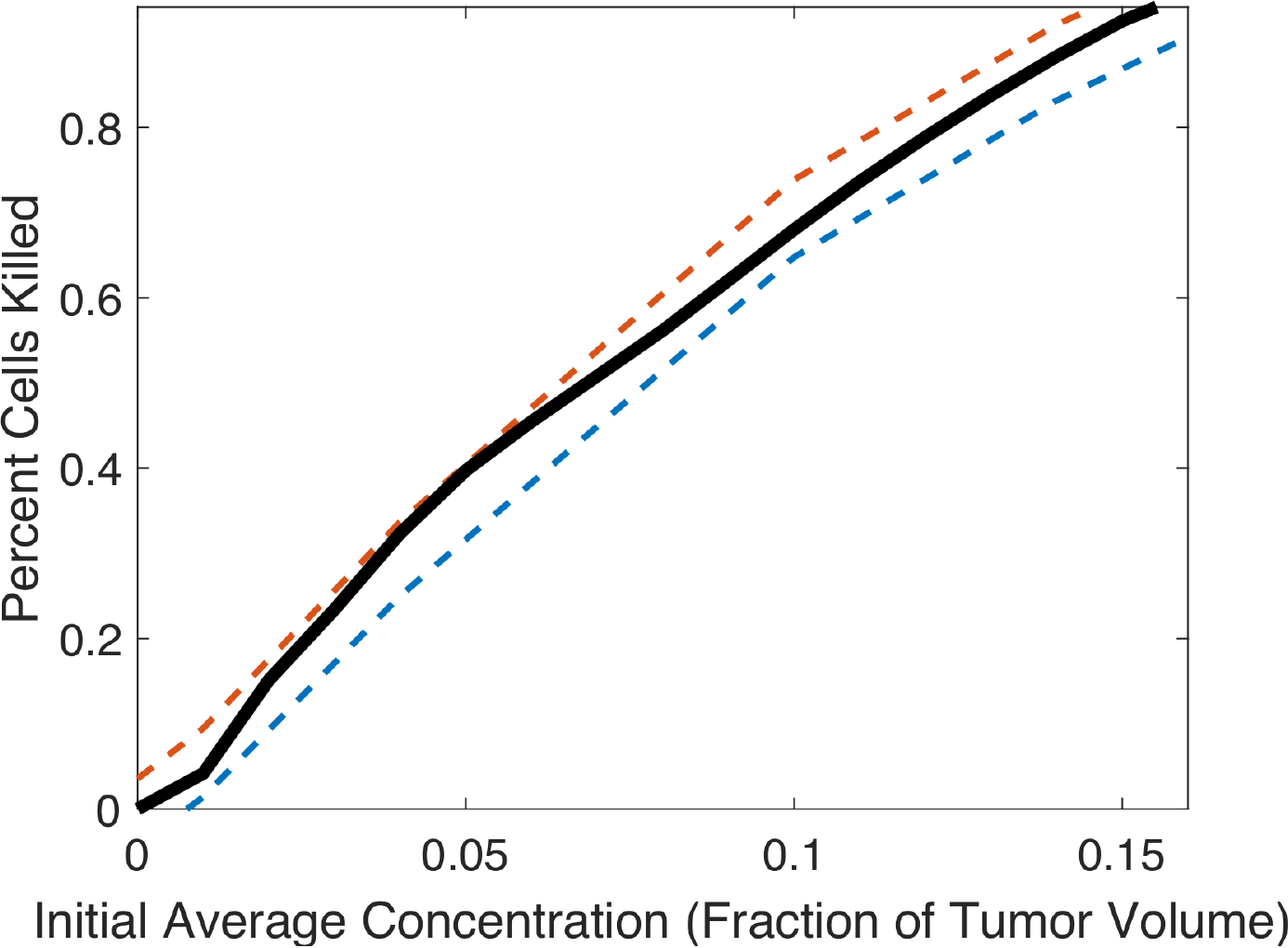}}
\stackinset{l}{4mm}{t}{-10mm}{\textbf{\small Selumetinib}}{\includegraphics[width = 0.19\textwidth]{ConfBand4}}
\stackinset{l}{4mm}{t}{-10mm}{\textbf{\small Vemurafenib}}{\includegraphics[width = 0.19\textwidth]{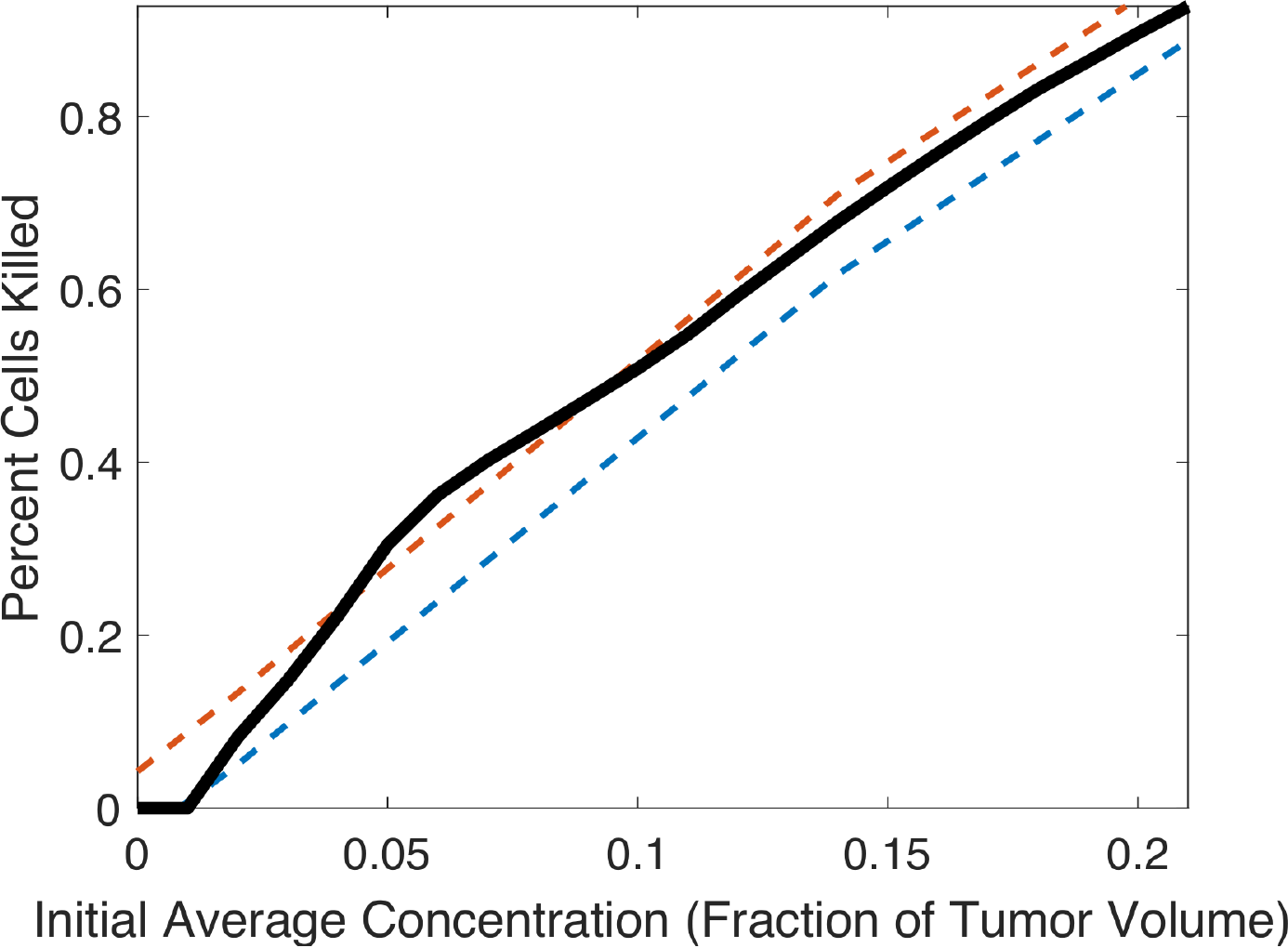}}

\bigskip
\stackinset{l}{-10mm}{t}{}{\textbf{\small \rotatebox{90}{COLO 858}}}{\qquad\qquad\qquad\qquad \includegraphics[width = 0.19\textwidth]{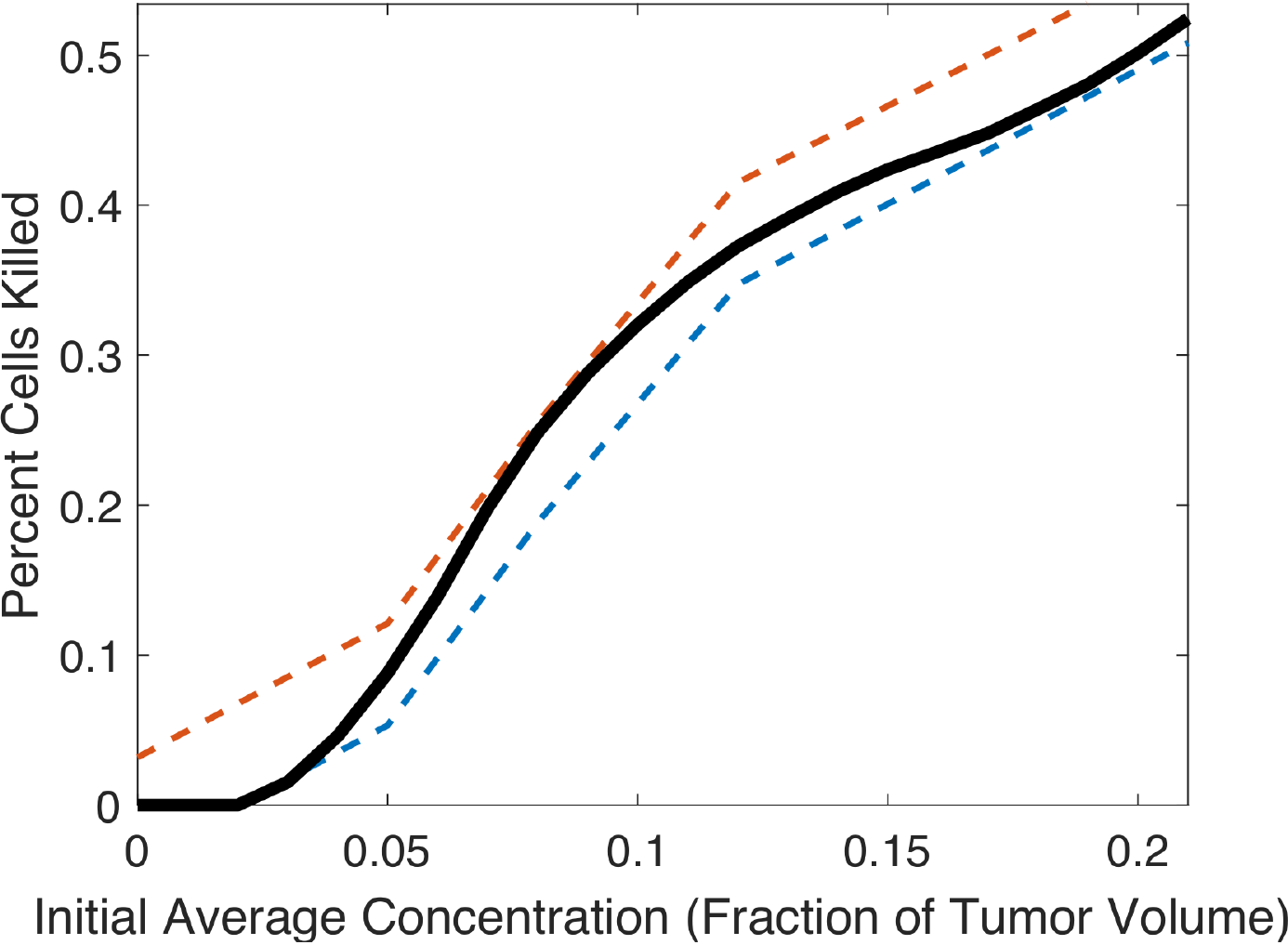}}
\includegraphics[width = 0.19\textwidth]{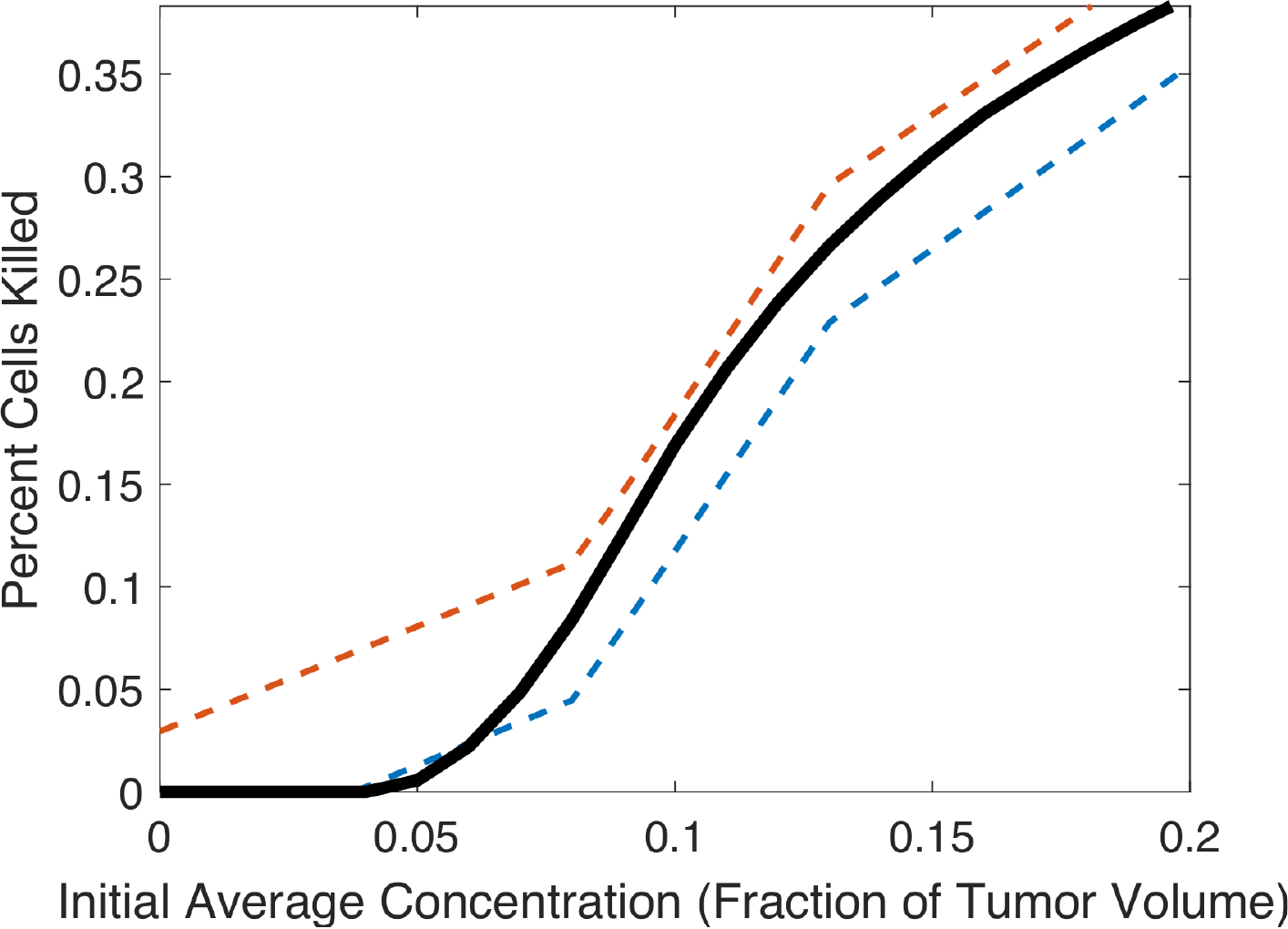}

\bigskip
\stackinset{l}{-10mm}{t}{1mm}{\textbf{\small \rotatebox{90}{WM-115}}}{\includegraphics[width = 0.19\textwidth]{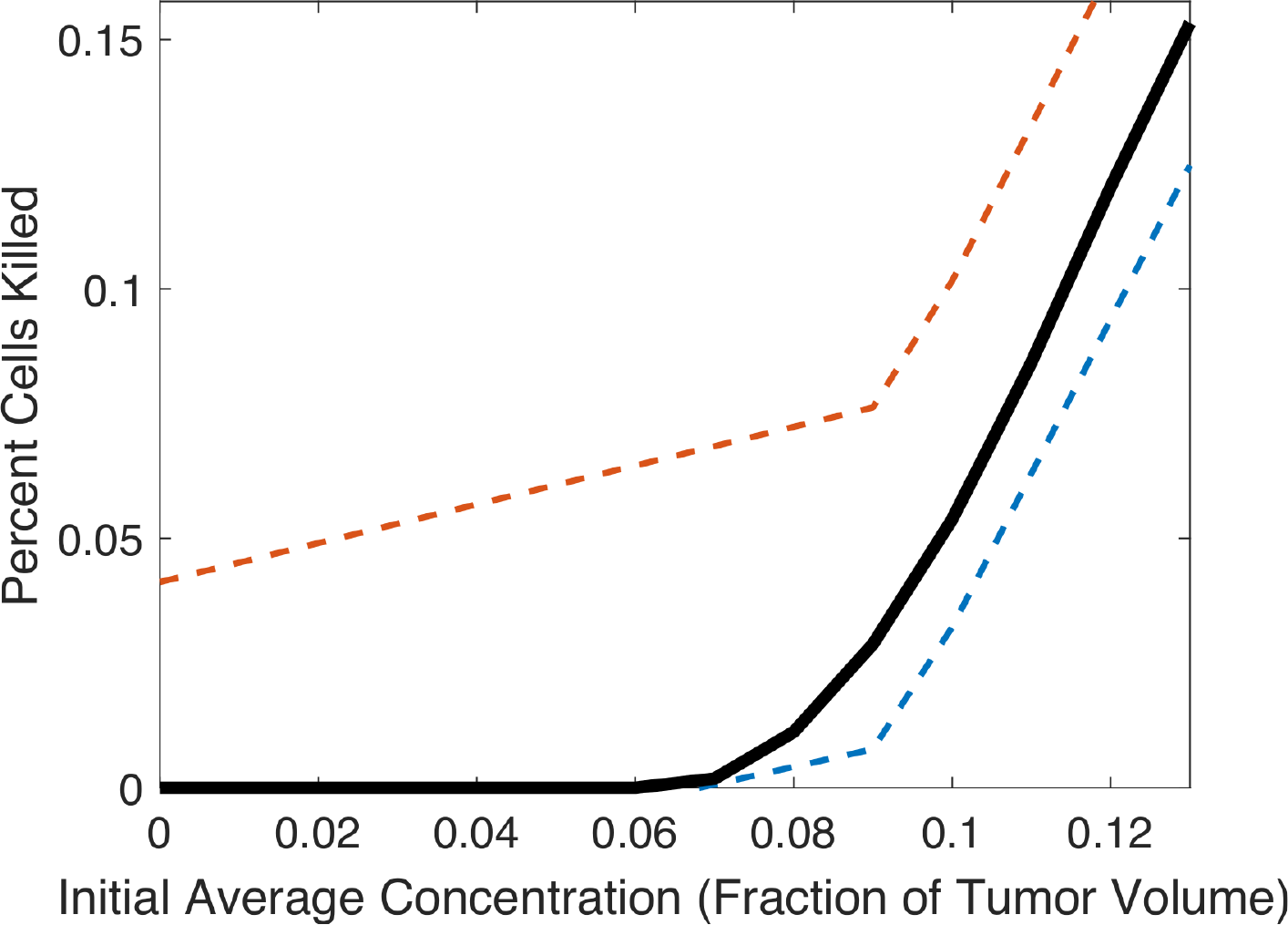}}
\qquad\qquad\qquad\qquad \includegraphics[width = 0.19\textwidth]{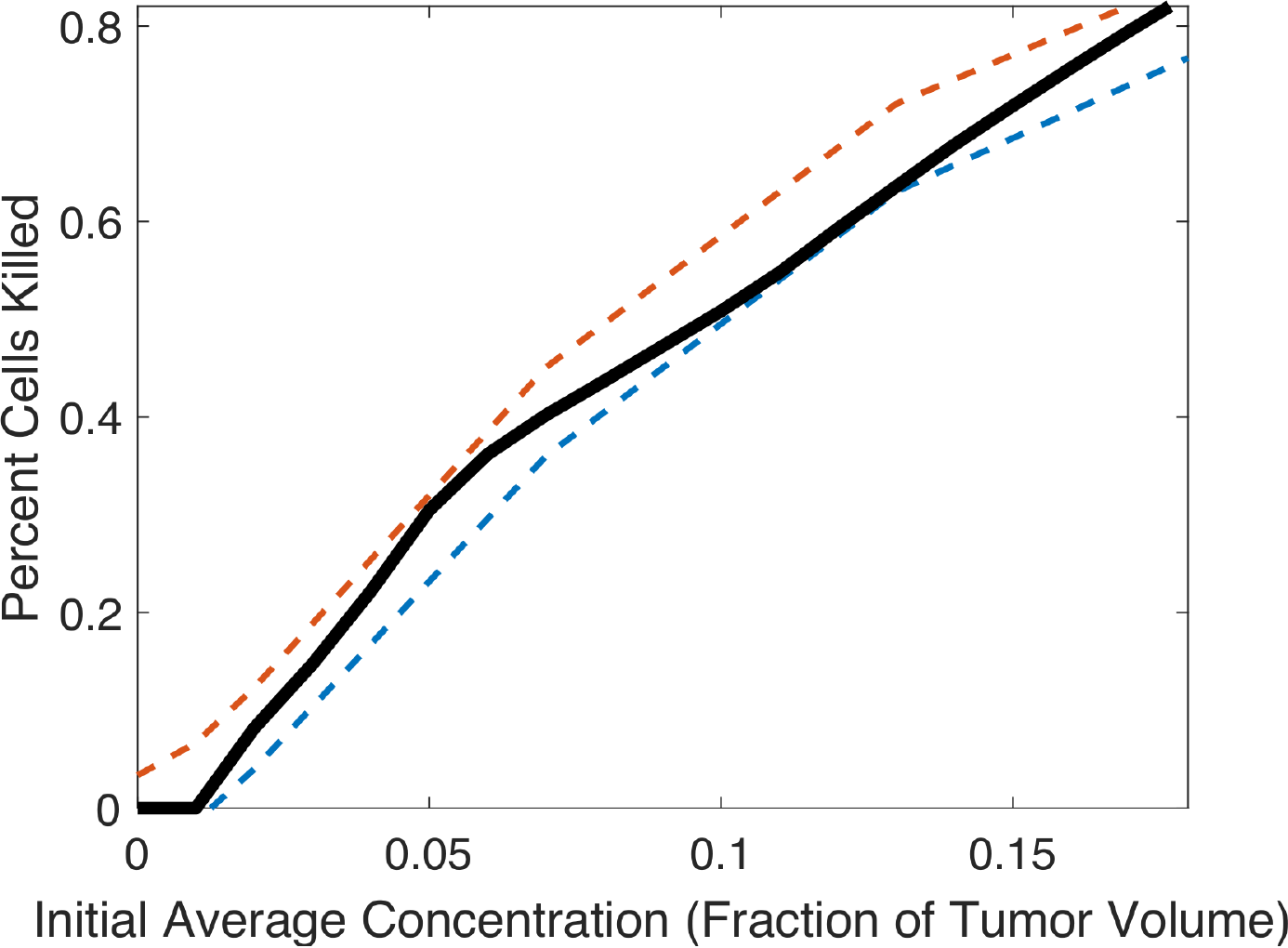}

\bigskip
\stackinset{l}{-10mm}{t}{1mm}{\textbf{\small \rotatebox{90}{M27-mel}}}{\qquad\qquad\qquad\qquad\includegraphics[width = 0.19\textwidth]{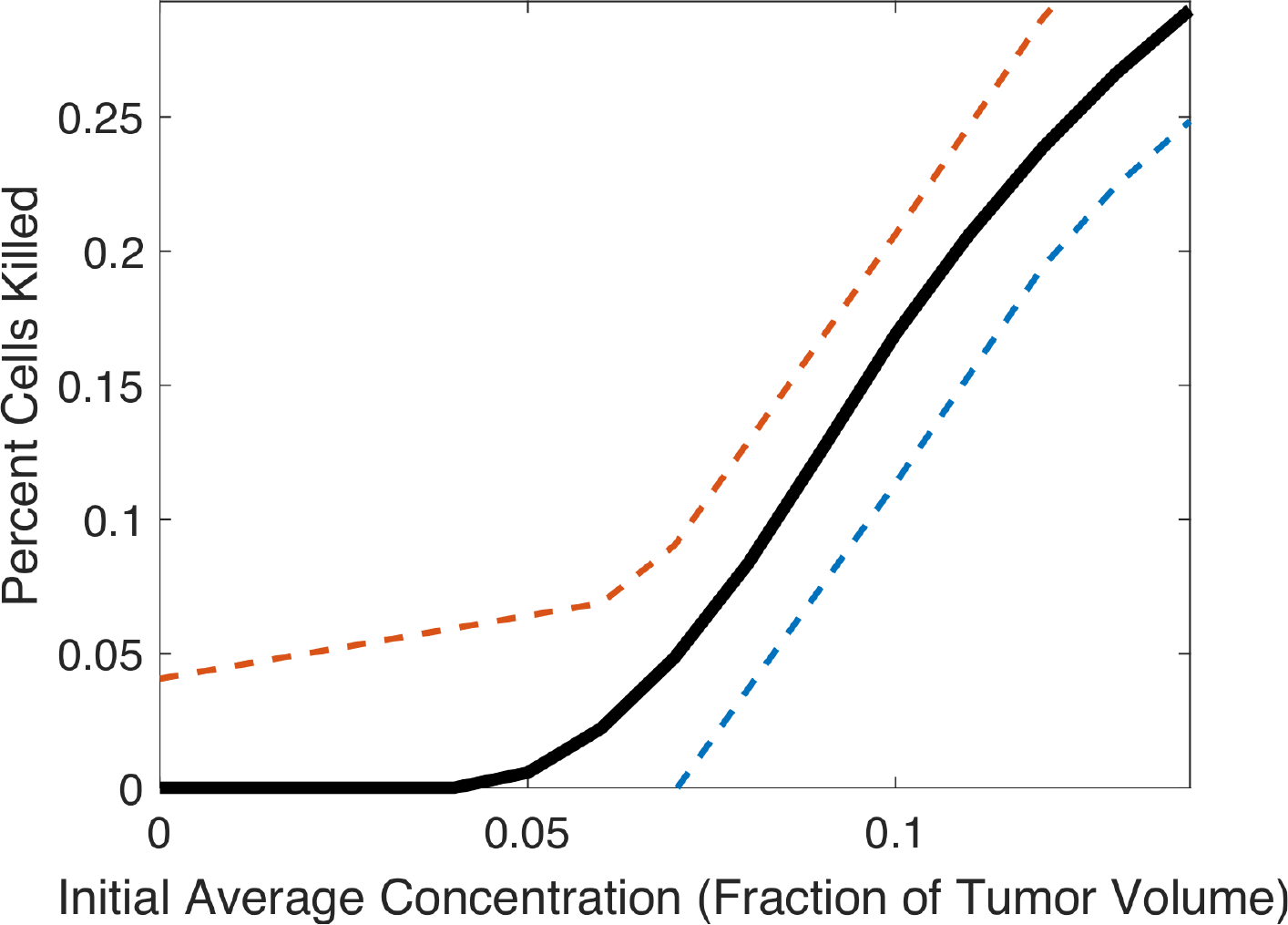}}
\includegraphics[width = 0.19\textwidth]{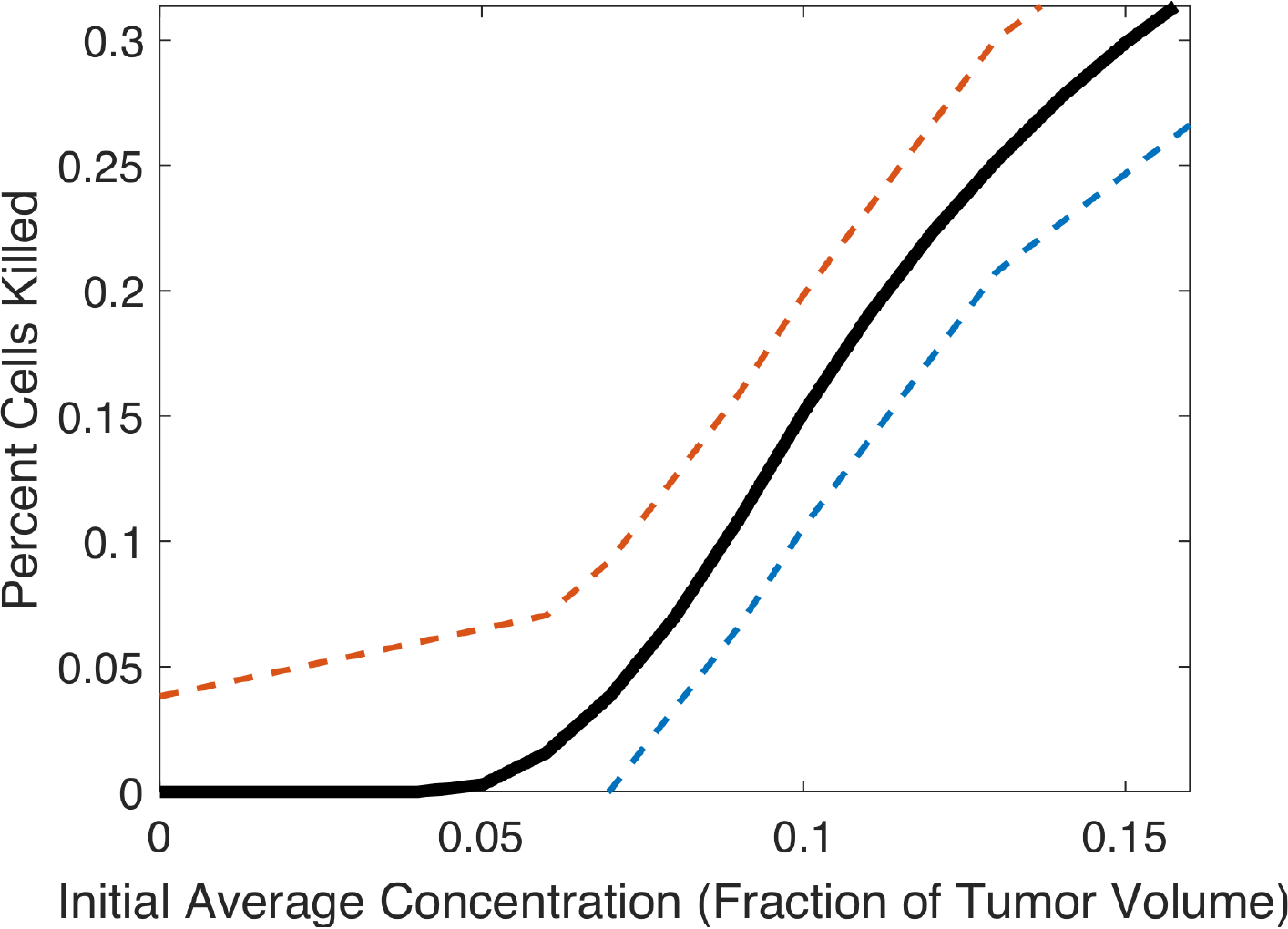}
\includegraphics[width = 0.19\textwidth]{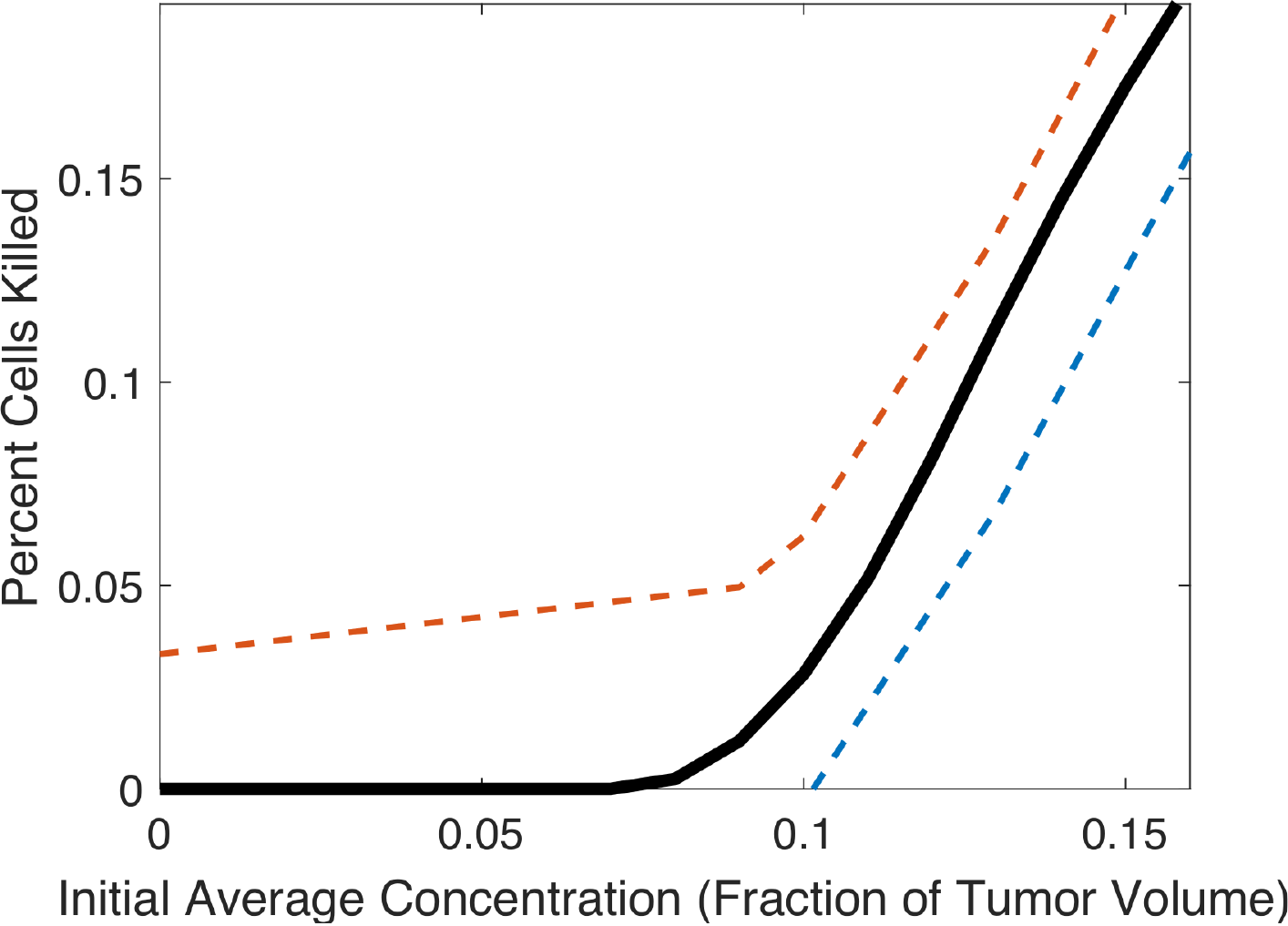}

\caption{Matrix of dose-response curves within $95\%$ confidence bands.  The columns denote the drugs
and the rows denote the cell lines for the data pairs.  Dose-response curves are produced in accordance
with the techniques delineated in Sec. \ref{Sec: Comparison}.}
\label{Fig: ConfBandMatrix}
\end{figure}

In this appendix, we present the remainder of the confidence band plots from Sec. \ref{Sec: Comparison}.
For each of the curves, an initial dose of $1 \mu\text{M}$ is mapped to an initial concentration
of $1/12$ tumor volume.  Equation 
\ref{Eq: time-threshold} gives us the relation between the three time points (24, 48, and 72 hours) and the
thresholds ($u_{24}$, $u_{48}$, and $u_{72}$).  Then three other initial doses are matched by varying the dose
until a minimum distance between the 72 hour time point and the curves is achieved.  We choose to study
the 72 hour time points since the least amount of noise is observed at this exposure time and therefore
more closely matches the Hill equation \cite{Hill1910}.  We need one extra point in order to fit a four
parameter sigmoidal curve to the empirical data points.  Notice that at an initial dose of zero the response
must be around zero, and hence we use that point as well.  Using a bootstrap scheme, fifty sigmoidal curves
fitting the data, with a prescribed probability spread at each dose point, are produced (shown in Fig. \ref{Fig: ConfBand_Illustration} in Sec. \ref{Sec: Comparison}).  These curves give us an artificial replication study.
From the curves we find the 95\% confidence intervals at each matched dose point, and connect a piece-wise
linear curve to the lower and upper quantiles of these intervals, which yields a confidence band.  Then the dose-response
curves from the model are plotted.  Since the data set is quite noisy for low initial doses, not every data pair
would fit a sigmoidal curve, but many did in fact work as shown in Fig. \ref{Fig: ConfBandMatrix}.

\end{document}